\documentclass[11pt,preprint]{aastex}

\usepackage{booktabs,graphicx,dcolumn,amsmath,paralist,mhchem,longtable}
\usepackage{dcolumn,xcolor,braket,cleveref,rotating}

\newcommand{\abinitio}{\emph{ab initio}}
\newcommand{\Duo}{{\sc Duo}}

\newcommand{\Marvel}{{\sc Marvel}}
\newcommand{\TiO}{\ce{^{48}Ti^{16}O}}

\newcommand{\cm}{cm$^{-1}$}

\newcommand{\X}{X $^3\Delta$}
\newcommand{\E}{E $^3\Pi$}
\newcommand{\D}{D $^3\Sigma^-$}
\newcommand{\A}{A $^3\Phi$}
\newcommand{\B}{B $^3\Pi$}
\newcommand{\C}{C $^3\Delta$}

\usepackage{array}
\newcolumntype{H}{>{\setbox0=\hbox\bgroup}c<{\egroup}@{}}
\newcommand{\Sa}{a $^1\Delta$}
\newcommand{\Sd}{d $^1\Sigma^+$}
\newcommand{\Se}{e $^1\Sigma^+$}
\newcommand{\Sb}{b $^1\Pi$}
\newcommand{\SGamma}{g $^1\Gamma$}

\newcommand{\Sc}{c $^1\Phi$}
\newcommand{\Sf}{f $^1\Delta$}

\newcommand{\mc}{\multicolumn}
\newcolumntype{d}{D{.}{.}{-1}}

\begin{document}

\title{\Marvel\ analysis of the measured high-resolution rovibronic spectra of \TiO }
\author{Laura K. McKemmish,$^1$ Thomas Masseron,$^2$ \\
Samuel Sheppard,$^3$ Elizabeth Sandeman,$^3$ Zak Schofield,$^3$ \\
Tibor Furtenbacher,$^4$ Attila G. Cs\'asz\'ar,$^4$ \\ Jonathan Tennyson,$^1$  Clara Sousa-Silva.$^1$}
\affil{$^1$Department of Physics and Astronomy, University College London, London, WC1E 6BT, UK \\
$^2$Institute of Astronomy, University of Cambridge, Madingley Road, Cambridge, CB3 0HA, UK \\
$^3$Highams Park School, Handsworth Avenue, Highams Park, London, E4 9PJ, UK \\
$^4$Institute  of Chemistry, 
Lor\'and E\"otv\"os University and MTA-ELTE Complex Chemical Systems Research Group,
H-1518 Budapest 112, Hungary}
\email{laura.mckemmish@gmail.com}

\date{\today}

\begin{abstract}
 Accurate, experimental rovibronic energy levels, with associated labels and 
uncertainties, are reported for 11 low-lying electronic states of the diatomic 
\TiO\ molecule, determined using the \Marvel\ (Measured Active 
Rotational-Vibrational Energy Levels) algorithm. All levels are based on lines 
corresponding to critically reviewed and validated high-resolution experimental 
spectra taken from 24 literature sources. The transition data are in the 2 $-$ 
22,160 \cm{} region. Out of the 49,679 measured transitions, 43,885 are 
triplet-triplet, 5710 are singlet-singlet and 84 are triplet-singlet 
transitions.  A careful analysis of the resulting experimental spectroscopic 
network (SN) allows 48,590 transitions to be validated. 
 The transitions determine 93 vibrational band origins of 
 \TiO\, including 71 triplet and 22 singlet ones.  There are 276 (73) 
triplet-triplet (singlet-singlet) band-heads derived from \Marvel\ experimental 
energies, 123 (38) of which have never been assigned in low or high resolution 
experiments. The highest $J$ value, where $J$ stands for the total angular 
momentum, for which an energy level is validated is 163. The number of 
experimentally-derived triplet and singlet \TiO\ rovibrational energy levels is 
8682 and 1882, respectively. The lists of validated lines and levels for  \TiO\  
are deposited in the Supporting Information to this paper. 
\end{abstract}


\section{Introduction}
Currently, any in-depth discussion on molecular data requirements with 
astronomers working on cool stars or hot Jupiter exoplanets highlights one 
molecule: TiO \citep{15HoDeSn.TiO,16FoRoDo.exo,jt631}. TiO is the major near-infrared (IR) and visible absorber in M-type stars 
\citep{00AlHaS1.TiO,02Loxxxx.TiO} and, potentially, hot Jupiter exoplanets 
\citep{08DeVide.TiO}. Despite line lists from the late twentieth century generated by  
\citet{75Collinsthesis.TiO}, \citet{76CoFaxx.TiO}, \citet{92Plxxxx.TiO}, 
\citet{94Joxxxx.TiO}, \citet{98Scxxxx.TiO} and  \citet{98Plxxxx.TiO}, and the recent VALD updates \citep{VALD3}, the new 
very high resolution observations, e.g., of exoplanetary atmospheres, cannot 
usually be modelled sufficiently accurately \citep{15HoDeSn.TiO}.

Exoplanets provide two major topical applications of high quality spectroscopic data for 
TiO. 

First, detecting potentially habitable Earth-sized exoplanets using transits is 
expected to be easier around M-dwarf stars than other stellar hosts due to the 
higher transit depth and faster transit times. However, characterising these 
planets requires high accuracy modelling of M-dwarf stellar spectra, which is 
significantly complicated by the strong molecular absorption of these cooler 
stars \citep{jt143,00AlHaS1.TiO}. Compared to main-group closed-shell molecules  like \ce{H2O} and \ce{CO}, the 
spectra of  transition metal diatomic species such as 
TiO  are significantly less well determined by either experimental or theoretical studies \citep{jt632}. 
In particular,  high accuracy spectral 
modelling requires a thorough and accurate analysis of experimental data.

Second, TiO opacity is  expected to be very important in modelling hot Jupiter 
exoplanets without clouds \citep{08FoLoMa.TiO}. 
However, due to the tidal interaction with their respective stars, there can be 
large differences in the day and night temperatures in hot Jupiters, giving rise 
to extreme conditions. This suggests that cloud cover is abundant on hot Jupiters, 
a supposition supported by observations \citep{15NiSiBu.TiO,16SiFoNi.TiO}. 
Thus far studies of the 
presence of TiO in hot Jupiter exoplanets have given mixed results. Evidence for TiO  on 
WASP-121b was reported by \citet{16EvSiWa.TiO}. Likely absence of TiO on WASP-19b was reported  
by \citet{13HuSiPo.TiO} and on WASP-12b by \citet{13SiLeFo.TiO}. 
It is predicted that the presence of TiO/VO in the atmospheres of hot Jupiter 
exoplanets is likely to cause a thermal inversion in the atmosphere 
\citep{16EvSiWa.TiO}; \citet{15HaMaMa.TiO} present an HST (Hubble Space Telescope) 
spectrum of WASP-33b consistent with emission from TiO. HST has been used to 
perform almost all of these observations; the upcoming launch of JWST (James 
Webb Space Telescope) will significantly increase the quality of the observed 
spectra. It is imperative to ensure that the quality of the available TiO 
line list is sufficiently high to allow these new spectra to be used optimally. Furthermore, the use of cross-correlation techniques
allows ground-based telescopes to detect molecules \citep{14deBiBr.TiO}.  The inaccuracies in current 
TiO line lists prevent the use of this technique for  TiO 
\citep{15HoDeSn.TiO}. 

 Historically, the detection of TiO in M-giants by \citet{1904Fowler.TiO} was 
one of the earliest molecular detections in stellar astrophysics, predating modern 
quantum mechanics. The very high experimental interest in this, from a chemical 
perspective, unusual molecule over the last century, as documented thoroughly in 
this manuscript \textbf{(\Cref{tab:datasources,tab:databandheads,tab:intensities,tab:otherrefs}, see below)}, is a 
direct consequence of this early identification in stellar bodies. TiO,
together with \ce{C2} \citep{jt637}, has provided a major
motivating factor for the development of theory and methods in the field of 
rovibronic spectroscopy. 
The references collated in this paper tell a fascinating story of how scientists 
tackled the complexity of transition metal diatomic spectra without 
significant computational power and thus without accurate \abinitio\ 
predictions. Questions like whether the singlet or triplet state was the true 
ground state did  not have obvious answers. The triplet ground state was mis-identified twice \citep{29Lowater.TiO,51Phxxxx.TiO} before finally being assigned 
correctly as \X{} by \citet{69Phxxxx.TiO}.The dominant electronic configuration of the \X{} ground electronic state can be written as $(core)(9\sigma)^1(1\delta)^1$, where $9\sigma$ and $1\delta$ are essentially the $4s$ and $3d$ orbitals of \ce{Ti^{2+}}, respectively. The singlet-triplet gap was estimated,
e.g., by \citet{52Phxxxx.TiO}, then eventually measured using formally spin-forbidden 
transitions first by \citet{83KoKuGu.TiO} and then more accurately by 
\citet{95KaMcHe.TiO}. This manuscript considers and collates all the available and 
assigned TiO experimental spectroscopic frequency data. We then use the Measured Active 
Rotational-Vibrational Energy Levels (\Marvel) 
algorithm  \citep{jt412,07CsCzFu.marvel,12FuCsi.method}, described in detail below, to extract the highest accuracy collation 
of TiO rovibronic energy levels ever produced. The experimentally-derived energy levels are all given 
uncertainties. The procedure is active in that future experimental data can be added to the collation and used 
to produce updated experimentally-derived energy levels in a  straightforward 
manner. 

\section{Theory}
\subsection{\Marvel}

The \Marvel\ approach \citep{jt412,07CsCzFu.marvel,12FuCsi.method} is a sophisticated methodology that allows 
extraction of experimental energy levels, and associated uncertainties, from a 
(usually large) set of experimental transition frequencies.
The methodology is similar to traditional approaches based on the Ritz principle, such as 
`combination differences', but is a more sophisticated, computational, 
near-black-box approach. The \Marvel\ program takes as input formatted assigned 
transitions. The program then constructs the experimental spectroscopic networks (SNs) 
\citep{11CsFuxx.marvel,12FuCsxx.marvel,14FuArMe.marvel,16ArPeFu.marvel,16CzFuAr.marvel} which contains all inter-connected transitions. For each 
SN, the assigned transition data is then inverted to find the 
energy levels. The uncertainties of the transition frequencies weight this 
inversion process using a robust reweighting procedure advocated by \citet{Watson03} allowing \Marvel\ to yield the uncertainty of each extracted 
energy level. 
For a detailed description of the approach, algorithm and program, we refer 
readers to \citet{12FuCsi.method}. \Marvel\ was originally developed and
used by an IUPAC Task Group (TG) studying water spectra \citep{jt562} and
applied to various water isotopologues \citep{jt454,jt482,jt539,jt576}. The energy
levels these studies yielded will provide the major source of water transition 
frequencies in the upcoming 2016 update of HITRAN \citep{HITRAN2016}. 
The naming convention for data sources employed here follows the one proposed
by this IUPAC TG.
Other
molecules for which rovibrational energy levels have been determined using \Marvel\ include 
H$_3^+$ \citep{13FuSzMa.marvel}, H$_2$$^{12}$C$^{12}$C$^{16}$O \citep{11FaMaFu.marvel}, H$_2$D$^+$ and D$_2$H$^+$ \citep{13FuSzFa.marvel} and $^{14}$NH$_3$ \citep{jt608}.
The only previous use of \Marvel\ for rovibronic spectra is the 
recently published analysis of \ce{^{12}C2} \citep{jt637}.

 The \Marvel\ software takes as input assigned, measured
transitions, with estimated uncertainties, and outputs assigned energy levels 
together with recommended uncertainties. However, often there is no consistent 
set of energy levels that produce the input transitions within the estimated 
uncertainties. This can occur due to typographic or digitisation errors, 
mis-assignments and under-estimated uncertainties for the transitions. For this reason, the master 
list of \Marvel\ input transitions should be gradually increased with issues 
resolved as new transitions are added to the master file. \Marvel\ produces new 
recommended uncertainties. If these are less than twice the original 
uncertainties, we generally adopt these recommended uncertainties. 
If there is a 
very large difference in the recommended uncertainty, we look for typographic 
and digitisation errors; if none are found, we then assume mis-assignment and 
put a negative in front of the transition frequency, thus retaining the data but 
not utilising it in the \Marvel\ algorithm for future runs. Transitions 
initially discarded in this way can be reconsidered later in the process. For 
each band in each experimental source, we track the number of validated 
transitions (i.e., transitions for which all extracted energies of the full data 
set are consistent) against the number of total input transitions as well as the 
minimum, average and maximum uncertainty of transition frequencies. The minimum 
uncertainty is usually our initial input uncertainty based on the original 
experimental paper (or our best educated guess) as the current \Marvel\ code can 
automatically increase uncertainties, but not reduce them. Generally, if we find that the 
average uncertainty is significantly higher than the minimum uncertainty, we 
increase the minimum uncertainty of the whole data set, and rerun the \Marvel\ 
analysis. 

 It is important throughout and particularly at the final stage that the trends 
and patterns in the energy levels are validated using available means.
In previous studies this has often been against energies calculated theoretically;
here we are more reliant on trends such as reasonably
 systematic quadratic increase in energy with $J$, approximately 
linear increase with vibrational quantum number and so forth.
  Some of us are also part-way through constructing a spectroscopic model of TiO 
using the \Duo{} software \citep{jt609}; this also allowed a preliminary validation 
of energy levels against a realistic theoretical model.

\begin{figure}
\includegraphics[angle=-90,width=\textwidth]{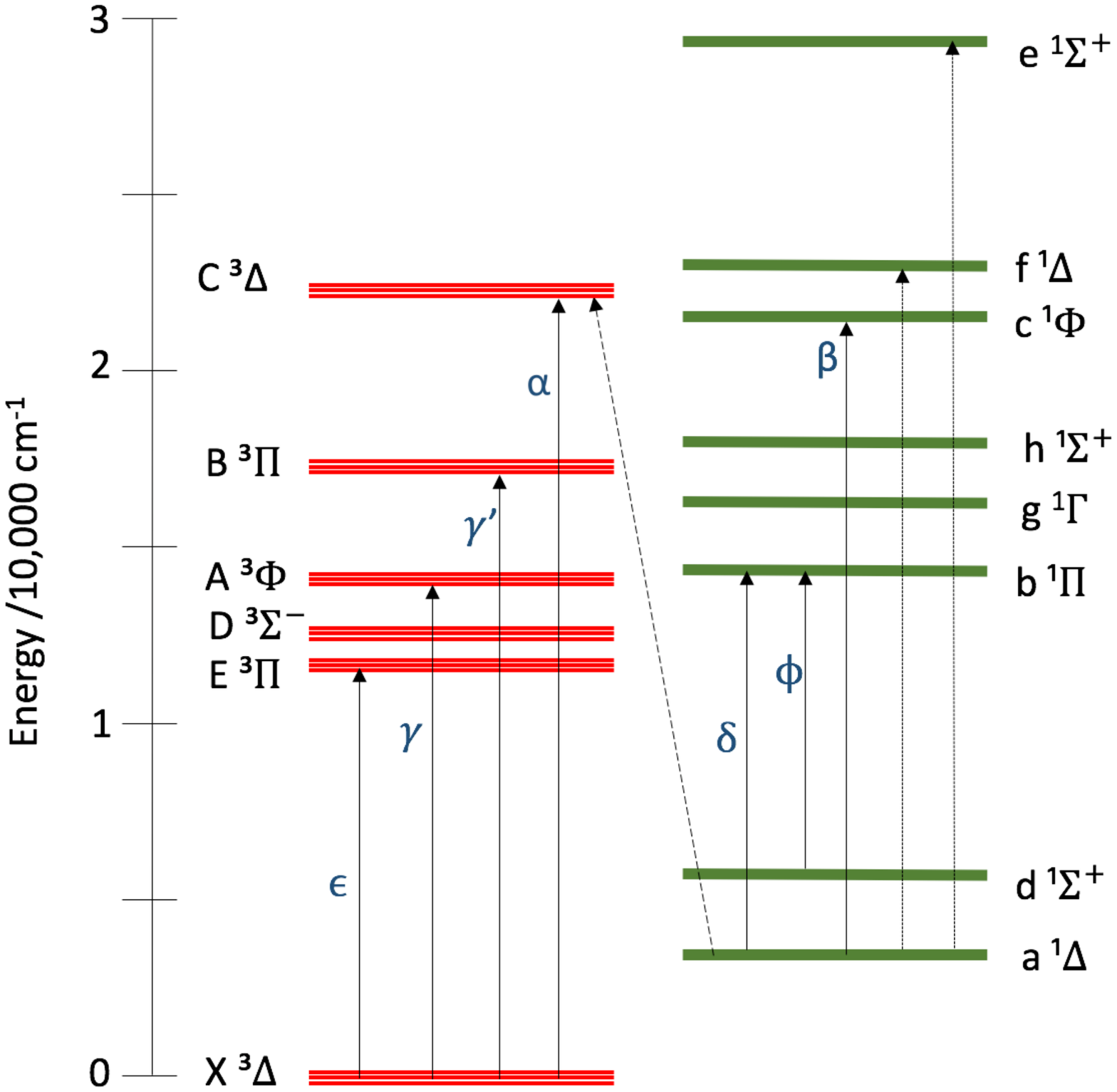}
\caption{\label{fig:fig1} The band system of TiO showing the bands considered in this work. The long-dashed line represents an experimentally-observed intercombination band. The short-dashed lines represent experimentally observed transitions that have not been named. There are three fine-structure components for the triple $\Pi$, $\Delta$, and $\Phi$ states (for the ground electronic state \X$_1$ is the lowest-energy component). }
\end{figure}

\subsection{Electronic structure and spectroscopy of TiO}
Like other transition-metal-containing diatomic species, TiO has a large number 
of low-lying electronic states which contribute significantly to the level 
density of the recorded spectra in the near-IR and in the visible. Those states 
with excitation energies below 23,000 \cm{}, and other well-characterised 
experimental electronic states are shown in \Cref{fig:fig1}, which
also gives the observed bands linking these states. The triplet ground 
state has allowed excitations to the \E{}, \A{}, \B{} and \C{} states. At the 
temperatures of the planetary atmospheres where TiO is thought to be abundant (i.e. 1500 
to 3000 K), significant absorption  also occurs from thermal population of the 
\Sa{} and \Sd{} states to higher singlet states, \Sb{}, \Sc{}, \Sf{} and \Se{}. 
 
\subsection{Quantum numbers and selection rules}
\Marvel\   uses quantum numbers solely as part of the labels used to uniquely 
identify each rovibronic state and the corresponding energy level. 
The three most obvious descriptors to use for the rovibronic states of
TiO are the electronic state, $state$, the total angular momentum quantum number, 
$J$,  and the vibrational quantum number, $v$. 
We find these descriptors to be relatively unambiguous, despite the fact that the 
vibrational quantum numbers are not good quantum numbers. 
For the triplet energy levels, we further need to give 
information about the coupling of the electronic angular momenta; 
we choose to do this in the Hund's coupling case (a) formulation \citep{16Bernat.method}.
For Hund's coupling case (a) the $\Omega$ quantum number is the 
sum of the quantum numbers describing  the axial component of the 
electron orbital angular momentum {\bf L}, $\Lambda$, 
and that of the electron spin angular momentum {\bf S}, $\Sigma$,
{\it i.e.}, $\Omega=\Lambda+\Sigma$.
Coupling case (a) is a good representation whenever $A\Lambda$ is much greater than
$BJ$, where $A$ (which can be both positive and negative) is the spin-orbit coupling
constant and $B$ is the rotational constant.
For the X$^3\Delta$ ground electronic state of TiO $A=50.7$ \cm; 
thus, of the three fine-structure
components $^3\Delta_{\Omega}$ the lowest state is $^3\Delta_{1}$.
Transitions within all three fine-structure states have been observed
experimentally (Table 6, \textit{vide infra}).
Note that Hund's coupling case (a) becomes less appropriate as $J$
increases (in this study energy levels with rather large $J$ values occur).
For singlet states, the component of the total electronic angular momentum along
the internuclear axis, described by the $\Omega$ quantum number,
is equal to $\Lambda$, as for singlet states $\Sigma=0$.

For some states the parity affects the final energy significantly enough 
to be experimentally observable; usually these state are of $\Pi$ symmetry. 
In these cases we will append the parity to the electronic state label.  
The parity of the energy level can be specified as 
(e/f) \citep{75Zare}.
For electronic dipole allowed transitions, the selection rules are 
e$\leftrightarrow$e and f$\leftrightarrow$f for P and R branches ($\Delta J = \pm 1$)
and e$\leftrightarrow$f for Q branches ($\Delta J = 0$).
For $\Pi$ states with experimental evidence of the splitting of the states, 
we distinguish between the $e$ and $f$ parity states. 
For the \B{} and \E{} states  the two parity states cannot be unambiguously assigned
 as $e$ and $f$; therefore, following the recommendations of ~\citep{75Zare} we  
retain the $a$ and $b$ designations \citep{55Mulliken} employed in the original manuscripts. 
For the \Sb{} state, the \Sb{} -- \Sd{}  transitions occur from the \Sd{} state
of well-defined parity $e$, which fixes
the parity of the observed levels of the associated \Sb{} state.

\begin{center}
\tiny

\begin{longtable}{p{1.2cm}p{1.8cm}crrrHclllll}
\caption{\label{tab:datasources}Data sources and their characteristics for \TiO, including the number of measured (A) and validated (V) transitions (Trans.). See \Cref{subsec:comments} for comments. }\\
\toprule
Tag &  Ref & & & Range (\cm{}) & J Range  & Method & Trans. (A/V) & \mc{3}{c}{Uncertainties (\cm{})} & Comments\\
& & &  & & & & &  Min & Av & Max \\
\midrule
\endfirsthead

\multicolumn{5}{c}%
{{ \tablename\ \thetable{} -- continued from previous page}} \\
\toprule
Tag &  Ref & & & Range (\cm{}) & J Range  & Method & Trans. (A/V) & \mc{3}{c}{Uncertainties (\cm{})} & Comment \\
& & &  & & & & &  Min & Av & Max \\
\midrule
\endhead

\bottomrule
\multicolumn{5}{c}{{Continued on next page}} \\
\endfoot

\bottomrule
\endlastfoot
50Phillips & \citet{50Phxxxx.TiO} & \Sb \enspace-- \Sa & 0 - 0 &   11106 - 11284 & 8 - \enspace94 &&376/373 &0.1 & 0.11 & 0.46 & (1a) \\* 
 & & \Sc \enspace-- \Sa & 0 - 0 &   17761 - 17858 & 9 - \enspace92 &&149/149 &0.1 & 0.11 & 0.42\\ 
\vspace{-0.5em}\\ 
50Phillips-ext & \citet{50Phxxxx.TiO}& \Sc \enspace-- \Sa & 0 - 0 &   17596 - 17860 & 2 - 101 &&178/178 &0.2 & 0.2 & 0.2 & (1a), (1d)\\* 
 & & \Sc \enspace-- \Sa & 1 - 1 &   17485 - 17760 & 2 - 100 &&283/207 &0.2 & 0.24 & 0.55\\* 
 & & \Sc \enspace-- \Sa & 2 - 2 &   17419 - 17654 & 2 - 100 &&252/182 &0.2 & 0.25 & 0.55\\ 
\vspace{-0.5em}\\ 
51Phillips & \citet{51Phxxxx.TiO} & \A \enspace-- \X & 0 - 0 &   13662 - 14172 & 5 - 119 &&765/763 &0.1 & 0.11 & 0.49 & (1a), (1b)\\* 
 & & \A \enspace-- \X & 0 - 1 &   12779 - 13173 & 8 - \enspace95 &&642/632 &0.1 & 0.11 & 0.48\\* 
 & & \A \enspace-- \X & 1 - 0 &   14579 - 15031 & 6 - \enspace90 &&638/635 &0.1 & 0.11 & 0.51\\ 
\vspace{-0.5em}\\ 
69Phillips & \citet{69Phxxxx.TiO} & \B \enspace-- \X & 0 - 0 &   16041 - 16233 & 2 - \enspace61 &&340/340 &0.1 & 0.11 & 0.39 & (1a), (1c)\\ 
\vspace{-0.5em}\\ 
71PhDa & \citet{71PhDaxx.TiO} & \Se \enspace-- \Sd & 0 - 0 &   24098 - 24302 & 1 - \enspace50 &&80/78 &0.05 & 0.051 & 0.075 & (1a)\\ 
\vspace{-0.5em}\\ 
71Phillips & \citet{71Phxxx1.TiO} & \B \enspace-- \X & 0 - 0 &   16216 - 16259 & 0 - \enspace36 &&192/138 &0.1 & 0.24 & 0.53 & (1a)\\ 
\vspace{-0.5em}\\ 
72Linton & \citet{72Lixxx1.TiO} & \Sf \enspace-- \Sa & 0 - 0 &   18879 - 19076 & 2 - \enspace66 &&111/109 &0.05 & 0.074 & 0.19 & (1e)\\ 
\vspace{-0.5em}\\ 
72Lindgren & \citet{72Lixxxx.TiO} & \Se \enspace-- \Sd & 1 - 0 &   24857 - 25147 & 8 - \enspace60 &&91/91 &0.05 & 0.053 & 0.11 & (1f)\\ 
\vspace{-0.5em}\\ 
73Phillips & \citet{73Phxxxx.TiO} & \A \enspace-- \X & 0 - 0 &   13365 - 14172 & 2 - 171 &&1353/1353 &0.2 & 0.2 & 0.42 & (1a), (1d)\\* 
 & & \A \enspace-- \X & 0 - 1 &   12340 - 13173 & 1 - 162 &&1276/1276 &0.2 & 0.2 & 0.52\\* 
 & & \A \enspace-- \X & 0 - 2 &   11696 - 12183 & 2 - 120 &&800/795 &0.2 & 0.2 & 0.48\\*
 & & \A \enspace-- \X & 1 - 0 &   14140 - 15031 & 1 - 158 &&1263/1262 &0.2 & 0.2 & 0.28\\* 
 & & \A \enspace-- \X & 1 - 1 &   13177 - 14031 & 1 - 165 &&1308/1308 &0.2 & 0.2 & 0.34\\* 
 & & \A \enspace-- \X & 1 - 2 &   12456 - 13041 & 1 - 143 &&1099/1097 &0.2 & 0.2 & 0.46\\* 
 & & \A \enspace-- \X & 1 - 3 &   11527 - 12061 & 1 - 151 &&1000/984 &0.2 & 0.2 & 0.55\\* 
 & & \A \enspace-- \X & 2 - 0 &   14994 - 15882 & 1 - 164 &&1230/1227 &0.2 & 0.21 & 0.51\\* 
 & & \A \enspace-- \X & 2 - 1 &   13952 - 14882 & 1 - 149 &&1211/1207 &0.2 & 0.2 & 0.5\\* 
 & & \A \enspace-- \X & 2 - 3 &   12237 - 12911 & 1 - 148 &&1107/1103 &0.2 & 0.2 & 0.54\\* 
 & & \A \enspace-- \X & 2 - 4 &   11524 - 11940 & 1 - 125 &&838/795 &0.2 & 0.2 & 0.51\\* 
 & & \A \enspace-- \X & 3 - 1 &   14991 - 15725 & 1 - 147 &&1056/1053 &0.2 & 0.21 & 0.4\\* 
 & & \A \enspace-- \X & 3 - 2 &   13909 - 14735 & 1 - 151 &&1104/1099 &0.2 & 0.2 & 0.36\\* 
 & & \A \enspace-- \X & 3 - 4 &   12237 - 12782 & 1 - 131 &&908/891 &0.2 & 0.2 & 0.4\\* 
 & & \A \enspace-- \X & 3 - 5 &   11494 - 11820 & 1 - 125 &&868/833 &0.2 & 0.2 & 0.34\\* 
 & & \A \enspace-- \X & 4 - 2 &   14813 - 15570 & 1 - 136 &&1062/1049 &0.2 & 0.2 & 0.42\\* 
 & & \A \enspace-- \X & 4 - 3 &   13761 - 14589 & 1 - 149 &&1051/1038 &0.2 & 0.21 & 0.5\\* 
 & & \A \enspace-- \X & 4 - 5 &   12041 - 12655 & 1 - 134 &&991/973 &0.2 & 0.2 & 0.4\\* 
 & & \A \enspace-- \X & 5 - 3 &   14781 - 15417 & 2 - 136 &&1025/1016 &0.2 & 0.2 & 0.4\\*
\vspace{-0.5em}\\* 
& & \B \enspace-- \X & 0 - 0 &   15560 - 16259 & 1 - 141 &&1735/1560 &0.2 & 0.21 & 0.52\\* 
\vspace{-0.5em}\\* 
& & \C \enspace-- \X & 0 - 0 &   18298 - 19349 & 1 - 159 &&879/879 &0.2 & 0.2 & 0.27\\* 
 & & \C \enspace-- \X & 0 - 1 &   17327 - 18349 & 1 - 157 &&864/864 &0.2 & 0.2 & 0.48\\* 
 & & \C \enspace-- \X & 0 - 2 &   16661 - 17359 & 1 - 143 &&689/686 &0.2 & 0.2 & 0.48\\* 
 & & \C \enspace-- \X & 0 - 3 &   15929 - 16378 & 2 - 100 &&438/411 &0.2 & 0.21 & 0.53\\* 
 & & \C \enspace-- \X & 1 - 0 &   18926 - 20178 & 1 - 156 &&848/842 &0.2 & 0.2 & 0.27\\* 
 & & \C \enspace-- \X & 1 - 2 &   17369 - 18188 & 1 - 126 &&706/698 &0.2 & 0.2 & 0.47\\* 
 & & \C \enspace-- \X & 1 - 3 &   16660 - 17206 & 1 - 118 &&629/586 &0.2 & 0.21 & 0.53\\* 
 & & \C \enspace-- \X & 2 - 0 &   20292 - 20998 & 1 - 107 &&609/608 &0.2 & 0.2 & 0.35\\* 
 & & \C \enspace-- \X & 2 - 1 &   19081 - 19998 & 1 - 126 &&637/637 &0.2 & 0.2 & 0.38\\* 
 & & \C \enspace-- \X & 2 - 3 &   17707 - 18026 & 1 - \enspace88 &&346/343 &0.2 & 0.21 & 0.54\\* 
 & & \C \enspace-- \X & 2 - 4 &   16427 - 17054 & 1 - 112 &&536/512 &0.2 & 0.21 & 0.54\\* 
 & & \C \enspace-- \X & 3 - 0 &   21191 - 21809 & 1 - 111 &&584/582 &0.2 & 0.2 & 0.35\\* 
 & & \C \enspace-- \X & 3 - 1 &   19976 - 20809 & 2 - 120 &&630/622 &0.2 & 0.2 & 0.54\\* 
 & & \C \enspace-- \X & 3 - 5 &   16444 - 16902 & 1 - 117 &&464/445 &0.2 & 0.21 & 0.51\\* 
 & & \C \enspace-- \X & 4 - 0 &   22089 - 22610 & 1 - 101 &&456/444 &0.2 & 0.2 & 0.38\\* 
 & & \C \enspace-- \X & 4 - 1 &   20896 - 21611 & 1 - 105 &&509/497 &0.2 & 0.2 & 0.25\\* 
 & & \C \enspace-- \X & 4 - 2 &   20260 - 20620 & 2 - \enspace90 &&439/430 &0.2 & 0.2 & 0.35\\* 
 & & \C \enspace-- \X & 5 - 1 &   21898 - 22404 & 2 - \enspace83 &&361/358 &0.2 & 0.2 & 0.51\\* 
 & & \C \enspace-- \X & 5 - 2 &   20830 - 21414 & 2 - \enspace92 &&381/379 &0.2 & 0.2 & 0.51\\* 
 & & \C \enspace-- \X & 6 - 2 &   21794 - 22195 & 3 - \enspace73 &&321/319 &0.2 & 0.2 & 0.33\\* 
 & & \C \enspace-- \X & 6 - 3 &   20847 - 21214 & 4 - \enspace86 &&276/270 &0.2 & 0.2 & 0.44\\* 
 & & \C \enspace-- \X & 7 - 3 &   21654 - 21986 & 1 - \enspace67 &&293/293 &0.2 & 0.2 & 0.2\\ 
\vspace{-0.5em}\\ 
74LiSi & \citet{74LiSixx.TiO} & \Sb \enspace-- \Sa & 0 - 0 &   11198 - 11284 & 1 - \enspace43 &&158/158 &0.1 & 0.1 & 0.36\\ 
\vspace{-0.5em}\\ 
74Linton & \citet{74Lixxxx.TiO} & \Sc \enspace-- \Sa & 0 - 0 &   17715 - 17859 & 2 - \enspace74 &&189/189 &0.02 & 0.035 & 0.13 \\* 
 & & \Sc \enspace-- \Sa & 1 - 1 &   17634 - 17759 & 2 - \enspace72 &&177/169 &0.02 & 0.023 & 0.09\\* 
 & & \Sc \enspace-- \Sa & 2 - 2 &   17523 - 17658 & 2 - \enspace67 &&162/161 &0.02 & 0.023 & 0.1\\* 
 & & \Sc \enspace-- \Sa & 3 - 3 &   17443 - 17556 & 2 - \enspace69 &&152/152 &0.02 & 0.022 & 0.056\\ 
\vspace{-0.5em}\\ 
79HoGeMe & \citet{79HoGeMe.TiO} & \B \enspace-- \X & 0 - 0 &   15951 - 16259 & 1 - \enspace55 &&732/731 &0.008 & 0.013 & 0.087 & (1g) \\* 
 & & \B \enspace-- \X & 0 - 1 &   15002 - 15245 & 0 - \enspace50 &&586/586 &0.008 & 0.011 & 0.043\\*
 & & \B \enspace-- \X & 1 - 0 &   16862 - 17122 & 1 - \enspace56 &&664/602 &0.008 & 0.0095 & 0.064\\* 
 & & \B \enspace-- \X & 1 - 1 &   15835 - 16107 & 1 - \enspace55 &&546/367 &0.008 & 0.014 & 0.093\\ 
\vspace{-0.5em}\\ 
79GaDe	&	\citet{79GaDexx.TiO} & \X \enspace-- \X & 1 - 0 & 975 - \enspace1022 & 2 - \enspace22 & & 40/40 & 0.2 & 0.2 & 0.3 &  (1h) \\
\vspace{-0.5em}\\ 
80GaBrDa & \citet{80GaBrDa.TiO} & \Sb \enspace-- \Sd & 0 - 0 &    8775 - \enspace9062 & 1 - \enspace93 &&240/240 &0.01 & 0.011 & 0.074\\* 
 & & \Sb \enspace-- \Sd & 0 - 1 &    7757 - \enspace8049 & 0 - \enspace86 &&210/210 &0.01 & 0.011 & 0.041\\* 
 & & \Sb \enspace-- \Sd & 0 - 2 &    6952 - \enspace7046 & 7 - \enspace49 &&49/49 &0.01 & 0.01 & 0.014\\* 
 & & \Sb \enspace-- \Sd & 1 - 0 &    9598 - \enspace9972 & 0 - \enspace86 &&233/233 &0.01 & 0.016 & 0.32\\* 
 & & \Sb \enspace-- \Sd & 1 - 1 &    8773 - \enspace8960 & 0 - \enspace70 &&152/152 &0.01 & 0.012 & 0.078\\* 
 & & \Sb \enspace-- \Sd & 1 - 2 &    7758 - \enspace7957 & 2 - \enspace77 &&174/174 &0.01 & 0.015 & 0.11\\* 
 & & \Sb \enspace-- \Sd & 1 - 3 &    6826 - \enspace6964 & 1 - \enspace67 &&95/95 &0.01 & 0.013 & 0.084\\* 
 & & \Sb \enspace-- \Sd & 2 - 0 &   10712 - 10874 & 1 - \enspace60 &&123/123 &0.01 & 0.017 & 0.34\\* 
 & & \Sb \enspace-- \Sd & 2 - 1 &    9582 - \enspace9862 & 0 - \enspace72 &&171/171 &0.01 & 0.011 & 0.05\\* 
 & & \Sb \enspace-- \Sd & 2 - 3 &    7679 - \enspace7866 & 1 - \enspace75 &&117/117 &0.01 & 0.011 & 0.028\\* 
 & & \Sb \enspace-- \Sd & 3 - 1 &   10446 - 10755 & 0 - \enspace74 &&151/151 &0.01 & 0.015 & 0.096\\* 
 & & \Sb \enspace-- \Sd & 3 - 2 &    9558 - \enspace9708 & 46 - \enspace70 &&34/34 &0.01 & 0.019 & 0.073\\* 
 & & \Sb \enspace-- \Sd & 3 - 4 &    7646 - \enspace7776 & 0 - \enspace51 &&95/95 &0.01 & 0.014 & 0.17\\* 
 & & \Sb \enspace-- \Sd & 3 - 5 &    6708 - \enspace6802 & 2 - \enspace55 &&43/43 &0.01 & 0.01 & 0.01\\* 
 & & \Sb \enspace-- \Sd & 4 - 2 &   10397 - 10636 & 0 - \enspace66 &&153/153 &0.01 & 0.015 & 0.094\\* 
 & & \Sb \enspace-- \Sd & 4 - 3 &    9626 - \enspace9643 & 1 - \enspace32 &&32/32 &0.01 & 0.013 & 0.035\\ 
\vspace{-0.5em}\\ 

85BrGa & \citet{85BrGaxx.TiO} & \Sf \enspace-- \Sa & 0 - 0 &   18830 - 19077 & 2 - \enspace71 &&127/127 &0.044 & 0.044 & 0.056\\* 
 & & \Sf \enspace-- \Sa & 0 - 1 &   17841 - 18068 & 2 - \enspace69 &&116/116 &0.044 & 0.045 & 0.1\\* 
 & & \Sf \enspace-- \Sa & 1 - 0 &   19726 - 19945 & 2 - \enspace63 &&101/101 &0.044 & 0.044 & 0.057\\* 
 & & \Sf \enspace-- \Sa & 1 - 1 &   18744 - 18937 & 2 - \enspace60 &&93/93 &0.044 & 0.045 & 0.081\\* 
 & & \Sf \enspace-- \Sa & 1 - 2 &   17774 - 17937 & 3 - \enspace56 &&67/67 &0.044 & 0.046 & 0.13\\* 
 & & \Sf \enspace-- \Sa & 2 - 1 &   19748 - 19800 & 4 - \enspace24 &&27/27 &0.044 & 0.044 & 0.044\\ 
\vspace{-0.5em}\\ 
90StShJuRu & \citet{90StShJu.TiO} & \X \enspace-- \X & 0 - 0 &       2 - \enspace\enspace\enspace\enspace3 & 1 - \enspace\enspace3 &&2/2 &10$^{-5}$ & 10$^{-5}$ & 10$^{-5}$ & (1i) \\ 
\vspace{-0.5em}\\ 
91GuAmVe & \citet{91GuAmVe.TiO} & \B \enspace-- \X & 1 - 2 &   14848 - 15134 & 3 - \enspace48 &&171/170 &0.03 & 0.031 & 0.046 & (1j) \\* 
 & & \B \enspace-- \X & 1 - 3 &   13925 - 14129 & 6 - \enspace24 &&14/14 &0.03 & 0.031 & 0.05\\* 
 & & \C \enspace-- \X & 2 - 1 &   19930 - 19995 & 5 - \enspace23 &&9/9 &0.03 & 0.036 & 0.061\\* 
 & & \C \enspace-- \X & 2 - 2 &   18946 - 18992 & 15 - \enspace21 &&7/7 &0.03 & 0.03 & 0.03\\* 
 & & \C \enspace-- \X & 2 - 4 &   16965 - 17040 & 10 - \enspace31 &&23/23 &0.03 & 0.031 & 0.06\\* 
 & & \C \enspace-- \X & 2 - 5 &   16031 - 16088 & 5 - \enspace22 &&24/24 &0.03 & 0.03 & 0.03\\ 
\vspace{-0.5em}\\ 
91SiHa & \citet{91SiHaxx.TiO} & \E \enspace-- \X & 0 - 0 &   11801 - 11852 & 0 - \enspace15 &&111/109 &0.1 & 0.13 & 0.47\\ 
\vspace{-0.5em}\\ 
95KaMcHe & \citet{95KaMcHe.TiO} & \C \enspace-- \Sa & 2 - 0 &   17675 - 17738 & 2 - \enspace34 &&84/84 &0.01 & 0.013 & 0.089\\* 
 & & \C \enspace-- \X & 2 - 3 &   17969 - 18011 & 3 - \enspace26 &&42/42 &0.01 & 0.014 & 0.045\\* 
 & & \C \enspace-- \X & 2 - 4 &   16995 - 17040 & 3 - \enspace27 &&39/39 &0.01 & 0.01 & 0.019\\ 
\vspace{-0.5em}\\ 
96AmChLu & \citet{96AmChLu.TiO} & \Sc \enspace-- \Sa & 0 - 0 &   17711 - 17860 & 3 - \enspace97 &&114/114 &0.005 & 0.0052 & 0.0091 & (1k) \\ 
\vspace{-0.5em}\\ 
96BaMeMe & \citet{96BaMeMe.TiO} & \A \enspace-- \X & 0 - 0 &   14022 - 14172 & 1 - \enspace26 &&63/63 &0.0002 & 0.0003 & 0.00063\\ 
\vspace{-0.5em}\\ 
96RaBeWa & \citet{96RaBeWa.TiO} & \Sb \enspace-- \Sa & 0 - 0 &   10960 - 11284 & 1 - 108 &&405/404 &0.02 & 0.021 & 0.076\\* 
 & & \Sb \enspace-- \Sa & 1 - 1 &   11009 - 11186 & 1 - \enspace82 &&231/231 &0.02 & 0.021 & 0.05\\ 
\vspace{-0.5em}\\ 
98NaSaRoSt & \citet{98NaSaRo.TiO} & \X \enspace-- \X & 0 - 0 &       7 - \enspace\enspace\enspace12 & 6 - \enspace11 &&9/9 & 10$^{-7}$ & 10$^{-7}$ & 10$^{-7}$ & (1l) \\ 
\vspace{-0.5em}\\ 
99RaBeDuWa & \citet{99RaBeDu.TiO} & & & & & & & & & & (1m) \\*
 -Lab & & \A \enspace-- \X & 0 - 0 &   13863 - 14172 & 3 - \enspace89 &&291/285 &0.004 & 0.0044 & 0.02\\* 
 & & \A \enspace-- \X & 0 - 1 &   12918 - 13173 & 3 - \enspace66 &&368/355 &0.004 & 0.0054 & 0.09\\* 
 & & \A \enspace-- \X & 1 - 0 &   14725 - 15031 & 2 - \enspace72 &&243/239 &0.004 & 0.0047 & 0.023\\* 
 & & \A \enspace-- \X & 1 - 1 &   13729 - 14031 & 3 - \enspace72 &&409/392 &0.004 & 0.0047 & 0.026\\* 
 & & \A \enspace-- \X & 1 - 2 &   12809 - 13041 & 2 - \enspace68 &&382/377 &0.004 & 0.0049 & 0.031\\* 
 & & \A \enspace-- \X & 2 - 1 &   14592 - 14882 & 5 - \enspace66 &&360/354 &0.004 & 0.0049 & 0.022\\* 
 & & \A \enspace-- \X & 2 - 3 &   12680 - 12911 & 3 - \enspace54 &&268/267 &0.004 & 0.0046 & 0.017\\* 
 & & \A \enspace-- \X & 3 - 2 &   14478 - 14733 & 4 - \enspace59 &&241/241 &0.004 & 0.0044 & 0.022\\* 
 & & \A \enspace-- \X & 3 - 4 &   12588 - 12760 & 7 - \enspace52 &&138/137 &0.004 & 0.0042 & 0.012\\* 
 & & \A \enspace-- \X & 4 - 3 &   14336 - 14589 & 8 - \enspace59 &&244/243 &0.004 & 0.0043 & 0.012\\* 
\vspace{-0.5em}\\* 
-Sunspots (SS) & & \A \enspace-- \X & 0 - 0 &   13601 - 14071 & 30 - 110 &&132/132 &0.01 & 0.01 & 0.03\\* 
 & & \A \enspace-- \X & 0 - 1 &   12830 - 13123 & 11 - \enspace98 &&102/102 &0.01 & 0.012 & 0.044\\* 
 & & \A \enspace-- \X & 1 - 0 &   14673 - 14883 & 12 - \enspace83 &&57/57 &0.01 & 0.012 & 0.043\\* 
 & & \A \enspace-- \X & 1 - 1 &   13606 - 13936 & 26 - 107 &&94/94 &0.01 & 0.011 & 0.033\\* 
 & & \A \enspace-- \X & 1 - 2 &   12703 - 12958 & 11 - \enspace98 &&149/149 &0.01 & 0.011 & 0.046\\* 
 & & \A \enspace-- \X & 2 - 1 &   14671 - 14722 & 7 - \enspace66 &&4/4 &0.01 & 0.019 & 0.038\\* 
 & & \A \enspace-- \X & 2 - 2 &   13618 - 13817 & 16 - \enspace82 &&70/70 &0.01 & 0.012 & 0.042\\* 
 & & \A \enspace-- \X & 2 - 3 &   12660 - 12831 & 13 - \enspace83 &&77/76 &0.01 & 0.011 & 0.027\\ 
\vspace{-0.5em}\\ 
02KoHaMuSe & \citet{02KoHaMu.TiO} & \A \enspace-- \X & 0 - 2 &   12176 - 12182 & 3 - \enspace15 &&12/12 &0.01 & 0.016 & 0.025 & (1n) \\* 
 & & \E \enspace-- \X & 0 - 0 &   11796 - 11855 & 0 - \enspace35 &&348/347 &0.01 & 0.01 & 0.036\\* 
 & & \E \enspace-- \X & 1 - 0 &   12739 - 12760 & 0 - \enspace13 &&57/56 &0.01 & 0.01 & 0.024\\ 
\end{longtable}

\end{center}

\subsection{Collation of data sources}
The collated data sources used in the rotationally-resolved \Marvel\ analysis 
are summarised in \Cref{tab:datasources}. In total, we  use 24 data 
sources, involving 11 electronic states with 49,679 transitions, 123 total 
(non-unique) vibronic bands and 84 total unique vibronic bands. The full list of 
compiled data converted to \Marvel\ format is in the Supplementary Information; 
an extract is given in \Cref{tab:marvelinput}.

\begin{table*}
\caption{\label{tab:marvelinput}
Extract from the 48Ti-16O.marvel.inp input file for \TiO.}
\footnotesize \tabcolsep=5pt
\renewcommand{\arraystretch}{1.2}
\begin{tabular}{lllcclccrrrrr}
\\
\toprule
  \mc{1}{c}{1}     &        \mc{1}{c}{2}        &   \mc{1}{c}{3}  &   \mc{1}{c}{4}  &  \mc{1}{c}{5}  &   \mc{1}{c}{6}  &   \mc{1}{c}{7}  &  \mc{1}{c}{8}  &   \mc{1}{c}{9} \\
    \midrule
    \multicolumn{1}{c}{$\tilde{\nu}$} & \multicolumn{1}{c}{$\Delta\tilde{\nu}$} &  \mc{1}{c}{State$^\prime$}    & $J^\prime$  &  \multicolumn{1}{c}{$v^\prime$}  &  \mc{1}{c}{State$^{\prime\prime}$}    & $J^{\prime\prime}$  &  \multicolumn{1}{c}{$v^{\prime\prime}$}  & \mc{1}{c}{ID}  \\
     \midrule
14463.63	&	0.2	&	A3Phi\_3	&	122	&	1	&	X3Delta\_2	&	122	&	0	&	73Phillips\_AX.18910	\\
14336.8	&	0.2	&	A3Phi\_3	&	122	&	1	&	X3Delta\_2	&	123	&	0	&	73Phillips\_AX.18914	\\
14634.87	&	0.2	&	A3Phi\_4	&	122	&	1	&	X3Delta\_3	&	121	&	0	&	73Phillips\_AX.18916	\\
14508.26	&	0.2	&	A3Phi\_4	&	122	&	1	&	X3Delta\_3	&	122	&	0	&	73Phillips\_AX.18918	\\
14380.56	&	0.2	&	A3Phi\_4	&	122	&	1	&	X3Delta\_3	&	123	&	0	&	73Phillips\_AX.18922	\\
14408.6	&	0.2	&	A3Phi\_2	&	123	&	1	&	X3Delta\_1	&	123	&	0	&	73Phillips\_AX.19008	\\
14281.54	&	0.2	&	A3Phi\_2	&	123	&	1	&	X3Delta\_1	&	124	&	0	&	73Phillips\_AX.19010	\\
14582.06	&	0.2	&	A3Phi\_3	&	123	&	1	&	X3Delta\_2	&	122	&	0	&	73Phillips\_AX.19012	\\
9635.433	&	0.01	&	b1Pi	&	3f	&	4	&	d1Sigma+	&	3	&	3	&	80GaBrDa.65	\\
9640.637	&	0.012	&	b1Pi	&	22e	&	4	&	d1Sigma+	&	21	&	3	&	80GaBrDa.662	\\
9637.572	&	0.015	&	b1Pi	&	26e	&	4	&	d1Sigma+	&	25	&	3	&	80GaBrDa.802	\\
9639.478	&	0.033	&	b1Pi	&	4e	&	4	&	d1Sigma+	&	3	&	3	&	80GaBrDa.85	\\
9635.617	&	0.01	&	b1Pi	&	28e	&	4	&	d1Sigma+	&	27	&	3	&	80GaBrDa.868	\\
9635.162	&	0.01	&	b1Pi	&	4f	&	4	&	d1Sigma+	&	4	&	3	&	80GaBrDa.97	\\
16229.687	&	0.127596	&	B3Pi\_0	&	5b	&	0	&	X3Delta\_1	&	4	&	0	&	69Phxxxx.1	\\
16231.492	&	0.213806	&	B3Pi\_0	&	14b	&	0	&	X3Delta\_1	&	13	&	0	&	69Phxxxx.10	\\
16197.913	&	0.1	&	B3Pi\_0	&	46a	&	0	&	X3Delta\_1	&	45	&	0	&	69Phxxxx.100	\\
16195.911	&	0.1	&	B3Pi\_0	&	47a	&	0	&	X3Delta\_1	&	46	&	0	&	69Phxxxx.101	\\
16193.918	&	0.1	&	B3Pi\_0	&	48a	&	0	&	X3Delta\_1	&	47	&	0	&	69Phxxxx.102	\\
16191.766	&	0.1	&	B3Pi\_0	&	49a	&	0	&	X3Delta\_1	&	48	&	0	&	69Phxxxx.103	\\
16189.615	&	0.1	&	B3Pi\_0	&	50a	&	0	&	X3Delta\_1	&	49	&	0	&	69Phxxxx.104	\\
\bottomrule
\end{tabular}

\begin{tabular}{ccl}
\\
             Column       &    Notation                 &      \\
\midrule
   1 &   $\tilde{\nu}$             &    Transition frequency (in \cm) \\
   2 & $\Delta\tilde{\nu}$        &   Estimated uncertainty in transition frequency (in \cm) \\
   3 &  State$^\prime$ &  Electronic state of upper energy level, including $\Omega$ for triplet states, where $\Omega=\Lambda+\Sigma$;  \\
 & &  $\Lambda$ and $\Sigma$ are projections of the total angular momentum and  the electron spin angular\\
    & & momentum on the internuclear axis, respectively, of the upper level  \\
   4 &  $J^\prime$               &       Total angular momentum of upper  level and rotationaless parity for $\Pi$ states   \\
   5 & $v^\prime$ & Vibrational quantum number  of upper  level \\
   6 &  State$^{\prime\prime}$ &  Electronic state of lower energy level, including $\Omega$ for triplet states \\
   7 &  $J^{\prime\prime}$               &       Total angular momentum of lower  level and rotationaless parity for $\Pi$ states   \\
   8 & $v^{\prime\prime}$ & Vibrational quantum number  of lower  level \\
   9 & ID & Unique ID for transition, with reference key for source (see \Cref{tab:datasources}) and counting number \\
\bottomrule
\end{tabular}
\end{table*}

\begin{table*}
\footnotesize
\caption{\label{tab:databandheads}  TiO references that contain experimental measurements of band positions (often bandheads). See \Cref{subsec:comments_bh} for comments. \# refers to the number of bandheads provided. }
\begin{tabular}{llHHlrc}
\\
\\ \toprule
Tag & Ref & Used? & Comment & System & \# &  Comment \\
\bottomrule
28Lowater & \citet{28Lowater.TiO} & & & various, some unassigned & 144 & (3a) \\
29Christya & \citet{29Chxxx1.TiO} & & & A-X, C-X & 62 & (3b) \\
37WuMe & \citet{37WuMexx.TiO}  & Got version, in German & & b-a, b-d & 7 & (3a) \\

57GaRoJu & \citet{57GaRoJu.TiO} & ? & Original data not found, however, 69LiNi provides assignment of band from this spectra. & b-a & 1 & (3a)\\
69LiNi & \citet{69LiNixx.TiO} & No & Bandheads only & c-a  & 4 & (3a) \\
69Lockwood & \citet{69Loxxxx.TiO} & No & 7 bands identified (including some new) from Mira spectra.  & b-d, b-a & 7 & \\
69Phillips & \citet{69Phxxxx.TiO}  & No &Bandheads only & B-X & 32 \\

72PhDa & \citet{72PhDaxx.TiO} &No & C-X bandheads only in paper; data used in this paper may be attributed to 73Phillips in Marvel TiO file (as the raw data is obtained from a tape processed by Kurucz in 1981). & C-X & 22 & (3c) \\
76ZyPa & \citet{76ZyPaxx.TiO}  & No &Bandheads only & B-X & 20 \\
77LiBrb& \citet{77LiBrx1.TiO} &No& Bandheads only & E-X & 45 & (3c)\\
82DeVore & \citet{82DeVore.TiO} & No & Bandheads only & f-a & 8 & \\
\bottomrule
\end{tabular}
\end{table*}

\begin{table*}
\footnotesize
\caption{\label{tab:intensities}  TiO references that contain measurements relevant to the verification of the dipole moments, e.g. lifetimes, transition intensities (relative or absolute) and dipole moment measurements.}

\begin{tabular}{llHll}
\\
\\ \toprule
Tag & Ref & Used? & Type & Bands/ States  \\
\bottomrule
54Phillips	&	\citet{54Phxxxx.TiO} & No & Relative intensity &  C-X   \\

70LiNi	&	\citet{70LiNixx.TiO} & No & Relative intensity & C-X, c-a \\
71PrSuPe	&	\citet{71PrSuPe.TiO} & No & Intensity & A-X, C-X \\
72Dube	&	\citet{72Duxxxx.TiO} & No & Intensity  & c-a \\
74PrSuPe	&	\citet{74PrSuPe.TiO} & No & Intensity & A-X, C-X 	\\
74FaWoBe & \citet{74FaWoBe.TiO}  & No & Intensity & C-X \\
75Zyrnicki	&	\citet{75Zyxxxx.TiO} & No &  Intensity  & c-a \\
76FeBiDa	&	\citet{76FeBiDa.TiO} & No & Lifetime   & \Sc{} ($v$=0) \\
77FeDa	&	\citet{77FeDaxx.TiO} & No & Lifetime & \Sc{} ($v$=0) \\
78FeDa	&	\citet{78FeDaxx.TiO} & No & Lifetime & \C$_3$ ($v$=2, $J$=17,87)		\\
78StLi	&	\citet{78StLixx.TiO} & 	No & Lifetime   &  \C{} ($v$=0, 1, 2)\\
79RaRaRa	&	\citet{79RaRaRa.TiO} & No & Intensity  	& B-X	\\ 
86DaLiPh	&	\citet{86DaLiPh.TiO} & No & Intensity  &  c-a, b-a, b-d, B-X, A-X and C-X \\
89StSh	&	\citet{89StShxx.TiO} & No & Dipole moment & X 		\\
92CaSc	&	\citet{92CaScxx.TiO} & No & Lifetime & 	\B$_1$ ($v$=0)	\\
92DoWe	&	\citet{92DoWexx.TiO} & No & Lifetime & \A$_2$ ($v$=0), \B$_0$ ($v$=0), \C$_1$ (v=0) 		\\
95HeNaCo	&	\citet{95HeNaCo.TiO} & No & Lifetime &  A, B, C, c, f and E   	\\
98Lundevall	&	\citet{98Luxxxx.TiO} & 	No & Lifetime & \E{} ($v$=0)	\\
03StVi & \citet{03StVixx.TiO} & No & Dipole moment  & X, E, A and B    \\
03NaMiIt & \citet{03NaMiIt.TiO} & No & Intensity & C-X  \\
04NaSaIt & \citet{04NaSaIt.TiO} & No & Intensity &  C-X  \\
\bottomrule
\end{tabular}
\end{table*}

\clearpage

\begin{center}
\footnotesize
\begin{longtable}{llHp{10cm}H}
\caption{\label{tab:otherrefs}  TiO references that are not  used in the rotationally-resolved \Marvel\ or band-head analysis and do not focus on intensity determination. This list concentrates on sunspot observations analysed specifically for TiO, experimental studies or analysis of experimental studies.}
\\
\\ \toprule
Tag & Ref & Used? & Comment & Other Use?  \\
\midrule
\endfirsthead

\multicolumn{5}{c}%
{{ \tablename\ \thetable{} -- continued from previous page}} \\
\toprule
Tag & Ref & Used? & Comment & Other Use?  \\
\midrule
\endhead

\bottomrule
\multicolumn{5}{c}{{Continued on next page}} \\
\endfoot

\bottomrule
\endlastfoot

1904Fowler & \citet{1904Fowler.TiO} & & No explicit assignment \\
26King & \citet{26King.TiO} & &  No rotationally resolved data \\
27BiCh & \citet{27BiChxx.TiO} & & Paper not available online \\
28ChBi & \citet{28ChBixx.TiO} & & No rotationally resolved data\\
29Lowater & \citet{29Lowater.TiO} & & No absolute band position data \\

29Christyb & \citet{29Chxxxx.TiO} & & Summary of 29Christya \\
36Budo	&	\citet{36Buxxxx.TiO} & Got version, in German & Combination differences only		\\
37Dobron	&	\citet{37Doxxxx.TiO} & & Source not available, but the measurements are unlikely to be accurate enough for use \\
52Phillips	&	\citet{52Phxxx2.TiO} & No & Identification of ground state symmetry, no new data \\
59Pettera & \citet{59Pettera.TiO} & No &Source not available, but the measurements are unlikely to be accurate enough for use in \Marvel\ & No\\
59Petterb & \citet{59Petterb.TiO} &No &Source not available, but the measurements are unlikely to be accurate enough for use  in \Marvel\ & No\\
61PeLi & \citet{61PeLixx.TiO} &No &Figures only, no numerical data \\
62Petter & \citet{62Petterson.TiO} & No & Source not available, but the measurements are unlikely to be accurate enough to use in \Marvel; contains d-b data \\
68Makita & \citet{68Makita.TiO} & No & Sunspot data with 63 lines only \\
70PaPa & \citet{70PaPaxx.TiO} & No&Bandheads only, and very high energy bands considered \\
71McThWe & \citet{71McThWe.TiO} &No &Inert neon matrix used, bandheads only \\
72BaGuPiDe	&	\citet{72BaGuPi.TiO} & No & Dissociation energy only & Yes		\\
72PaHs & \citet{72PaHsxx.TiO} & No &Bandheads only in UV \\
73Engvold & \citet{73Enxxxx.TiO} & No & Fitting to sunspot spectral, newer data available & \\
74Phillips	&	\citet{74Phxxxx.TiO} & No & Prediction of \X{} energy levels based on combination differences of other observed data	\\
75BrBr & \citet{75BrBrxx.TiO}   & No & Inert neon matrix used, bandheads only \\
75Collins	&	\citet{75Coxxxx.TiO} & No & Analysis only	\\
76Hilden	&	\citet{76Hixxxx.TiO} & No & No spectroscopic data, only dissociation energy & ?? 		\\
77DuGo & \citet{77DuGoxx.TiO}  &No & No rotationally resolved data; bandheads for highly excited state only \\
77LiBra & \citet{77LiBrxx.TiO} &No& Original measurement of C-a transition frequency, no tabulated rotationally resolved data \\
83KoKuGu	&	\citet{83KoKuGu.TiO} & 		Laura to look for & Measurement of singlet-triplet energy gap \\
84DyGrJoLe	&	\citet{84DyGrJo.TiO} & No &	Limited data on bandheads that is available elsewhere \\
85CaCrDu & \citet{85CaCrDu.TiO} &No& No relevant data \\

93FlScJu & \citet{93FlScJu.TiO} & & Analysis of hyperfine structure in $^{47}$Ti$^{16}$O \\
94WiRoVa	&	\citet{94WiRoVa.TiO} & No & Transitions observed in inert argon matrix 		\\
95AmAzLu	&	\citet{95AmAzLu.TiO} & Yes & Original transition data unfortunately not found: B-X (1,0) band at high sub-Doppler resolution (0.002 \cm{}) up to J=96 according to paper	\\
97BaMeMe	&	\citet{97BaMeMe.TiO} & No (Check) & Contains bands from very high $^3\Pi$ electronic states that give evidence of \D{} state at 12 284  \cm{} above \X{}, with a vibrational frequency around 968 \cm{}	\\
97LudAAmVe	&	\citet{97LudAAm.TiO} & 	Yes & Reanalysis of data from 96AmChLu	\\
98VeLuAm & \citet{98VeLuAm.TiO} & Yes & Reanalysis of data from 96AmChLu and  95AmAzLu \\

00CoSiGl	&	\citet{00CoSiGl.TiO} & No & Low-resolution data demonstrating detection only &		\\
01HePeDu & \citet{01HePeDu.TiO} & No & Unassigned very high temperature spectra \\
02AmLuVe & \citet{02AmLuVe.TiO} &No& No data on the \ce{^{48}Ti^{16}O} isotopologue & \\
03NaItDa	&	\citet{03NaItDa.TiO} & No & No new experimental data \\
05ViStBr & \citet{05ViStBr.TiO} & No & Zeeman splitting data only, B-X (0-0) and A-X (0-0) \\
12WoPaHo & \citet{12WoPaHo.TiO} & No & Unresolved spectra & \\
13HuLuChLa & \citet{13HuLuCh.TiO} &No& TiO$^+$ spectra, some low-resolution TiO bands not  considered here \\

\end{longtable}
\end{center}

There are a number of data sources, particularly from the early-mid twentieth century, 
which provide data on positions of bands (usually band-heads, though sometimes 
this is unspecified). Often these early studies went to significantly higher 
vibrational levels than more modern experiments which have tended to focus on very high 
accuracy rotationally resolved lines. These two types of data are often quite 
complementary and together build a quite extensive understanding of the rovibronic energies 
of the molecule.  We have collated data sources with information on bands in 
\Cref{tab:databandheads}.

Another important type of data is measurements of the intensity of bands and the 
lifetimes of states. The sources of this data have been collated in 
\Cref{tab:intensities}. These data are not used here but 
will be used later to verify the dipole moment 
curves for the Duo spectroscopic model of TiO. 

There are a number of other studies of TiO spectra which we have not been used in this study for various 
reasons. These data sources are collated in \Cref{tab:otherrefs} with comments.

\subsection{Comments on the rotationally-resolved data sources (Table 1)}
\label{subsec:comments}

Many papers give uncertainties that we adopt unaltered and found to be 
reasonably consistent with all other TiO data (i.e. a relatively small number of 
transitions needed adjusted uncertainties or could not be verified), specifically:  
0.02 \cm{} (for unblended lines, up to 0.07 \cm{} for unblended lines) in 
74Linton, 0.008 \cm{} (unblended lines) for 79HoGeMe,  0.01 \cm{} in 80GaBrDa, 
0.044 \cm{} in 85BrGa, 0.03 \cm{} in 91GuAmVe, 0.1 \cm{} in 91SiHaxx, 0.01 
\cm{} in 95KaMcHe, 0.002 \cm{} in 96BaMeMe, 0.02 \cm{} in 96RaBeWa. Other comments related to Table 1 are as follows. 

\begin{description}
\item[(1a)] Data due to Phillips (50Phillips, 51Phillips, 69Phillips, 71PhDa, 
71Phillips, 73Phillips-AX, 73Phillips-BX and 73Phillips-CX)  are obtained from 
photographic plates. Originally, we used 0.045 \cm{} as the estimated 
uncertainty for these data. However, we found significant inconsistencies with 
this uncertainty and increased it to 0.1 \cm{} for data published in these
papers and 0.2 \cm{} for data found from external sources (though these data have been 
analysed within the published papers). 
\item[(1b)] 51Phillips incorrectly assigns that the $\gamma$ band to the a 
$^3\Delta$--$^3\Pi$ band; it is actually a $^3\Phi$--$^3\Delta$ band (the lowest 
state at that stage was believed to be X$^3\Pi$). We have modified the $state$ 
and $\Omega$ quantum numbers. 

\item[(1c)] 69Phillips incorrectly identifies the band as the unphysical 
\B$_{1}${} -- \X$_{0}$ in the data table only, rather than \B$_{0}${} -- 
\X$_{1}$ (as in the text). 

\item[(1d)] 50Phillips-ext and 73Phillips data were obtained from tapes given by 
Phillips to Kurucz in 1981 (these data are not in the original publication). It is not 
clear if the c-a data from this tape data has been published; we have chosen 
to link the data to the original Phillips c-a paper, i.e. 50Phillips-ext. The 
bandhead details from the A-X, B-X and C-X data are given in 73Phillips; thus we 
assign the tape data on these bands to this paper. The  tape data has 174 
transitions which have unphysical assignments, $J \le | \Omega-\Sigma|$; e.g. an 
\A{} energy level with J$<$2. There are 55 C-X, 112 A-X and 7 c-a unphysical 
transitions. There is some repetition between data in the 73Phillips compilation 
and earlier data, e.g. the 71Phillips B-X data. However, the tape compilation of data 
is significantly more extensive while the former has been published explicitly 
assigned. Therefore, we use both. Note that the number of unverified transitions 
from these data is significantly higher than other data sources; however, as the 
resulting energies were reasonable, we chose not to exclude these data sets. We note
that these data have been used to inform some of the available TiO linelists, particularly 
the recent update of the \citet{98Plxxxx.TiO} linelist for inclusion in the VALD database \citep{VALD3}.

\item[(1e)] 72Linton: obs-calc was given as 0.03 \cm{}; however, we found  
uncertainties of 0.05 \cm{} were more consistent with other measurements. 

\item[(1f)] 72Lindgren gives no uncertainties; we used 0.05 \cm{} (based on 
72Linton) which gave self-consistent results.

\item[(1g)] 79HoGeMe: a full set of data were obtained from Amiot (private communication, 2015). Only 
the 0-0 data were provided in the original paper.

\item[(1h)] 79GaDe provides rovibrational energy levels, but does not distinguish between the spectra of different spin components; we have used the median $S=0$, i.e. $\Omega=2$ for the associated energy levels.

\item[(1i)] 90StShJu: the stated uncertainty is  0.5 MHz, on the order of 
$10^{-5}$ \cm{}, which has been adopted.

\item[(1j)] 91GuAmVe data were obtained from Amiot (private communication, 2015). 

\item[(1k)] 96AmChLu state that the width of the lines under their experimental 
conditions was 0.005 \cm{}; we adopted this as the estimated uncertainty of the 
line position.

\item[(1l)] 98NaSaRo estimated uncertainty is 
8 kHz, equivalent to $10^{-7}$ \cm{}, which has been adopted. 

\item[(1m)] 99RaBeDu laboratory and sunspots (SS) measurements: the need for 
consistency with other measurements (and to maximise the number of validated 
transitions and minimize the need for increased uncertainties of some lines) 
meant that we doubled the uncertainties from the original paper from 0.02 and 
0.005 \cm{} for lab and sunspot data to 0.004 and 0.01 \cm{}. 

\item[(1n)] 02KoHaMc uncertainties estimates were given as 0.002 -- 0.005 \cm{}; 
however, 0.01 \cm{} seems to be a more reasonable estimate based on the overall 
\Marvel\ model. This value was adopted. 

\end{description}

\subsection{Comments on data sources for band-head information (Table 3)}

\label{subsec:comments_bh}

\begin{description}

\item[(3a)] 69LiNi suggests assignments for 2 bands in the 28Lowater data, 7 in 
the 37WuMe data and 1 in the 57GaRoJu data. 

\item[(3b)] 29Christya has rotationally-resolved data, but more recent higher 
resolution data sources are available, so we only used the bandhead 
information. 

\item[(3c)]  72PhDa and 77LiBrb: it is assumed that the wavelengths are taken 
in air at standard temperature and pressure; a refraction index of 1.00029 is 
used to convert to frequency in vacuum.
\end{description}

\section{\Marvel\ energy levels}

\begin{figure*}
\includegraphics[width=0.7\textwidth]{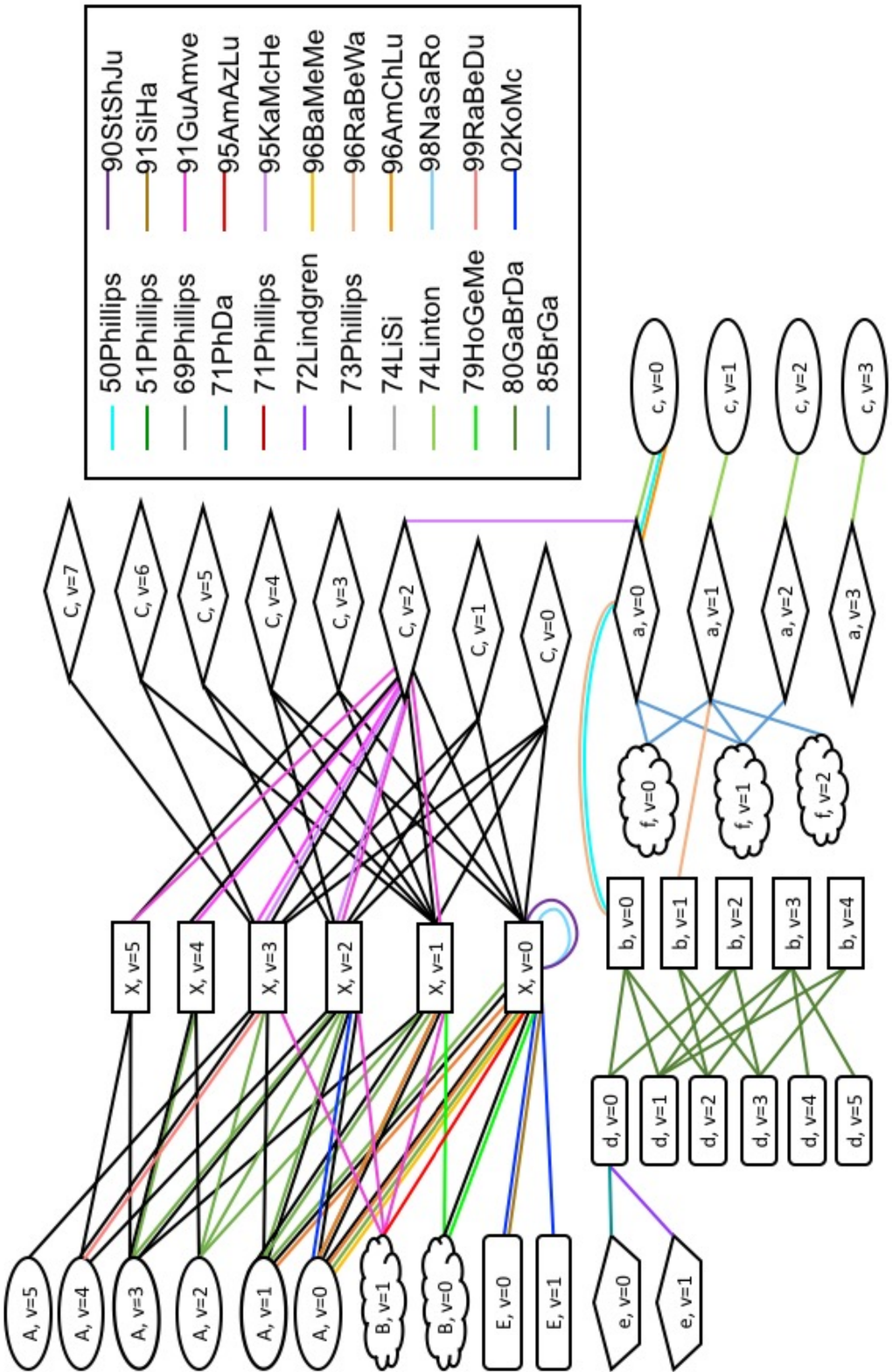}
\caption{\label{fig:vibronicSN}Vibronic structure of the \TiO\ spectroscopic network.}
\end{figure*}

\subsection{Spectroscopic Networks}
The vibronic structure of the spectroscopic network of the experimentally 
assigned TiO transitions is shown in \Cref{fig:vibronicSN}. Probably the most 
important observed transitions are the spin-forbidden \C{} -- \Sa{} transitions 
from \citet{95KaMcHe.TiO} that allows the relative energy of the triplet and 
singlet manifolds to be fixed.  
The figure makes clear that the \X{}, \A{} and \C{}  states, up to high 
vibrational energies, are well characterised. There are  a number of sources providing vibrational connections, 
though further observations of the vibrationally excited \C{}{} -- \X{} 
transitions with modern techniques would be beneficial. 

No transitions involving the \B{} state higher than $v=1$ have been assigned in 
rotationally-resolved spectra. The bond lengths of the \A{} and \B{} states are 
comparable and significantly larger than the bond length of the \X{} state; we 
thus expect that \B{}{} -- \X{} Franck Condon transitions with higher changes in 
vibrational quantum number should be observable like the \A{}{} -- \X{} 
transitions. Indeed, as discussed below, band-heads for these transitions have 
been assigned. 

The \E{} state is sparsely characterized and the key experiments by 
\citet{02KoHaMu.TiO} were only  performed after construction of the seminal TiO line lists 
of \citet{94Joxxxx.TiO}, \citet{98Plxxxx.TiO}, and 
\citet{98Scxxxx.TiO}. In particular, the observation of the $v=1$ band allow a 
reasonable Morse oscillator fit to the \E{} state potential energy curve that 
previously only was characterised by its ground vibrational level. 

Taken together, the experimental observations of the singlet states produce an 
almost completely connected network. For example, none of the \Sc{} -- \Sa{} 
transitions from \citet{74Lixxxx.TiO} involve a change in the vibrational quantum 
number due to the near parallel curves for the two states; by themselves these 
give no absolute vibrational energies. However, the \Sf{} -- \Sa{} transitions do often 
involve changes in the vibrational quantum number and allow the absolute vibrational 
energies of the \Sc{} and \Sa{} states to be extracted. These sorts of arguments 
are common in the singlet manifold; due to this there is only one band 
unconnected to the large TiO spectroscopic network: the transitions between the 
\Sc{} ($v$=3) and \Sa{} ($v$=3) states. This band is treated as a floating 
component in this study. Unlike in the triplet manifold, however, most 
transitions in the singlet manifold have only been measured once and often this 
is pre-1990s. Modern re-measurements would allow higher accuracy results for the 
singlet energy levels of TiO.

  \begin{table}
\caption{\label{tab:results} Extract from the 48Ti-16O.energies output file for \TiO. Energies and uncertainties are given in \cm{}. No indicates the number of transitions which contributed to the stated energy and uncertainty. }
\begin{center}
\renewcommand{\arraystretch}{1.2}
\begin{tabular}{lcclll}
\toprule
   \mc{1}{c}{State} & $J$ &  $v$  &   \mc{1}{c}{$\tilde{E}$}  &  \mc{1}{c}{Unc.} & \mc{1}{c}{No} \\
\midrule
X3Delta\_1	&	1	&	0	&	0.0	&	0.00001	&	36	\\
X3Delta\_1	&	2	&	0	&	2.111897	&	0.000007	&	50	\\
X3Delta\_1	&	3	&	0	&	5.279694	&	0.00001	&	60	\\
X3Delta\_1	&	4	&	0	&	9.505353	&	0.000199	&	59	\\
X3Delta\_1	&	5	&	0	&	14.78605	&	0.000199	&	58	\\
X3Delta\_1	&	6	&	0	&	21.121889	&	0.000001	&	65	\\
X3Delta\_1	&	7	&	0	&	28.513037	&	0.000001	&	61	\\
X3Delta\_1	&	8	&	0	&	36.959873	&	0.000001	&	68	\\
X3Delta\_1	&	9	&	0	&	46.463111	&	0.000001	&	70	\\
b1Pi	&	86f	&	0	&	18511.91059	&	0.008909	&	3	\\
A3Phi\_3	&	43	&	4	&	18513.79788	&	0.003993	&	10	\\
A3Phi\_2	&	47	&	4	&	18514.59668	&	0.003993	&	10	\\
A3Phi\_3	&	14	&	5	&	18514.86149	&	0.11547	&	3	\\
b1Pi	&	20e	&	4	&	18518.44328	&	0.005774	&	3	\\
A3Phi\_3	&	83	&	1	&	18520.93357	&	0.005725	&	18	\\
A3Phi\_4	&	39	&	4	&	18522.97952	&	0.003672	&	10	\\
A3Phi\_3	&	15	&	5	&	18529.54712	&	0.11547	&	3	\\
A3Phi\_2	&	24	&	5	&	18532.63495	&	0.11547	&	3	\\
B3Pi\_0	&	68b	&	0	&	18535.06106	&	0.11547	&	3	\\
B3Pi\_1	&	67b	&	0	&	18535.90423	&	0.11547	&	3	\\
B3Pi\_0	&	68a	&	0	&	18536.51772	&	0.11547	&	3	\\
B3Pi\_1	&	67a	&	0	&	18536.5709	&	0.11547	&	3	\\
B3Pi\_2	&	66a	&	0	&	18538.92466	&	0.141421	&	2	\\
B3Pi\_2	&	66b	&	0	&	18538.92466	&	0.141421	&	2	\\
b1Pi	&	21e	&	4	&	18539.41806	&	0.005774	&	3	\\
b1Pi	&	21f	&	4	&	18539.49211	&	0.01	&	1	\\
A3Phi\_2	&	75	&	2	&	18540.01583	&	0.057735	&	12	\\
B3Pi\_0	&	54b	&	1	&	18540.12798	&	0.008	&	1	\\
\bottomrule
\end{tabular}
  \end{center}
  
\end{table}

\subsection{\Marvel\ energy levels}

The final energy levels from the \Marvel\ analysis are collated in the 
Supplementary Information. An extract from this file, together with a 
description of each column, is provided in \Cref{tab:results}. 
The data of \Cref{tab:results} for the \X$_1$, \X$_2$, and \X$_3$ states, where the subscript corresponds to the three possible $\Omega$ values, confirm that the three fine-structure states have very slightly different ``rotational'' levels and that transitions have been observed within all three fine-structure states. Note also that only a  very small number of transitions within a fine-structure state have been measured, which calls for further experimental studies.

\begin{figure}
\includegraphics[width=\textwidth]{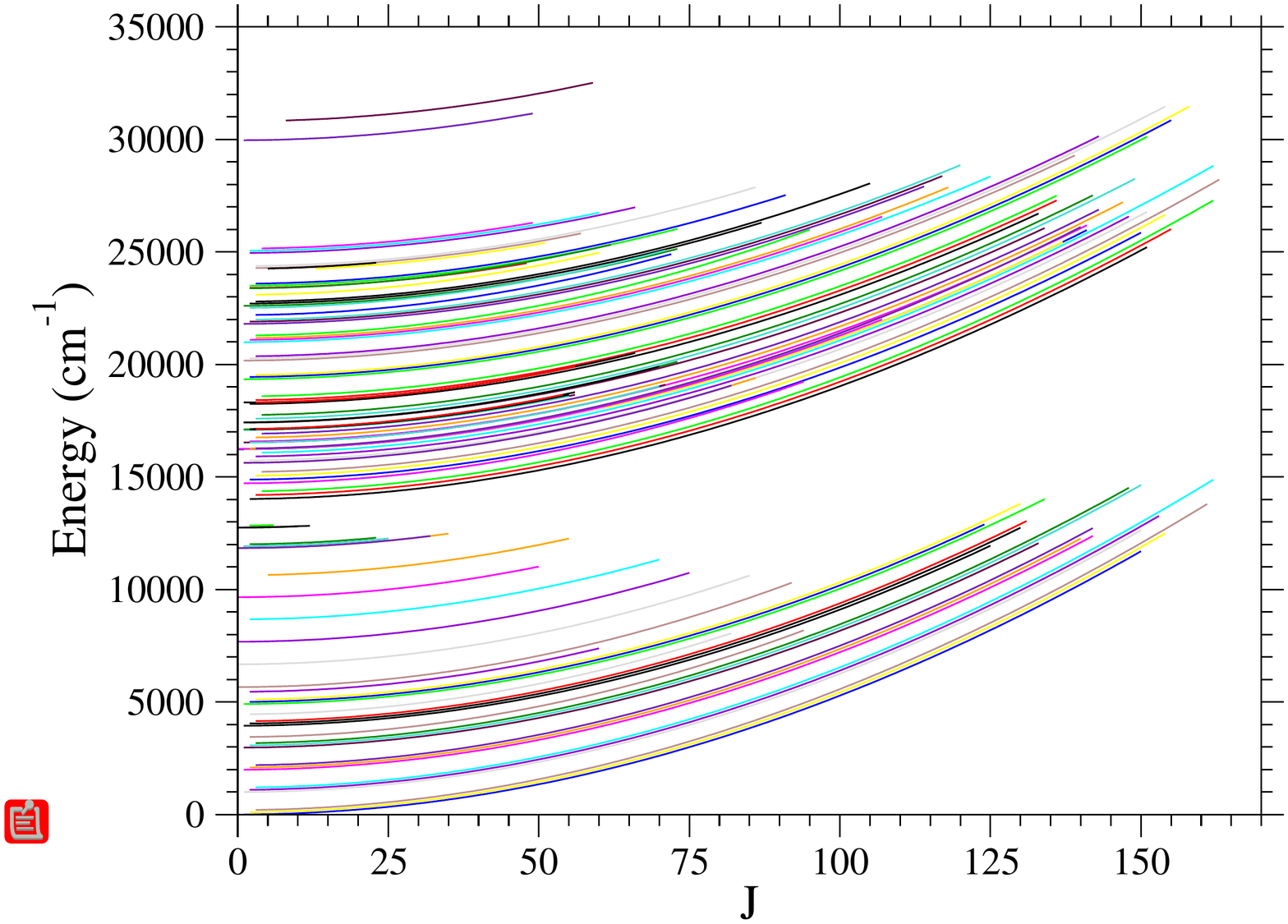}
\caption{\label{fig:EnvsJ}Summary of characterized energy levels. Different lines indicate different spin-vibronic states.}
\end{figure}

\Cref{fig:EnvsJ}  shows graphically the energy against the total angular momentum  for all different spin-vibronic states in the main spectroscopic network. The triplets can be identified by near parallel closely spaced lines. The vibrational levels of each electronic state are separated by approximately 1000 \cm{}. The fact that all curves are smooth quadratics provides confidence in the extracted \Marvel\ energy levels. 

\clearpage

\begin{center}
\footnotesize 
\begin{longtable}{lllclllllll}
\caption{\label{tab:E1}Summary of energy levels found through the \Marvel\ analysis.} \\ 
\toprule
 & $v$  & $p$ & $J$ Range   & \mc{3}{c}{Uncertainties (\cm{})} \\
&  &  & & Min & Aver. & Max \\
\midrule
\endfirsthead

\multicolumn{5}{c}%
{{ \tablename\ \thetable{} -- continued from previous page}} \\
\toprule
 & $v$  & $p$ &$J$ Range   & \mc{3}{c}{Uncertainties (\cm{})} \\
&  &  & & Min & Aver. & Max \\
\midrule
\endhead

\bottomrule
\multicolumn{5}{c}{{Continued on next page}} \\ 
\endfoot

\bottomrule
\endlastfoot
\vspace{-0.5em}\\ 
\X$_1$   &  0 &  & 1 - 150  & 0.0002 & 0.021 & 0.12\\*
 &  1 &  & 1 - 150  & 0.0013 & 0.026 & 0.14\\*
 &  2 &  & 1 - 142  & 0.0016 & 0.034 & 0.2\\*
 &  3 &  & 1 - 133  & 0.0016 & 0.039 & 0.2\\*
 &  4 &  & 1 - 125  & 0.0028 & 0.068 & 0.2\\*
 &  5 &  & 1 - 134  & 0.02 & 0.086 & 0.2\\
\vspace{-0.5em}\\ 
\X$_2$   &  0 &  & 2 - 154  & 0.0002 & 0.022 & 0.1\\*
 &  1 &  & 2 - 153  & 0.0012 & 0.025 & 0.2\\*
 &  2 &  & 2 - 140  & 0.0016 & 0.029 & 0.2\\*
 &  3 &  & 2 - 150  & 0.0016 & 0.04 & 0.2\\*
 &  4 &  & 2 - 130  & 0.0028 & 0.062 & 0.2\\*
 &  5 &  & 2 - 124  & 0.028 & 0.1 & 0.2\\
\vspace{-0.5em}\\ 
\X$_3$   &  0 &  & 3 - 161  & 0.00048 & 0.029 & 0.14\\*
 &  1 &  & 3 - 162  & 0.0013 & 0.036 & 0.14\\*
 &  2 &  & 3 - 142  & 0.0018 & 0.036 & 0.2\\*
 &  3 &  & 3 - 148  & 0.0017 & 0.055 & 0.26\\*
 &  4 &  & 3 - 131  & 0.0035 & 0.097 & 0.2\\*
 &  5 &  & 3 - 130  & 0.02 & 0.086 & 0.2\\
\vspace{-0.5em}\\ 
\A$_2$   &  0 &  & 2 - 151  & 0.0002 & 0.031 & 0.2\\*
 &  1 &  & 2 - 150  & 0.0015 & 0.031 & 0.14\\*
 &  2 &  & 2 - 151  & 0.0016 & 0.048 & 0.2\\*
 &  3 &  & 2 - 141  & 0.0018 & 0.05 & 0.2\\*
 &  4 &  & 2 - 134  & 0.0023 & 0.047 & 0.14\\*
 &  5 &  & 2 - 133  & 0.12 & 0.13 & 0.2\\
\vspace{-0.5em}\\ 
\A$_3$   &  0 &  & 3 - 155  & 0.0002 & 0.029 & 0.2\\*
 &  1 &  & 3 - 154  & 0.0013 & 0.029 & 0.2\\*
 &  2 &  & 3 - 148  & 0.0016 & 0.038 & 0.2\\*
 &  3 &  & 3 - 147  & 0.0018 & 0.053 & 0.2\\*
 &  4 &  & 3 - 149  & 0.0023 & 0.071 & 0.42\\*
 &  5 &  & 3 - 136  & 0.12 & 0.13 & 0.2\\
\vspace{-0.5em}\\ 
\A$_4$   &  0 &  & 4 - 162  & 0.00048 & 0.041 & 0.14\\*
 &  1 &  & 4 - 163  & 0.0014 & 0.045 & 0.2\\*
 &  2 &  & 4 - 162  & 0.0017 & 0.061 & 0.2\\*
 &  3 &  & 4 - 143  & 0.0023 & 0.07 & 0.2\\*
 &  4 &  & 4 - 142  & 0.0023 & 0.063 & 0.2\\*
 &  5 &  & 4 - 136  & 0.12 & 0.13 & 0.2\\
\vspace{-0.5em}\\ 
\B$_0$   &  0 & a & 0 - 141  & 0.0033 & 0.084 & 0.2\\*
 &  0 & b & 1 - 137  & 0.004 & 0.075 & 0.2\\*
 &  1 & a & 2 - 56  & 0.0033 & 0.0046 & 0.0081\\*
 &  1 & b & 1 - 55  & 0.0035 & 0.0058 & 0.014\\
\vspace{-0.5em}\\ 
\B$_1$   &  0 & a & 0 - 102  & 0.0032 & 0.046 & 0.2\\*
 &  0 & b & 0 - 107  & 0.0039 & 0.063 & 0.18\\*
 &  1 & a & 1 - 53  & 0.0023 & 0.0051 & 0.03\\*
 &  1 & b & 2 - 55  & 0.0036 & 0.0072 & 0.03\\
\vspace{-0.5em}\\ 
\B$_2$   &  0 & a & 2 - 140  & 0.0035 & 0.081 & 0.2\\*
 &  0 & b & 3 - 140  & 0.004 & 0.082 & 0.2\\*
 &  1 & a & 2 - 56  & 0.0033 & 0.0058 & 0.017\\*
 &  1 & b & 3 - 54  & 0.004 & 0.006 & 0.0094\\
\vspace{-0.5em}\\ 
\C$_1$   &  0 &  & 1 - 151  & 0.071 & 0.082 & 0.14\\*
 &  1 &  & 1 - 139  & 0.082 & 0.092 & 0.2\\*
 &  2 &  & 1 - 125  & 0.017 & 0.092 & 0.2\\*
 &  3 &  & 1 - 114  & 0.082 & 0.11 & 0.36\\*
 &  4 &  & 1 - 73  & 0.082 & 0.088 & 0.2\\*
 &  5 &  & 2 - 48  & 0.1 & 0.11 & 0.2\\*
 &  6 &  & 13 - 51  & 0.1 & 0.11 & 0.2\\*
 &  7 &  & 2 - 66  & 0.14 & 0.16 & 0.2\\
\vspace{-0.5em}\\ 
\C$_2$   &  0 &  & 2 - 155  & 0.071 & 0.09 & 0.2\\*
 &  1 &  & 2 - 154  & 0.082 & 0.1 & 0.2\\*
 &  2 &  & 2 - 107  & 0.028 & 0.082 & 0.2\\*
 &  3 &  & 2 - 117  & 0.082 & 0.094 & 0.2\\*
 &  4 &  & 2 - 87  & 0.082 & 0.089 & 0.2\\*
 &  5 &  & 2 - 73  & 0.1 & 0.12 & 0.36\\*
 &  6 &  & 3 - 57  & 0.1 & 0.11 & 0.2\\*
 &  7 &  & 2 - 60  & 0.14 & 0.15 & 0.2\\
\vspace{-0.5em}\\ 
\C$_3$   &  0 &  & 3 - 158  & 0.071 & 0.089 & 0.2\\*
 &  1 &  & 3 - 143  & 0.082 & 0.097 & 0.2\\*
 &  2 &  & 3 - 118  & 0.0036 & 0.067 & 0.2\\*
 &  3 &  & 3 - 120  & 0.082 & 0.1 & 0.2\\*
 &  4 &  & 3 - 105  & 0.082 & 0.094 & 0.2\\*
 &  5 &  & 3 - 91  & 0.1 & 0.11 & 0.33\\*
 &  6 &  & 3 - 86  & 0.1 & 0.12 & 0.23\\*
 &  7 &  & 4 - 49  & 0.14 & 0.15 & 0.2\\
\vspace{-0.5em}\\ 
\E$_0$   &  0 & a & 0 - 35  & 0.0057 & 0.0069 & 0.01\\*
 &  0 & b & 0 - 32  & 0.0057 & 0.0068 & 0.01\\*
 &  1 & a & 1 - 13  & 0.0058 & 0.0085 & 0.01\\*
 &  1 & b & 0 - 12  & 0.0058 & 0.0076 & 0.011\\
\vspace{-0.5em}\\ 
\E$_1$   &  0 & a & 1 - 25  & 0.0058 & 0.0066 & 0.01\\*
 &  0 & b & 1 - 25  & 0.0058 & 0.0067 & 0.01\\*
 &  1 & a & 2 - 6  & 0.01 & 0.01 & 0.01\\*
 &  1 & b & 2 - 6  & 0.01 & 0.01 & 0.01\\
\vspace{-0.5em}\\ 
\E$_2$   &  0 & a & 2 - 23  & 0.0058 & 0.0065 & 0.0071\\
\vspace{-0.5em}\\ 
\Sa  &  0 &  & 2 - 100  & 0.0024 & 0.0073 & 0.14\\*
 &  1 &  & 2 - 92  & 0.0063 & 0.034 & 0.32\\*
 &  2 &  & 2 - 60  & 0.011 & 0.013 & 0.022\\*
 &  3 &  & 5 - 59  & 0.011 & 0.014 & 0.021\\
\vspace{-0.5em}\\ 
\Sb  &  0 & e & 1 - 99  & 0.0038 & 0.0077 & 0.1\\*
 &  0 & f & 1 - 99  & 0.0051 & 0.0086 & 0.028\\*
 &  1 & e & 1 - 86  & 0.0034 & 0.0079 & 0.058\\*
 &  1 & f & 1 - 82  & 0.0046 & 0.0063 & 0.013\\*
 &  2 & e & 1 - 71  & 0.0041 & 0.0069 & 0.023\\*
 &  2 & f & 2 - 70  & 0.0058 & 0.0077 & 0.035\\*
 &  3 & e & 1 - 73  & 0.0045 & 0.0087 & 0.029\\*
 &  3 & f & 1 - 70  & 0.0058 & 0.01 & 0.056\\*
 &  4 & e & 1 - 66  & 0.0058 & 0.011 & 0.066\\*
 &  4 & f & 3 - 56  & 0.0071 & 0.0096 & 0.019\\
\vspace{-0.5em}\\ 
\Sc  &  0 &  & 3 - 101  & 0.0028 & 0.016 & 0.2\\*
 &  1 &  & 3 - 93  & 0.011 & 0.052 & 0.49\\*
 &  2 &  & 3 - 60  & 0.011 & 0.013 & 0.02\\*
 &  3 &  & 6 - 59  & 0.011 & 0.014 & 0.021\\
\vspace{-0.5em}\\ 
\Sd  &  0 &  & 0 - 92  & 0.0033 & 0.0045 & 0.01\\*
 &  1 &  & 0 - 85  & 0.0029 & 0.0047 & 0.028\\*
 &  2 &  & 0 - 75  & 0.0033 & 0.0054 & 0.01\\*
 &  3 &  & 2 - 70  & 0.0038 & 0.0065 & 0.02\\*
 &  4 &  & 0 - 50  & 0.0058 & 0.0088 & 0.024\\*
 &  5 &  & 2 - 55  & 0.0071 & 0.0094 & 0.01\\
\vspace{-0.5em}\\ 
\Se  &  0 &  & 1 - 49  & 0.035 & 0.041 & 0.053\\*
 &  1 &  & 8 - 59  & 0.035 & 0.04 & 0.078\\
\vspace{-0.5em}\\ 
\Sf  &  0 &  & 2 - 71  & 0.019 & 0.023 & 0.044\\*
 &  1 &  & 2 - 62  & 0.018 & 0.023 & 0.044\\*
 &  2 &  & 5 - 23  & 0.031 & 0.039 & 0.044\\
\end{longtable}
\end{center}

\Cref{tab:E1} tabulates the number of \Marvel\ energy levels that have been obtained for each spin-vibronic state, including the minimum, average and maximum uncertainty of the levels and the $J$ range covered. 
In the \X{}, \A{} and \C{} states, quite high vibrational excitations have been observed, which should facilitate high accuracy in the spectroscopically-refined potential energy curves (PEC) for these states. However, in the \E{} and \B{} states, only the ground and first excited vibrational states have available data. 
The \Sa{}, \Sb{}, \Sc{} and \Sd{} singlet states have been well characterised to moderate vibrational excitations which will permit good refinement of the PECs. The \Se{} and \Sf{} states have two and three vibrational levels characterised, respectively; this will permit reasonable first-order approximations to the PECs. Note, however, that the number of perturbing states at higher excitation energies is very large and the potential energy curves of the more highly excited states (particularly the \Se{} state) are likely to be stongly affected. 

\section{Discussion}

\begin{table}
\caption{\label{tab:tripletbandorigins}Triplet vibronic level origins from \Marvel\ data, and difference from \citet{98Scxxxx.TiO} line list data,  \TiO; $J_\text{min}$ = $\Omega$ unless otherwise specified; all numners are given in cm${-1}$ . }
\center
\begin{tabular}{HlHHrrrrrrr}
\\ \toprule
&$v$  & & &  \mc{1}{r}{\X$_1$} &  & \mc{1}{r}{\X$_2$} & &  \mc{1}{r}{\X$_3$}  \\
\hline
\X	 & 0 	 &   	 & 1 	 & 0.0000(2)    & +0.0000	 & 98.9039(2) & $-$0.036	 & 203.7006(5)  & $-$0.0229 \\
\X	 & 1 	 &   	 & 1 	 & 1000.019(5) & +0.003	 & 1098.922(6)  & $-$0.030 	 & 1203.711(6) & $-$0.015\\
\X	 & 2 	 &   	 & 1 	 & 1990.89(9) & $-$0.01 	 & 2089.790(4)  & $-$0.036 	 & 2194.579(5) & $-$0.026 \\
\X	 & 3 	 &   	 & 1 	 & 2972.55(9) & +0.02 	 & 3071.45(9) & $-$0.013 	 & 3176.235(7) & $-$0.004 \\
\X	 & 4 	 &   	 & 1 	 & 3945.2(1)& $-$0.2	 & 4044.1(1) 	 & $-$0.181 	 & 4148.70(1) & +0.01 \\
\X	 & 5 	 &   	 & 1 	 & 4908.3(1) & +0.0 	 & 5007.3(1) 	 & $-$0.123 	 & 5112.0(1) & $-$0.0 \\
\bottomrule
\vspace{-0.5em} \\

\\ \toprule
&$v$  & & &  \mc{1}{c}{\A$_2$} &  & \mc{1}{c}{\A$_3$} & &  \mc{1}{c}{\A$_4$}  \\
\hline
\A$_2$	 & 0 	 &   	 & 2 	 & 14021.6986(2) & +0.0369 	 & 14197.6325(2)   	 & +0.0302 	 & 14370.4654(5) & $-$0.0572 \\
\A$_2$	 & 1 	 &   	 & 2 	 & 14881.69(6)& $-$0.11 	 & 15057.388(3)  & +0.026	 & 15229.94(7) & +0.04 \\
\A$_2$	 & 2 	 &   	 & 2 	 & 15734.01(6) & +0.09 	 & 15909.39(6)  & $-$0.00 	 & 16081.55(6) & +0.06 \\
\A$_2$	 & 3 	 &   	 & 2 	 & 16578.51(6)  	 & +0.04 	 & 16753.58(6)  & $-$0.06 	 & 16925.45(6) & +0.02 \\
\A$_2$	 & 4 	 &   	 & 2 	 & 17414.91(7)  & $-$0.19 	 & 17589.85(7)  & $-$0.12	 & 17761.44(7) & $-$0.02\\
\A$_2$	 & 5 	 &   	 & 2 	 & 18243.4(2)  & +0.3	 & 18418.0(1)  & $-$0.0 & 18589.4(1) &-0.0 \\
\bottomrule

\vspace{-0.5em} \\

\\ \toprule
&$v$  & & &  \mc{1}{c}{\B$_0$} &  & \mc{1}{c}{\B$_1$} & &  \mc{1}{c}{\B$_2$}  \\
\hline
\B$_0$	 & 0 	 &  b 	 & 1 	 & 16225.767(6) 
& 1 	 & 16248.457(6) 
& 2 	 & 16267.360(6) \\
\B$_0$	 & 1 	 &  a 	 
 & 1 	 & 17089.313(8)$^{J=1}$ 
 & 1 	 & 17112.64(3) 
& 2 	 & 17131.681(8) \\
\bottomrule

\vspace{-0.5em} \\

\\ \toprule
&$v$  & & &  \mc{1}{c}{\C$_1$} &  & \mc{1}{c}{\C$_2$} & &  \mc{1}{c}{\C$_3$}  \\
\hline
\C$_1$	 & 0 	 &   	 & 1 	 & 19341.5(1)  	 & $-$0.7	 & 19442.3(1)  & +1.0 	 & 19537.2(1) & +0.9 \\
\C$_1$	 & 1 	 &   	 & 1 	 & 20170.1(1) 	 & $-$0.3 	 & 20271.3(1) 	 & +0.9 	 & 20365.5(1) & +0.8\\
\C$_1$	 & 2 	 &   	 & 1 	 & 20990.6(1)  & $-$0.8	 & 21091.4(1) 	 & +0.8 	 & 21181.262(4) & $-$0.051\\
\C$_1$	 & 3 	 &   	 & 1 	 & 21802.4(2)  	 & $-$1.7  	 & 21902.8(1) & +0.3 	 & 21993.3(1) & +0.1 \\
\C$_1$	 & 4 	 &   	 & 1 	 & 22605.3(1)  & $-$3.0 	 & 22704.6(1)	 & $-$0.2 	 & 22797.0(1) & +0.1\\
\C$_1$	 & 5 	 &   	 & 2 	 & 23401.9(2)$^{J=2}$ & 	 & 23497.1(2) 	 & $-$0.2	 & 23591.9(2) & $-$0.3 \\
\C$_1$	 & 6 	 &   	 & 13 	 & 24252.0(1)$^{J=13}$ &  	 & 24283.2(2)$^{J=3}$ & 3 	 & 24376.7(2) \\
\C$_1$	 & 7 	 &   	 & 2 	 & 24952.4(2)$^{J=2}$ &  	 & 25053.7(2)  & 4 	 & 25155.5(2)$^{J=4}$ \\
\bottomrule
\vspace{-0.5em} \\

\\ \toprule
&$v$ & & &  \mc{1}{c}{\E$_0$} &  & \mc{1}{c}{\E$_1$} & &  \mc{1}{c}{\E$_2$}  \\
\hline
\E$_0$	 & 0 	 &  a 	 & 0 	 & 11838.204(5)  
 & 1 	 & 11924.082(5) &
 2 	 & 12013.724(5) \\
\E$_0$	 & 1 	 &  a 	 & 1 	 & 12752.166(4) &
 2 	 & 12838.667(5) \\
\bottomrule
\end{tabular}
\end{table}

\begin{table}
\caption{\label{tab:singletbandorigins}Singlet vibronic level origins in \cm{} for \TiO; $J_\text{min}$ = $\Lambda$ (= $\Omega$) unless otherwise specified.}
\center
\begin{tabular}{llHlrrrrr}
\\ \toprule
& $v$ & p & $J$ &\Marvel\ & \citet{98Scxxxx.TiO}  \\
\hline
\Sa & 0 	 &   	 & 2 	 & 3446.481(8) & $-$0.044 \\
	 & 1 	 &   	 & 2 	 & 4455.67(2) & $-$0.03\\
	 & 2 	 &   	 & 2 	 & 5455.83(2) & +0.022\\
	 \vspace{-0.5em} \\

\Sb	 & 0 	 &  b 	 & 1 	 & 14717.055(9) & +3.016 \\
	 & 1 	 &  b 	 & 1 	 & 15628.21(1) & +3.175\\
	 & 2 	 &  a 	 & 1 	 & 16530.741(6) & +3.176\\
	 & 3 	 &  b 	 & 1 	 & 17424.48(1) & +3.14\\
	 & 4 	 &  a 	 & 1 	 & 18309.459(7) & +2.995 \\
\vspace{-0.5em} \\
\Sc	 & 0 	 &   	 & 3 	 & 21290.11(1) & +0.20 \\
	 & 1 	 &   	 & 3 	 & 22199.59(2) & $-$0.145 \\
 & 2 	 &   	 & 3 	 & 23099.06(1) & $-$0.127  \\
 \vspace{-0.5em} \\

\Sd	 & 0 	 &   	 & 0 	 & 5661.92(1) & +0.03 \\
	 & 1 	 &   	 & 0 	 & 6675.304(7) & $-$0.08 \\
 & 2 	 &   	 & 0 	 & 7678.78(1) & $-$0.04 \\
	 & 3 	 &   	 & 2 	 & 8675.824(7)& $-$0.080\\
 & 4 	 &   	 & 0 	 & 9656.64(1)  & $-$0.07\\
	 & 5 	 &   	 & 5 	 & 10646.90(1) & $-$0.07  \\
\vspace{-0.5em} \\

\Se	 & 0 	 &   	 & 1 	 & 29960.98(5)\\
	 & 1 	 &   	 & 8 	 & 30839.17(5)\\
\vspace{-0.5em} \\

\Sf	 & 0 	 &   	 & 2 	 & 22515.29(3) \\
	 & 1 	 &   	 & 2 	 & 23384.44(4) \\
	 & 2 	 &   	 & 5 	 & 24260.42(3)\\
\bottomrule
\end{tabular}
\end{table}

\subsection{Vibronic Band Origins}
The triplet and singlet vibronic band origins from the \Marvel\ data are given 
in \Cref{tab:tripletbandorigins} and \Cref{tab:singletbandorigins}, 
respectively. In most cases, the level given is the lowest possible $J$ for that 
spin-vibronic state; however, there are some cases (e.g., high vibrational states 
of the \C{} state) where this level was not observed. These \Marvel\ data will 
soon be used with high level \abinitio\ data to construct a full spectroscopic 
model of \TiO; this can be used to predict the lowest $J$ energy levels for all 
states, as well as higher vibrational levels not accessed by 
rotationally-resolved \TiO\ data. 

The \C$_{3}$ ($v$=2) origin and the \Sc{} ($v$=0) origin are separated by about 
120 \cm{} and are spin-orbit coupled; the resulting perturbations have been 
extensively studied, see \citet{03NaItDa.TiO}. 
The vibronic band origins are consistent with the spectroscopic parameters 
(term energies, vibrational frequencies and spin-orbit couplings) extracted 
previously from individual experiments using model Hamiltonians.

\begin{table*}
\footnotesize
\caption{\label{tab:bh_AX} Triplet \A{} -- \X{} R-branch band-heads  in \cm{} for \TiO.}
\begin{tabular}{lrrrrrrrrrrrrrrr}
\\ \toprule
$v$'-$v$" 	&	 \mc{3}{c}{\A$_2$ -- \X$_1$ (c)} 	&	 \mc{3}{c}{\A$_3$ -- \X$_2$ (b)} 	&	 \mc{3}{c}{\A$_4$ -- \X$_3$ (a)} \\
\cmidrule(r){2-4} \cmidrule{5-7} \cmidrule(l){8-10}
	&	 $J$ 	&	 MARVEL 	&	 Low-res obs. 	&	 $J$ 	&	 MARVEL 	&	 Low-res obs. 	&	 $J$ 	&	 MARVEL 	&	 Low-res obs. \\
\hline
0-0	&	20	&	14030.258	& 14030.1 [1]	&	18	&	14105.342	& 14104.7 [1]	&	17	&	14171.984	&	14171.4 [1] &	\\
0-1	&	23	&	13031.547	&	&	20	&	13106.365	&	&	19	&	13172.872	&	&	\\
0-2	&	26	&	12042.400	&	&	23	&	12116.854&	&	21	&	12183.165	&	&	\\
0-3*	&	31	&	11063.037	&	&	26	&	11136.944	&	&	24	&	11203.004	&	&	\\
0-4*	&	37	&	10093.892	&	&	31	&	10166.938	&	&	28	&	10232.599	&	&	\\
0-5*	&	46	&	9135.714	&	&	38	&	9207.438	&	&	34	&	9272.352	&	&	\\
1-0	&	18	&	14889.145	& 14889.4 [1]&	16	&	14964.137	& 14963.6 [1] &	15	&	15030.610	& 15030.1 [1]	&	\\
1-1	&	20	&	13890.137	& 13889.6 [1]	&	18	&	13964.949	& 13964.5 [1]	&	17	&	14031.319	& 14030.1 [1]	&	\\
1-2	&	23	&	12900.552	&	&	20	&	12975.114&	&	19	&	13041.355	&	&	\\
1-3	&	26	&	11920.557	&	&	23	&	11994.751&	&	21	&	12060.818	&	&	\\
1-4*	&	30	&	10950.355	&	&	26	&	11024.037	&	&	24	&	11089.856	&	&	\\
1-5*	&	37	&	9990.374	&	&	32	&	10063.272	&	&	28	&	10128.667	&	&	\\
2-0	&	16	&	15740.491	& 15743.1 [1]	&	15	&	15815.347	& 15814.7 [1]	&	14	&	15881.637	&	&	\\
2-1	&	18	&	14741.273	& 14741.3 [1]	&	16	&	14815.991	&	&	15	&	14882.195	&	&	\\
2-2	&	20	&	13751.408	&	&	18	&	13825.938	&	&	17	&	13892.056	&	&	\\
2-3	&	22	&	12770.968	&	&	20	&	12845.272	&	&	18	&	12911.259	&	&	\\
2-4	&	26	&	11800.133	&	&	23	&	11874.095&	&	21	&	11939.912	&	&	\\
2-5*	&	30	&	10839.158	&	&	25	&	10912.584	&	&	24	&	10978.171	&	&	\\
3-0*	&	15	&	16584.161	&	&	14	&	16658.838	&	&	12	&	16724.832	&	&	\\
3-1	&	16	&	15584.788	&	15586.3 [1] &	15	&	15659.365	& 15658.9 [1]	&	14	&	15725.306	&	&	\\
3-2	&	18	&	14594.705	& 14594.0 [1]	&	16	&	14669.158	& 14669.1 [1]	&	15	&	14735.027	&	&	\\
3-3*	&	19	&	13613.992	&	&	18	&	13688.271	&	&	16	&	13754.043	&	&	\\
3-4	&	22	&	12642.744	&	&	20	&	12716.789	&	&	18	&	12782.465	&	&	\\
3-5	&	26	&	11681.134	&	&	23	&	11754.775 &	&	21	&	11820.357&	&	\\
4-0*	&	13	&	17420.027	&	&	12	&	17494.548	&	&	12	&	17560.413	&	&	\\
4-1*	&	14	&	16420.538	&	&	13	&	16494.964	&	&	13	&	16560.785	&	&	\\
4-2	&	16	&	15430.316	& 15430.2 [1]	&	15	&	15504.628	& 15505.4 [1]	&	14	&	15570.404	&	&	\\
4-3	&	17	&	14449.410	&	&	16	&	14523.591	& 14522.8 [1]	&	15	&	14589.288	& 14588.0 [1]	&	\\
4-4*	&	19	&	13477.864	&	&	17	&	13551.898	&	&	16	&	13617.525	&	&	\\
4-5	&	22	&	12515.842	&	&	20	&	12589.596	&	&	18	&	12655.156	&	&	\\
5-0*	&	12	&	18248.069	&	&	11	&	18322.338	&	&	10	&	18387.978	&	&	\\
5-1*	&	13	&	17248.458	&	&	12	&	17322.646	&	&	12	&	17388.282	&	&	\\
5-2*	&	15	&	16258.090	& 16258.9 [1]	&	13	&	16332.224	&	&	12	&	16397.805	&	&	\\
5-3	&	15	&	15277.035	& 15276.6 [1]	&	14	&	15351.024	& 15350.6 [1]	&	14	&	15416.585	&	&	\\
5-4*	&	17	&	14305.324	&	&	16	&	14379.181	&	&	15	&	14444.694	&	&	\\
5-5*	&	19	&	13343.000	&	&	19	&	13416.662	&	&	16	&	13482.091	&	&	\\
\bottomrule
\end{tabular}

[1] 28Lowater \citep{28Lowater.TiO}, [2] 29Christya \citep{29Chxxx1.TiO}, [3] 72PhDa \citep{72PhDaxx.TiO}

* \Marvel\ predicted band-heads
\end{table*}

\begin{table*}
\footnotesize
\caption{\label{tab:bh_BX} Triplet \B -- \X{} R-branch  band-heads in \cm{} for \TiO.}
\begin{tabular}{lrrp{2cm}rrp{2cm}rrp{2cm}HHH}
\\ \toprule
v'-v" 	&	 \mc{3}{c}{\B$_0$ -- \X$_1$} 	&	 \mc{3}{c}{\B$_1$ -- \X$_2$} 	&	 \mc{3}{c}{\B$_2$ -- \X$_3$} \\
\cmidrule(r){2-4} \cmidrule{5-7} \cmidrule(l){8-10}
	&	 $J$ 	&	 MARVEL 	&	 Low-res obs. 	&	 $J$ 	&	 MARVEL 	&	 Low-res obs. 	&	 $J$ 	&	 MARVEL 	&	 Low-res obs. \\
\hline
0-0	&	12	&	16233.187	&16233 [1]	&	17	&	16160.243	&	16160 [2]&	28	&	16085.853	& 16085 [2]	&	\\
	&		&		&16233 [2]	&		&		&	16160 [2]&		&		& 16085 [2]	&	\\
0-1	&	13	&	15233.618	& 15218 [1]	&	19	&	15161.155	& 15156 [2]	&	32	&	15088.458	& 15081 [1]	&	\\
0-2*	&	15	&	14243.289	&	&	22	&	14171.535	&	&	36	&	14101.011	&	&	\\
0-3*	&	16	&	13262.269	&	&	26	&	13191.512	&	&	41	&	13123.750	&	&	\\
0-4*	&	18	&	12290.645	&	&	31	&	12221.408	&	&	47	&	12157.203	&	&	\\
0-5*	&	22	&	11328.578	&	&	38	&	11261.867	&	&	57	&	11202.350&	&	\\
1-0	&	12	&	17096.309	&17098 [1]	&	15	&	17023.495	& 17022 [2]	&	25	&	16947.583	&16950 [1]	&	\\
	&		&		& 17095 [2]	&		&		& 17022 [2]	&		&		&16950 [2]	&	\\
1-1	&	12	&	16096.673	&16081 [1]	&	17	&	16024.203	& 16022  [1]	&	28	&	15949.664	& 15930 [1]	&	\\
&		&		& 16096 [2]	&		&		&  16023 [2]	&		&	15949.664	& 15949 [2]	&	\\
1-2	&	14	&	15106.267	&	&	19	&	15034.244	&	&	31	&	14961.413	&	&	\\
1-3	&	15	&	14125.142	&	&	22	&	14053.757	&	&	35	&	13983.062	&	&	\\
1-4*	&	17	&	13153.331	&	&	25	&	13082.904	&	&	41	&	13014.997	&	&	\\
1-5*	&	19	&	12190.954	&	&	30	&	12121.999	&	&	48	&	12057.532	&	&	\\
2-0* & & & 17952 [1] & & & 17881 [1] & & & 17804 [1] \\
2-1* & & & 16931 [1] & & & 16877 [1] & & & 16799 [1] \\
 & & &  & & & 16881 [2] & & &  16804 [2] \\
2-2* & & & 15961 [2] & & & 15887 [1] & & & 15814 [2] \\
 & & & & & & 15887 [2] & & & \\
3-0* & & & & & & 18727 [1] & \\
3-1* & & & 17804 [1] & & & 17722 [1] & & & 17650 [1] \\
3-2* & & & 16799 [1] & & & 16717 [1] & & & 16654 [1] \\
 & & &  & & &  16736 [2] & & &  16663 [2] \\
4-2* & & & 17650 [1] & & & 17579 [1] & & & 17502 \\
4-3* & & & 16654 [1] & & & 16574 [1] & & & 16504 [1] \\
 & & &  & & & 16596 [2] & & & 16521 [2] \\
5-4* & & & & & & 16332 [2] & & & 16382 [2] \\
\bottomrule
\end{tabular}

[1] 69Phillips \citep{69Phxxxx.TiO}, [2] 76ZyPa \citep{76ZyPaxx.TiO}

* \Marvel\ predicted band-heads
\end{table*}
\clearpage

\begin{center}
\footnotesize
\begin{longtable}{lrrp{2cm}rrp{2cm}rrp{2cm}p{2cm}}
\caption{\label{tab:bh_CX2} \C{} -- \X{} R-branch band-heads for \TiO.} \\
\toprule
v'-v" 	&	 \mc{3}{c}{\C$_1${} -- \X$_1$} 	&	 \mc{3}{c}{\C$_2${} -- \X$_2$} 	&	 \mc{3}{c}{\C$_3${} -- \X$_3$} & 	 \mc{1}{c}{\C{} -- \X} \\
\midrule
\cmidrule(r){2-4} \cmidrule(r){5-7} \cmidrule(r){8-10} \cmidrule(r){11-11}
	&	 $J$ 	&	 MARVEL 	&	 Low-res obs. 	&	 $J$ 	&	 MARVEL 	&	 Low-res obs. 	&	 $J$ 	&	 MARVEL 	&	 Low-res obs. &  Low-res obs.\\

\midrule
\endfirsthead

\multicolumn{5}{c}%
{{ \tablename\ \thetable{} -- continued from previous page}} \\
\toprule
v'-v" 	&	 \mc{3}{c}{\C$_1${} -- \X$_1$} 	&	 \mc{3}{c}{\C$_2${} -- \X$_2$} 	&	 \mc{3}{c}{\C$_3${} -- \X$_3$} & 	 \mc{1}{c}{\C{} -- \X} \\
\midrule
\cmidrule(r){2-4} \cmidrule(r){5-7} \cmidrule(r){8-10} \cmidrule(r){11-11}
	&	 $J$ 	&	 MARVEL 	&	 Low-res obs. 	&	 $J$ 	&	 MARVEL 	&	 Low-res obs. 	&	 $J$ 	&	 MARVEL 	&	 Low-res obs. &  Low-res obs.\\
\midrule
\endhead

\bottomrule
\multicolumn{5}{c}{{Continued on next page}} \\
\endfoot

\bottomrule
\endlastfoot

\hline
0-0	&	11	&	19347.333	&	19347 [2] &	11	&	19349.241	& 19349 [2]	&	11	&	19339.917	&	19340 [2] & 19348 [3]	\\
0-1	&	12	&	18347.688	& 18347 [2]&	12	&	18349.571	& 18349 [2] &	12	&	18340.234	& 18339 [2] & 18350 [3]	\\
0-2	&	13	&	17357.230	& 17358 [1]	&	13	&	17359.1176& 17361 [1]&	13	&	17349.818	& 17350 [1]	& 17359 [3]	\\
0-3	&	14	&	16376.034	&	&	13	&	16377.898	&	&	14	&	16368.658	&	& 16378 [3]	\\
0-4*	&	15	&	15404.171	&	&	15	&	15405.969	&	&	15	&	15396.793	&	&	\\
0-5*	&	17	&	14441.577	&	&	16	&	14443.352	&	&	17	&	14434.300	&	&	\\
1-0	&	10	&	20175.638	&	 20177 [2] &	10	&	20177.880	& 20178 [2] &	10	&	20167.862	& 20168 [2]	& 20176 [3]	\\
1-1*	&	11	&	19175.912	&	&	11	&	19178.139	&	&	12	&	19168.148	&	&	\\
1-2	&	12	&	18185.422	&	&	12	&	18187.640	&	&	12	&	18177.671	&	& 18186 [2]	\\
	&		&		&	&	&		&	&		&		&	& 18186 [3]	\\
1-3	&	13	&	17204.151	& 17204 [1]	&	13	&	17206.354	& 17207 [2,3]	&	13	&	17196.403	&	17192 [1]&	\\
1-4*	&	14	&	16232.174	&  16231 [1]	&	14	&	16234.350	&	&	14	&	16224.432	&	&	\\
	&		&		& 16232 [2]	&		\\
1-5*	&	15	&	15269.512	&	&	15	&	15271.589	&	&	16	&	15261.827	& 15264 [1]	&	\\
2-0	&	10	&	20995.714	& 20995 [2]	&	9	&	20997.734	&  20997 [2]	&	10	&	20983.435	& 20983 [2]	& 20998 [3]	\\
2-1	&	10	&	19995.958	& 19995 [2]	&	9	&	19997.910	& 19996 [2]&	11	&	19983.702	& 19984 [2]	& 19998 [3]\\
2-2	&	10	&	19005.366	&	&	11	&	19007.348	&	&	12	&	18993.166	&	&	\\
2-3	&	12	&	18024.053	&	&	11	&	18025.981	&	&	13	&	18011.876	&	& 18026 [3]	\\
2-4	&	13	&	17051.956	& 17051 [1]	&	12	&	17053.891	& 17055 [2]	&	13	&	17039.866	&	& 17054 [3]	\\
2-5	&	13	&	16089.177	&	&	14	&	16091.005	&	&	15	&	16077.182	&	& 16086 [2]\\

3-0	&	9	&	21807.075	& 21806 [2]	&	9	&	21808.677	& 21809 [2]	&	10	&	21795.164	& 21795 [2]& 21809 [3]	\\
3-1	&	9	&	20807.262	& 20807 [2]	&	9	&	20808.853	& 20810 [2]	&	10	&	20795.391	&	20796 [2] & 20810 [3]	\\
3-2*	&	11	&	19816.669	&	&	11	&	19818.257	&	&	10	&	19804.784	&	&	\\
3-3*	&	11	&	18835.356	&	&	11	&	18836.890	&	&	11	&	18823.417	&	& 18835 [2]	\\
3-4*	&	11	&	17863.148	&	&	11	&	17864.735	&	&	12	&	17851.329	&	& 17859.4 [2]\\
3-5	&	12	&	16900.260	&	&	12	&	16901.808	&	&	13	&	16888.491	&	& 16901 [3]	\\
3-6* & & & & &  & 15949 [1] \\
 & & & & &  & 15950 [2] \\
4-0	&	8	&	22609.714	&	&	8	&	22610.389	& 22610 [2]&	9	&	22598.630	& 22598 [2]	&   22608 [3]	\\
4-1	&	10	&	21609.903	&	&	9	&	21610.534	& 21610 [2]&	9	&	21598.807	&	& 21611 [3]	\\
4-2	&	10	&	20619.300	&	&	10	&	20619.912	&	&	10	&	20608.153	&	20611 [2]& 20624 [2]	\\
	&		&	&	&	&	&	&	&	&	&  20621  [3]	\\
4-3*	&	11	&	19637.898	&	&	10	&	19638.495	&	&	11	&	19626.797	&	&	\\
4-4*	&	11	&	18665.690	&	&	11	&	18666.320	&	&	11	&	18654.639	&	& 18655 [2]	\\
4-5*	&	11	&	17702.743	&	&	11	&	17703.359	&	&	11	&	17691.749	&	&	\\
5-0*	&	8	&	23404.326	&	&	8	&	23402.807	&	&	7	&	23392.984	&	& 23413 [2]	\\
5-1	&	8	&	22404.467	&	&	8	&	22402.937	& 22403 [2]	&	9	&	22393.112	&	& 22405 [3]\\
5-2	&	9	&	21413.793	& 21414 [2]	&	10	&	21412.258	& 20412 [2]	&	10	&	21402.472	&	20402 [2] &	\\
5-3*	&	11	&	20432.387	&	20433 [2] &	10	&	20430.841	& 20431 [2]	&	10	&	20421.063	& 20423 [2]	&	\\
5-4*	&	11	&	19460.179	&	&	10	&	19458.568	&	&	10	&	19448.840	&	&	\\
5-5*	&	11	&	18497.232	&	&	10	&	18495.551	&	&	11	&	18485.936	&	&	\\
6-0*	&	&	&	&	7	&	24185.806	&	&	8	&	24177.683	&	&	\\
6-1*	&	&	&	&	7	&	23185.891	&	&	8	&	23177.807	&	& 23169 [2]\\
6-2	&	&	&	&	9	&	22195.159	&22196 [3]	&	9	&	22187.097	& 22187 [2]	& 	\\
6-3	&	&	&	&	9	&	21213.672	&	&	9	&	21205.630	&	&	\\
6-4*	&	&	&	&	9	&	20241.353	&	&	10	&	20233.358	&	&	\\
6-5*	&	&	&	&	10	&	19278.337	&	&	10	&	19270.397	&	&	\\
7-0*	&	7	&	24954.533	&	&	7	&	24958.629	&	&	7	&	24952.632	&	&	\\
7-1*	&	7	&	23954.634	&	&	7	&	23958.714	&	&	8	&	23952.739	&	& 23951[2]	\\
7-2*	&	7	&	22963.885	&	&	7	&	22967.964	&	&	8	&	22962.022	&	&	 22963 [2] \\
7-3	&	8	&	21982.354	&	&	8	&	21986.421	& 21986 [3]	&	8	&	21980.502	& 21981 [2]	&	\\
7-4*	&	10	&	21010.106	& 	&	8	&	21014.077	&	&	10	&	21008.202	&	21008 [2]&	 21017 [2]\\
7-5*	&	10	&	20047.088	&	&	9	&	20050.967	&	&	10	&	20045.240	&	&	\\
\end{longtable}

[1] 28Lowater \citep{28Lowater.TiO}, [2] 29Christya \citep{29Chxxx1.TiO}, [3] 72PhDa \citep{72PhDaxx.TiO}

* \Marvel\ predicted band-heads
\end{center}

\begin{table*}
\footnotesize
\caption{\label{tab:bh_EX} \E -- \X{} R-branch band-heads in \cm{} for \TiO.}
\begin{tabular}{llrp{2cm}HHp{2cm}HHp{2cm}H}
\\ \toprule
v'-v" 	&	 \mc{3}{c}{\E$_0$ -- \X$_1$} 	&	 \mc{3}{c}{\E$_1$ -- \X$_2$} 	&	 \mc{3}{c}{\E$_3$ -- \X$_3$} \\
\cmidrule(r){2-4} \cmidrule(r){5-7} \cmidrule(r){8-10} 
	&	 $J$ 	&	 MARVEL 	&	 Low-res obs. 	&	 $J$ 	&	 MARVEL 	&	 Low-res obs. 	&	 $J$ 	&	 MARVEL 	&	 Low-res obs. &  Low-res obs.\\
\hline
0-0 & 26 & 11854.767 & 11856 [1] & & & 11842 [1] & && 11828 [1] \\
0-1 & 32 &  10856.099 & 10857 [1] &  &&  10845 [1] & &&  10831 [1] \\
1-0 & & & 12774 [1] & && 12760 [1] & && 12743 [1] \\
1-1* & & & 11768 [1] & && 11753 [1] & & &11739 [1] \\
1-2* & & & 10777 [1] & && 10766 [1] & & &10752 [1] \\
2-1* & & & 12674  [1] & && 12658 [1] & & &12643 [1] \\
2-2* & & & 11679 [1] & && 11667 [1] & & &11652 [1] \\
2-3* & & & 10701 [1] & & &10689 [1] & && 10675 [1] \\
3-2* & & & 12578 [1] & & &12564 [1] & && 12548 [1] \\
3-3* & & & 11588 [1] & & &11576 [1] & && 11564 [1] \\
3-4* & & & 10623 [1] & & &10607 [1] & && 10594 [1] \\
4-3* & & & 12478 [1] & & &12462 [1] & && 12448 [1] \\
4-4* & & & 11504 [1] & & &11487 [1] & && 11474 [1] \\
4-5* & & & 10544 [1] & & &10521 [1] & && 10509 [1] \\
5-4* & & & 12371 [1] & & & 12356 [1] & & & 12342 [1]\\
\bottomrule

\end{tabular}

[1] 77LiBr \citep{77LiBrxx.TiO}

* \Marvel\ predicted band-heads
\end{table*}

\begin{center}
\footnotesize
\begin{longtable}{llrrlll}
\caption{\label{tab:bh_singlet}  Singlet R-branch  band-heads in \cm{} for \TiO.} \\
\toprule
& v'-v" & $J$ & MARVEL & Low-res obs. \\
\midrule
\endfirsthead

\multicolumn{5}{c}%
{{ \tablename\ \thetable{} -- continued from previous page}} \\
\toprule
& v'-v" & $J$ & MARVEL & Low-res obs. \\
\midrule
\endhead

\bottomrule
\multicolumn{5}{c}{{Continued on next page}} \\
\endfoot

\bottomrule

\endlastfoot

\Sb{} -- \Sa{} & 0-0 & 22 & 11284.109 &  & \\
 & 0-1* & 25 & 10276.404 & 10280 [1] 10282 [3] & \\
 & 0-2* & 28 & 9278.175 &  & \\
 & 1-0 & 19 & 12194.027 & 12194 [2] & \\
 & 1-1* & 22 & 11185.966 &  11186 [1] & \\
 & 1-2* & 26 & 10187.226 & 10187 [1] 10191 [3]  & \\
 & 2-0* & 17 & 13095.493 &  & \\
 & 2-1* & 19 & 12087.207 &  12092 [1] & \\
 & 2-2* & 22 & 11088.132 & 10099 [1] 10103 [3] & \\
 & 3-0* & 16 & 13988.405 &  & \\
 & 3-1* & 17 & 12979.947 &  & \\
 & 3-2* & 19 & 11980.629 & 11981 [1] & \\
 & 3-4* & & & 10011 [1], 10015 [3] \\
 & 4-0* & 15 & 14872.663 &  & \\
 & 4-1* & 16 & 13864.061 &  & \\
 & 4-2* & 17 & 12864.569 &  & \\
  \vspace{-0.5em} \\

\Sb{} -- \Sd{}  & 0-0 & 15 & 9061.930 & 9064 [1]  & \\
 & 0-1 & 16 & 8049.405 &  & \\
 & 0-2 & 18 & 7046.835 & 7046.343 [6] & \\
 & 0-3* & 21 & 6054.266 &  & \\
 & 0-4* & 24 & 5071.780 &  & \\
 & 0-5* & 27 & 4099.283 &  & \\
 & 1-0 & 14 & 9972.462 &  9972 [1], 9976 [3], 9972.424 [6] & \\
 & 1-1 & 15 & 8959.784 & 8962 [1], 8959.789 [6]  & \\
 & 1-2 & 16 & 7957.084 & 7967.036 [6]  & \\
 & 1-3 & 18 & 6964.277 & 6964.220 [6] & \\
 & 1-4* & 21 & 5981.463 &  & \\
 & 1-5* & 24 & 5008.744 &  & \\
 & 2-0 & 13 & 10874.420 & 10874.381 [6] & \\
 & 2-1 & 14 & 9861.679 & 9867 [3], 9861.640 [6] & \\
 & 2-2* & 15 & 8858.820 &  & \\
 & 2-3 & 17 & 7865.838 & 7865.786 [6] & \\
 & 2-4* & 19 & 6882.782 & 6882.550 [6] & \\
 & 2-5* & 21 & 5909.698 &  & \\
 & 3-0* & 12 & 11767.709 &  & \\
 & 3-1 & 12 & 10754.909 & 10754.867 [6] & \\
 & 3-2 & 14 & 9751.932 &  9756 [3], 9651.879 [6] & \\
 & 3-3* & 15 & 8758.826 &  & \\
 & 3-4 & 17 & 7775.659 & 7775.519 [6]  & \\
 & 3-5* & 19 & 6802.248 & 6802.185 [6] & \\
 & 4-0* & 11 & 12652.286 &  & \\
 & 4-1* & 11 & 11639.391 &  & \\
 & 4-2 & 13 & 10636.341 &  10636.312 [6] & \\
 & 4-3 & 14 & 9643.116 & 9643.049 [6] & \\
 & 4-4* & 15 & 8659.749 &  & \\
 & 4-5* & 16 & 7686.202 &  & \\

\Sc{} -- \Sa{} & 0-0 & 36 & 17859.641 &  17859 [4] & \\
 & 0-1* & 46 & 16855.359 &  & \\
 & 1-0* & 30 & 18765.794 &  18767 [5] & \\
 & 1-1 & 36 & 17759.615 &  17759 [4] & \\
 & 1-2* & 44 & 16763.966 &  16770 [4] & \\
 & 2-0* & 24 & 19662.833 &  & \\
 & 2-1* & 29 & 18655.669 & 18658 [5]  & \\
 & 2-2 & 35 & 17658.308 & 17658 [4] & \\
& 3-2* & & & 18549 [5] \\
& 3-3 & & & 17556 [4] \\
& 3-4* & & & 16566 [4] \\
& 4-3* & & & 18438 [5] \\
& 4-4* & & & 17455 [4]\\
 \vspace{-0.5em} \\

\Sf -- \Sa&0-0&15&19076.916 &19075.4 [7]&\\
&0-1*&17&18068.396&18068.4 [7]&\\
&0-2*&18&17069.021&17072.1 [7]&\\
&1-0*&14&19945.353&&\\
&1-1&15&18936.706& 18918.3 [7]&\\
&1-2*&17&17937.144&17918.7 [7]&\\
&2-0*&14&20809.072&&\\
&2-1*&14&19800.392&19785.5 [7]&\\
&2-2&17&18800.792&18763.9 [7] &\\
&2-3 & & & 17775.9 [7] \\
\vspace{-0.5em} \\
\Se -- \Sd&0-0&9&24302.257&&\\
&0-1*&9&23289.220&&\\
&0-2*&9&22285.939&&\\
&0-3*&11&21292.407&&\\
&0-4*&11&20308.720&&\\
&0-5*&12&19334.758&&\\
&1-0&9&25146.767&&\\
&1-1*&10&24133.737&&\\
&1-2*&10&23130.521&&\\
&1-3*&10&22137.051&&\\
&1-4*&10&21153.289&&\\
&1-5*&10&20179.273&&\\
\end{longtable}
[1] 37WuMe \citep{37WuMexx.TiO}, [2] 57GaRoJu \citep{57GaRoJu.TiO}, [3] 69Lockwood \citep{69Loxxxx.TiO}, [4] 28Lowater \citep{28Lowater.TiO}, [5] 69LiNi \citep{69LiNixx.TiO}, [6] 80GaBrDa \citep{80GaBrDa.TiO}, [7] 82DeVore \citep{82DeVore.TiO}

* \Marvel\ predicted band-heads

\end{center}

\subsection{Prediction of Unmeasured Lines}
The \Marvel\ spin-rovibronic states for which we have assigned energies will be 
involved in more transitions than were used in their generation. The tabulation 
and analysis of these potential transitions provides key information which can 
be used to assist assignment of new spectra. We have produced a list of all 
transitions between \Marvel\ energy levels that obey the following selection 
rules: $|\Delta J| \le 1$, $|\Delta \Lambda| \le 1$ and $\Delta S = 0$. This data is provided in the Supplementary 
Information.

\subsection{Band-heads}

\Cref{tab:bh_AX,tab:bh_BX,tab:bh_CX2,tab:bh_EX,tab:bh_singlet}  tabulate the 
\Marvel-derived band-heads for each spin$-$vibronic state and compare these 
band-heads against low-resolution observations of band-heads from the references 
tabulated in \Cref{tab:databandheads}. Additionally, there are some band-heads 
that have been experimentally observed and assigned and involve some spin 
vibronic states not studied in any high-resolution study that are thus not in 
the \Marvel\ analysis. These will be very useful to verify the final \Duo\ 
spectroscopic model for \TiO\ in a future study. Further, we tabulate the 
approximate $J$ for the bandhead based on the transition frequencies derived from \Marvel\ energy levels; this can be used to help suggest a $J$ value 
associated with these other experimentally observed band-heads.  

\Cref{tab:bh_AX} provides the \A -- \X{} R-band-heads. Agreement between the 
low-resolution and \Marvel\ band-heads is generally within 2 \cm{}. 

\Cref{tab:bh_BX} gives the \B -- \X{} R-band-heads: 5 have been observed in 
rotationally-resolved spectra, 6 have positions predicted by \Marvel\ data and 9 
other band-heads have been observed in low-resolution non-rotationally-resolved 
observations. Of the 28 low-resolution band-heads observed by 69Phillips, 9 were also 
calculated using \Marvel\ data. Most agree with our calculations to around a few \cm{}, 
but there are clearly some mis-assignments for the 15,930 and 16,081 \cm{} 
band-heads. The higher vibrational levels of the \B{} state have yet to be 
observed in a rotationally-resolved study, but there is significant band-head 
information that can be very valuable in fitting the \B{} state PEC for the 
final spectrosopic model of \TiO. Further high-resolution rotationally-resolved 
studies would be welcome.

\Cref{tab:bh_CX2} tabulates \C -- \X{} R-band-heads. There is very extensive 
coverage both rotationally-resolved and low-resolution band-head observations. 
There is good agreement (within a couple of \cm{}) between almost all \Marvel\ 
and low-resolution observations. band-heads from transitions with large $\Delta 
v$ can be predicted from \Marvel\ data despite not being directly observed due 
to either congestion in the spectra and/or low intensity due to small 
Franck$-$Condon factors. 

\Cref{tab:bh_EX} tabulates \E -- \X{} R-band-heads. The coverage of high 
vibrational levels of the \E{} state in the low-resolution observed band-heads is 
much more extensive than any rotationally-resolved data and will be valuable for 
the future \Duo\ model. Again, high resolution studies of these bands would be 
valuable. 

For the singlet states (band-heads shown in \Cref{tab:bh_singlet}), the 
rotationally resolved data in combination with the \Marvel\ predicted band-heads 
are generally more extensive and accurate than the low-resolution 
observations. The key exception is probably the \Sc{} -- \Sa{} data, for which 
low-resolution data exist involving vibrational levels up to $v=4$, including 
non-vertical transitions (i.e. $\Delta v \ne 0$). The agreement between the \Marvel\ energies and the low-resolution observations is generally high, except for the \Sf{} -- \Sa{} data. The bandhead assignments from \citet{82DeVore.TiO} involving higher vibrational quantum numbers do not agree with the \Marvel\ data obtained mostly from the rotationally-resolved study of \citet{85BrGaxx.TiO}. The difference between these two assignments is in the vibrational frequency of the \Sf{} level; it is likely that the higher resolution rotationally-resolved data we have used is the correct assignment. 


\subsection{Comparison with \citet{98Scxxxx.TiO}}
\begin{figure}
\includegraphics[width=0.5\textwidth]{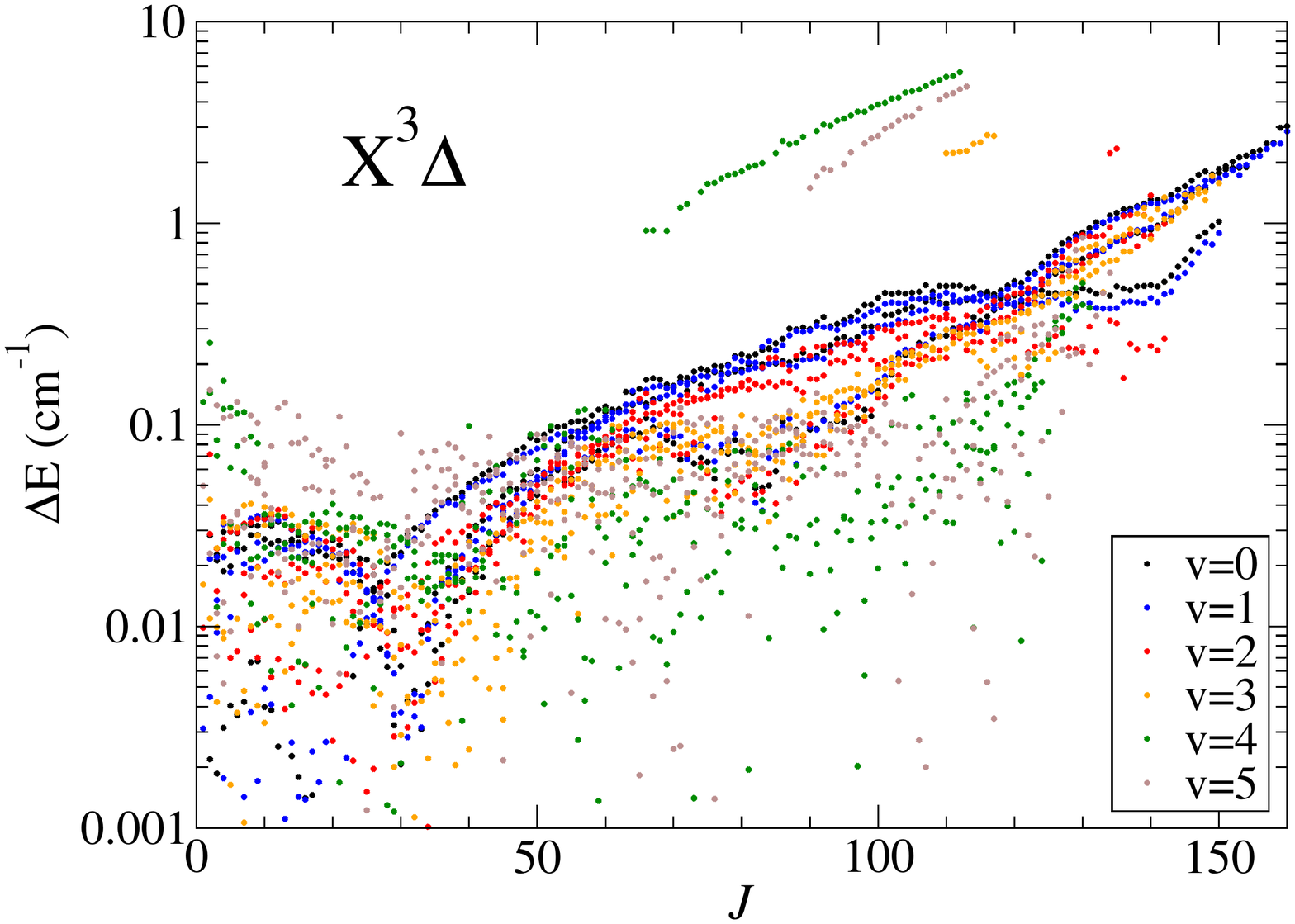}
\includegraphics[width=0.5\textwidth]{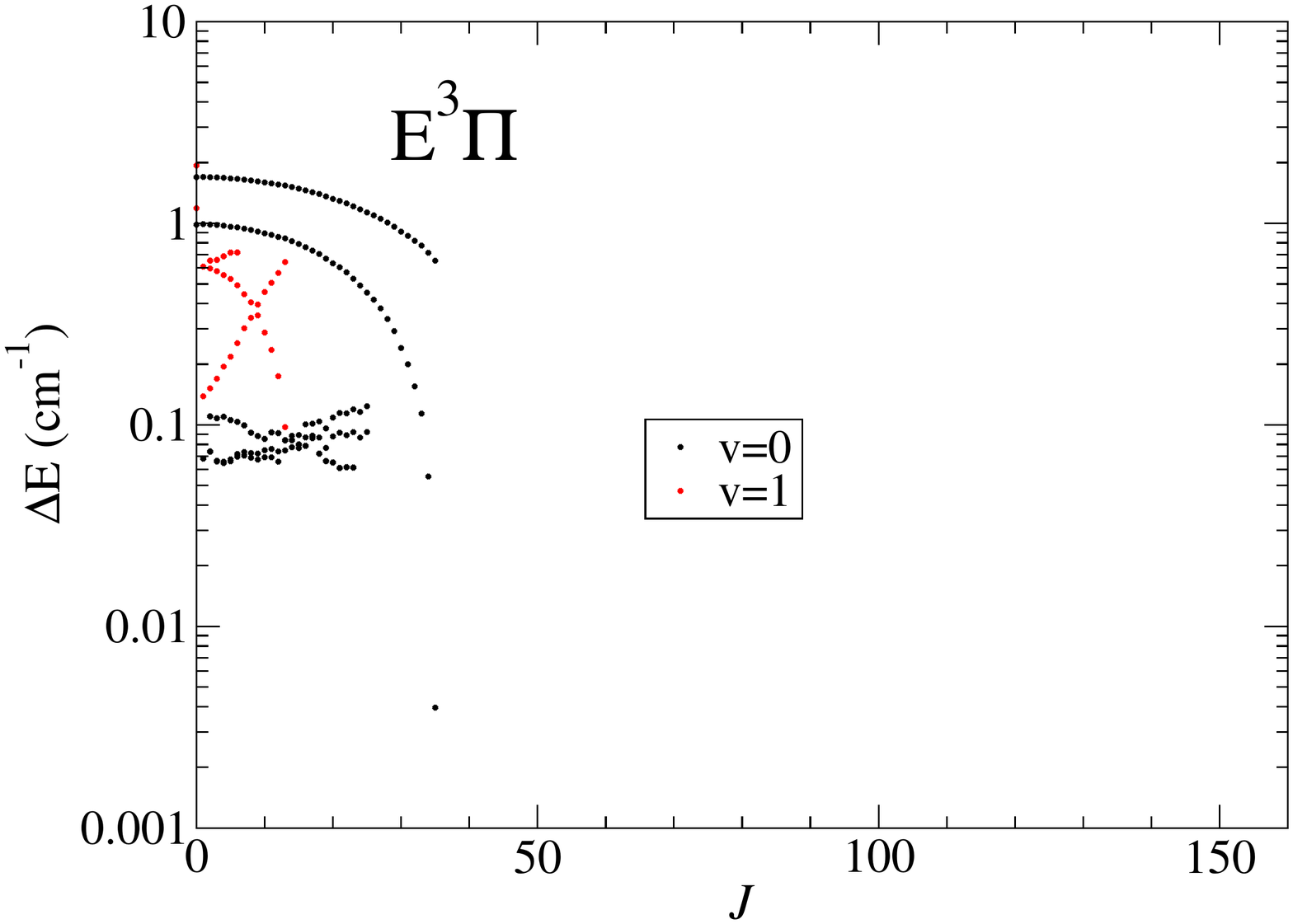}
\includegraphics[width=0.5\textwidth]{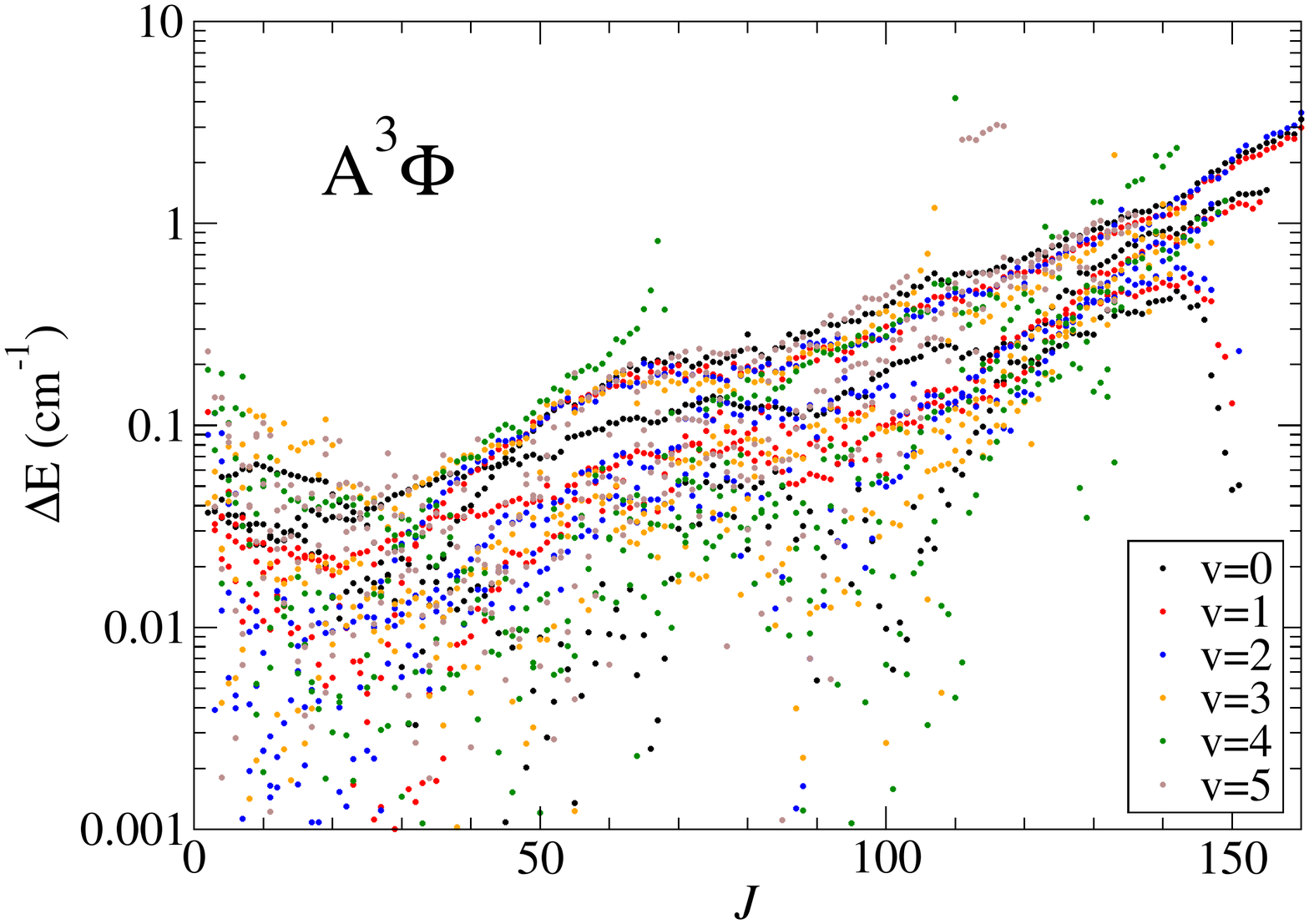}
\includegraphics[width=0.5\textwidth]{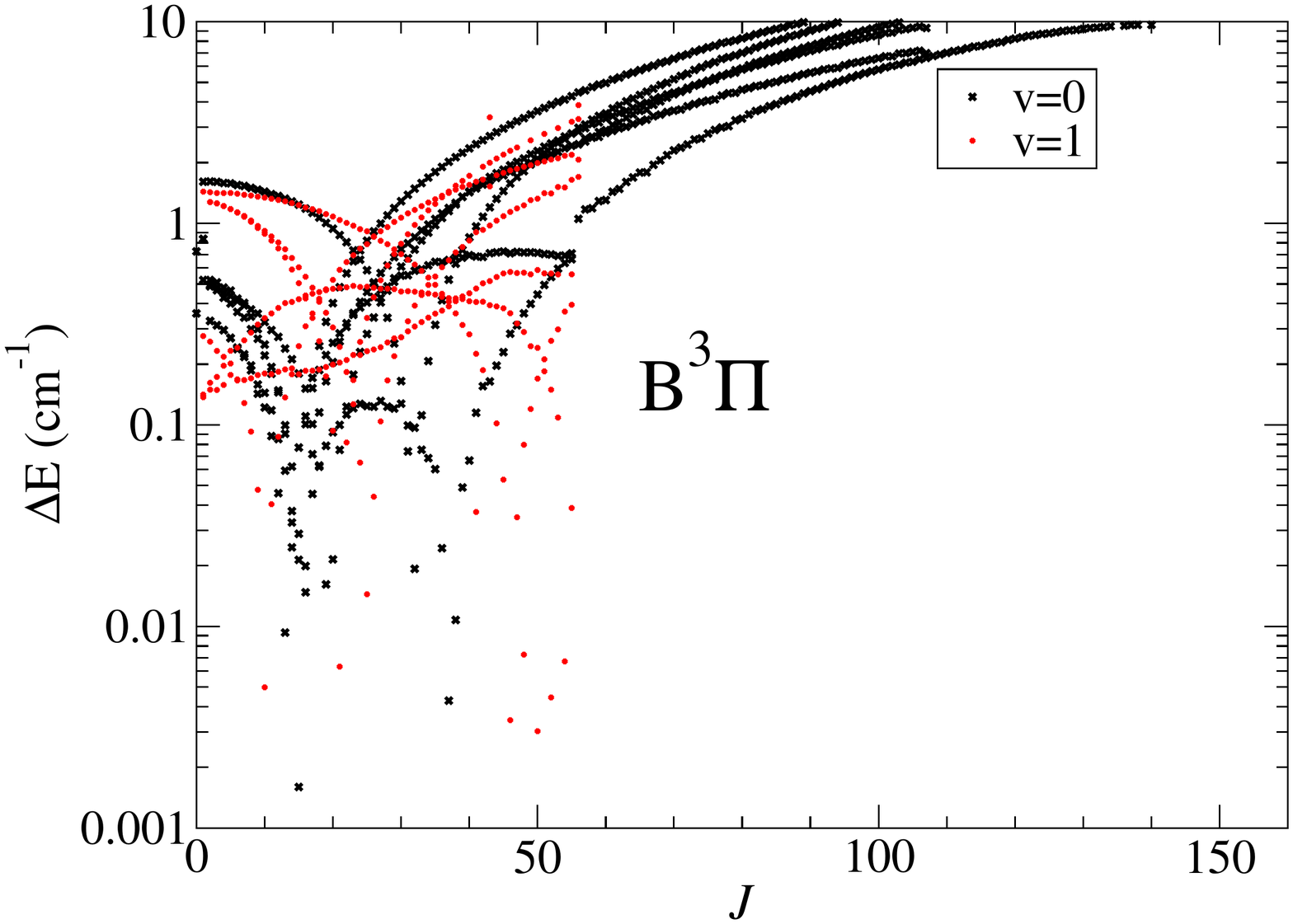}
\includegraphics[width=0.5\textwidth]{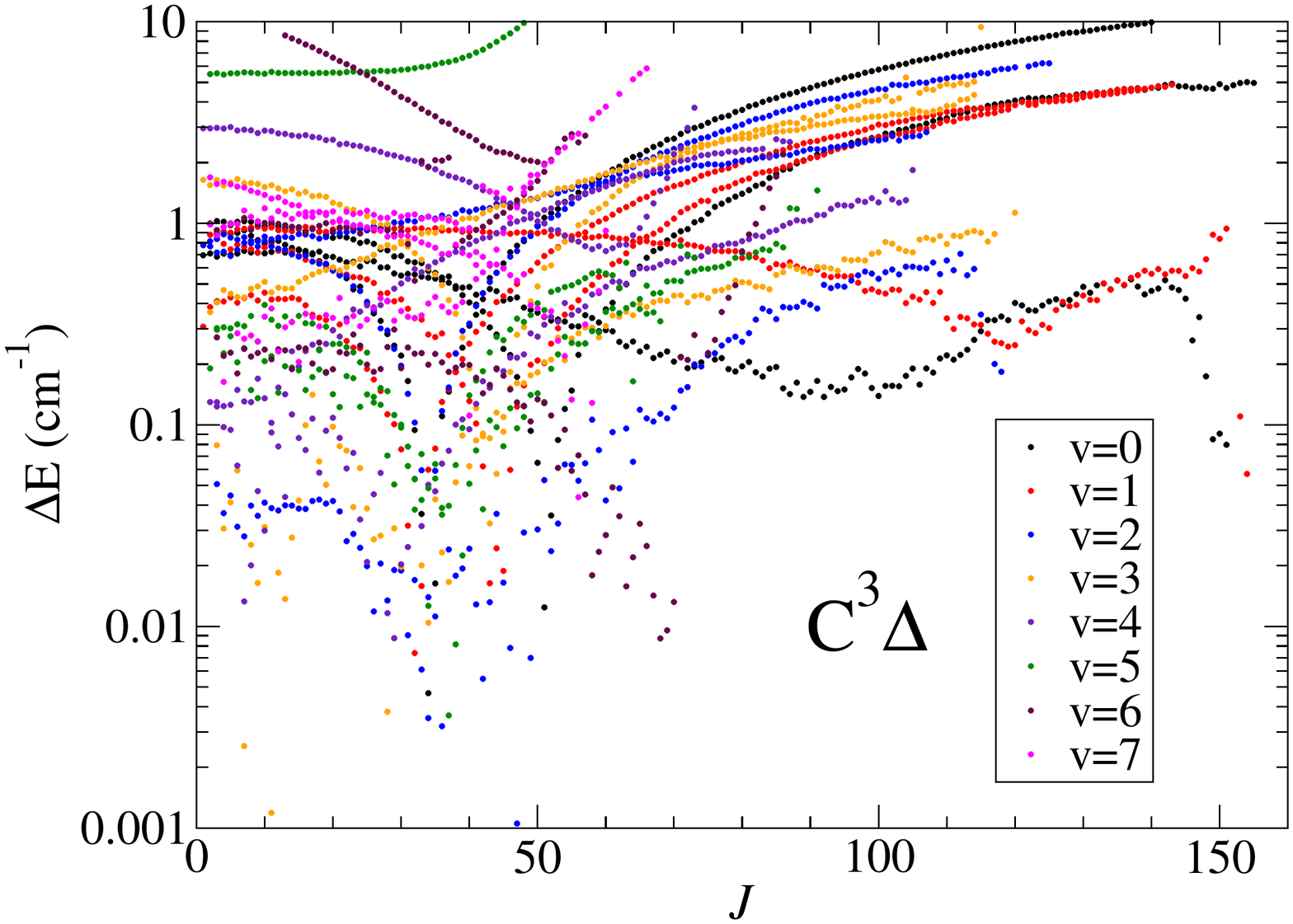}
\caption{\label{fig:Sch1} Visual comparison of the absolute energy difference between the  \Marvel\ experimentally-derived energy levels and those in the \citet{98Scxxxx.TiO} linelist for triplet states. Note the logarithmic vertical axis. }
\end{figure}

\begin{figure}
\includegraphics[width=0.5\textwidth]{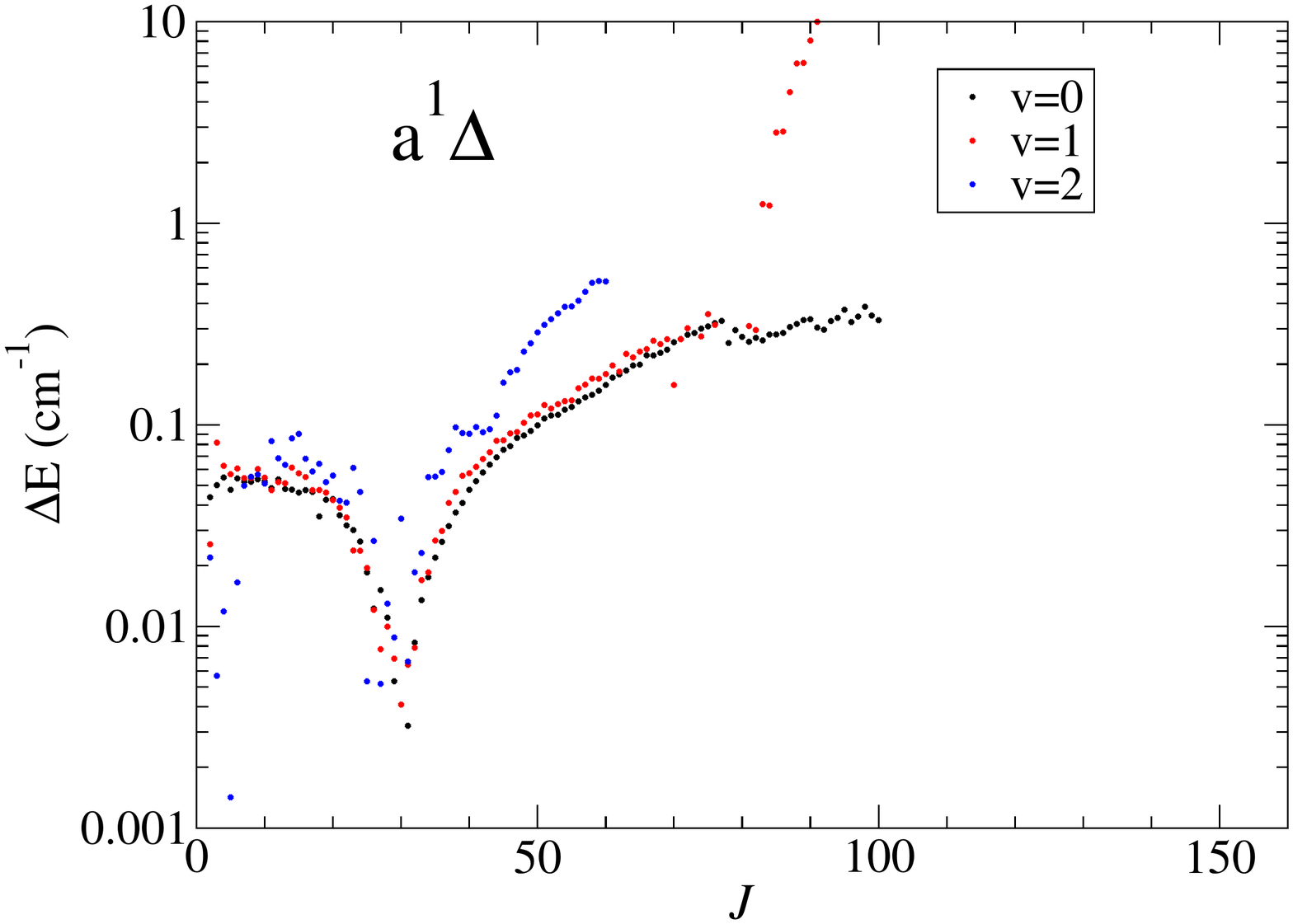}
\includegraphics[width=0.5\textwidth]{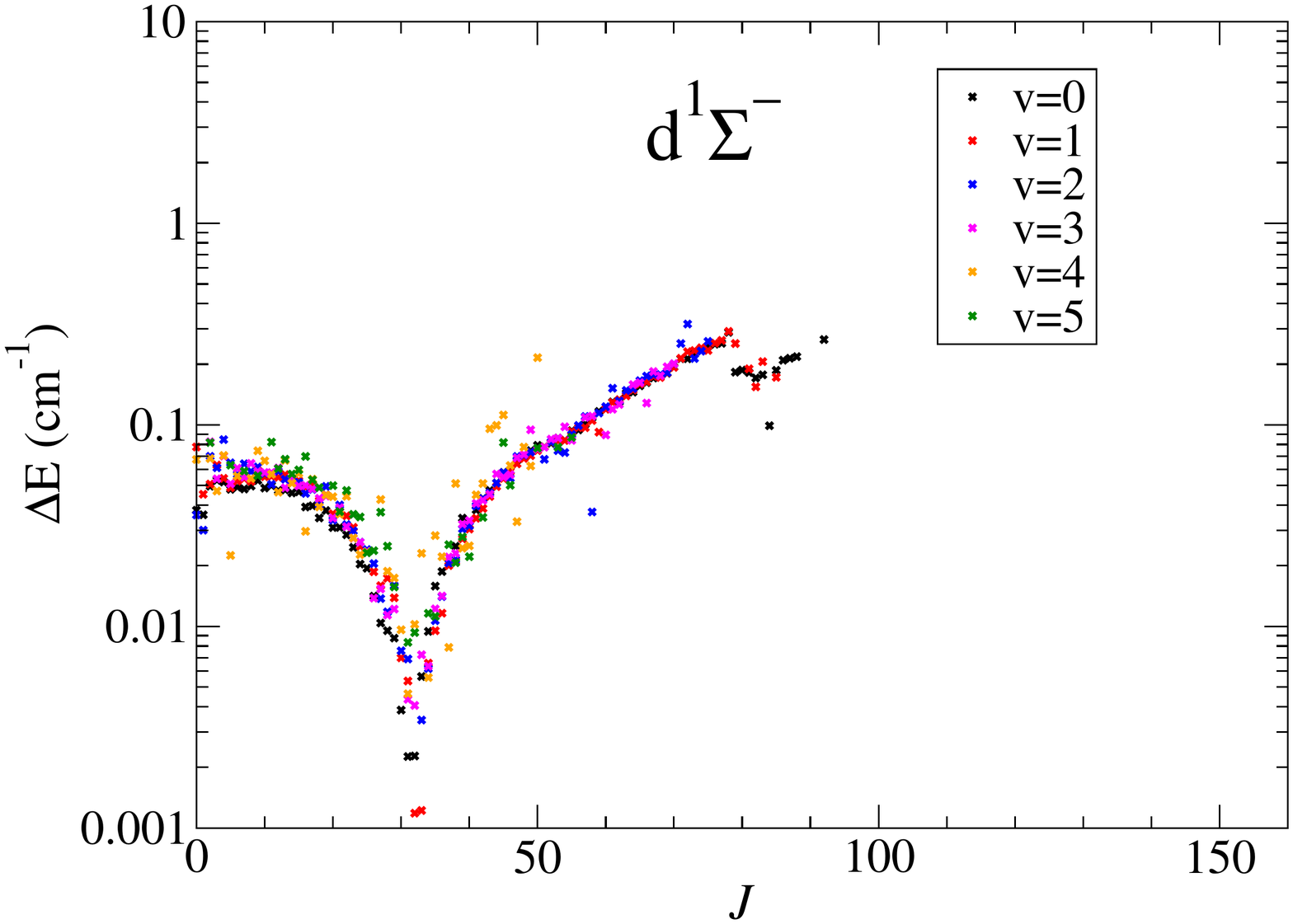}
\includegraphics[width=0.5\textwidth]{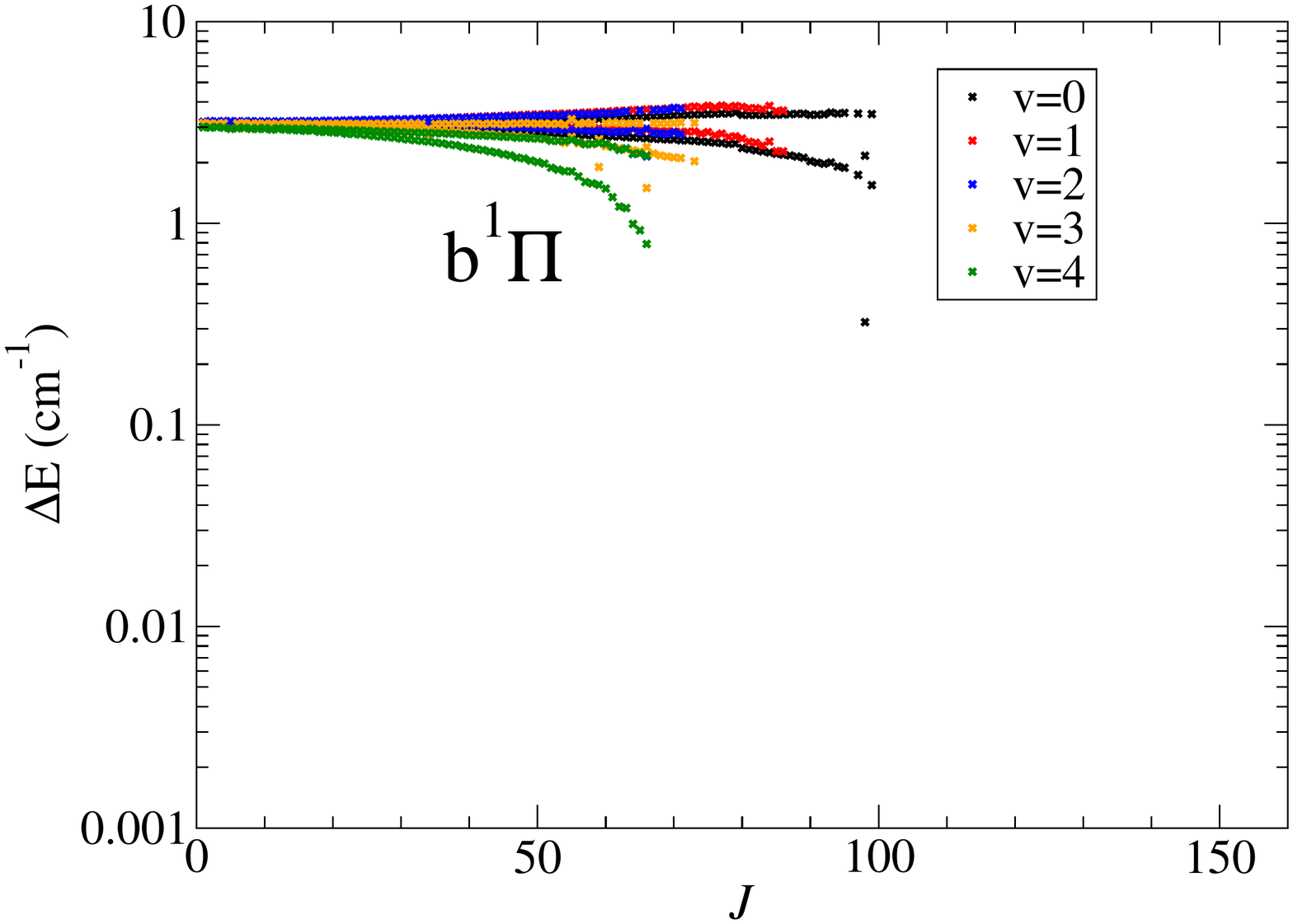}
\includegraphics[width=0.5\textwidth]{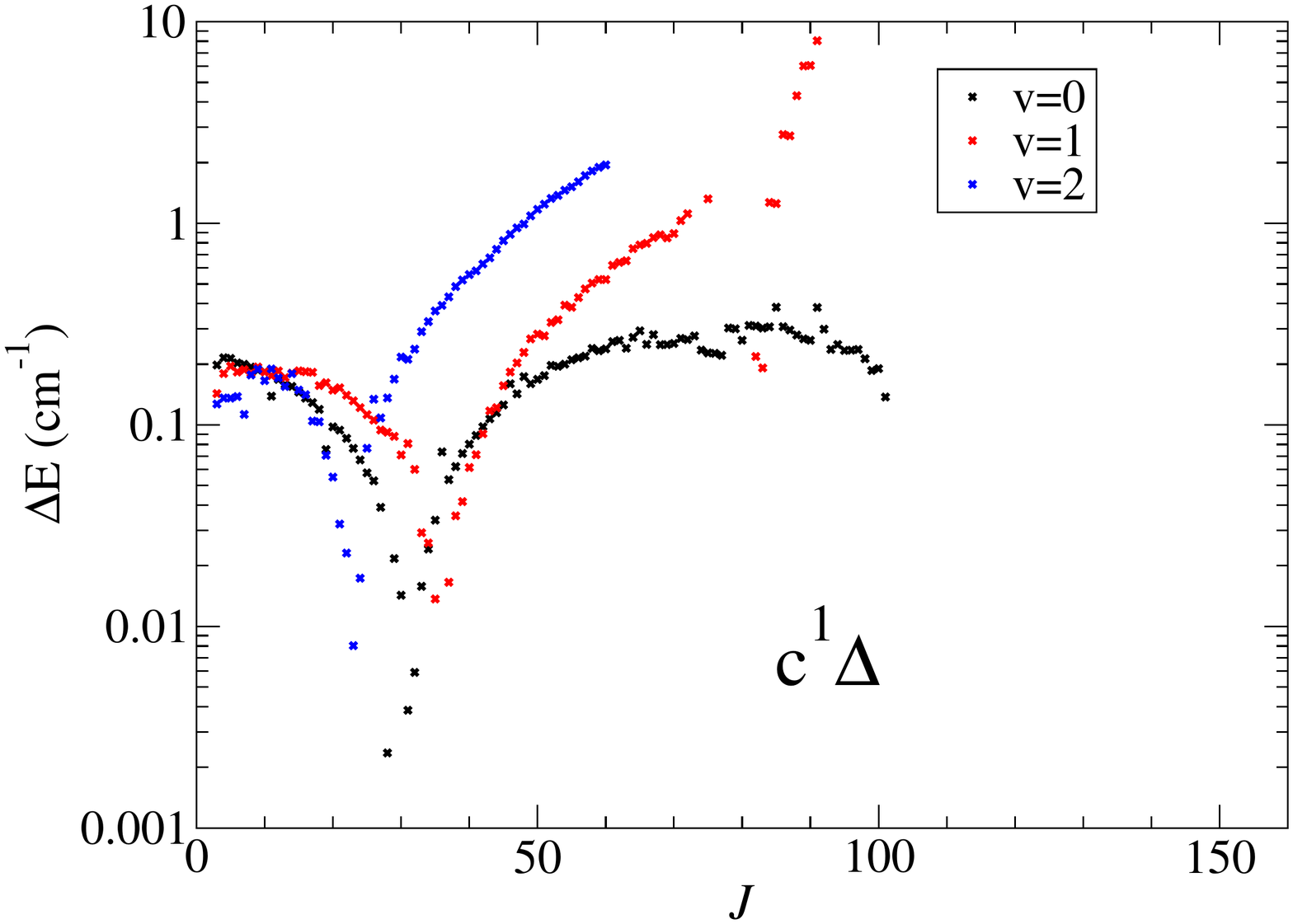}
\includegraphics[width=0.5\textwidth]{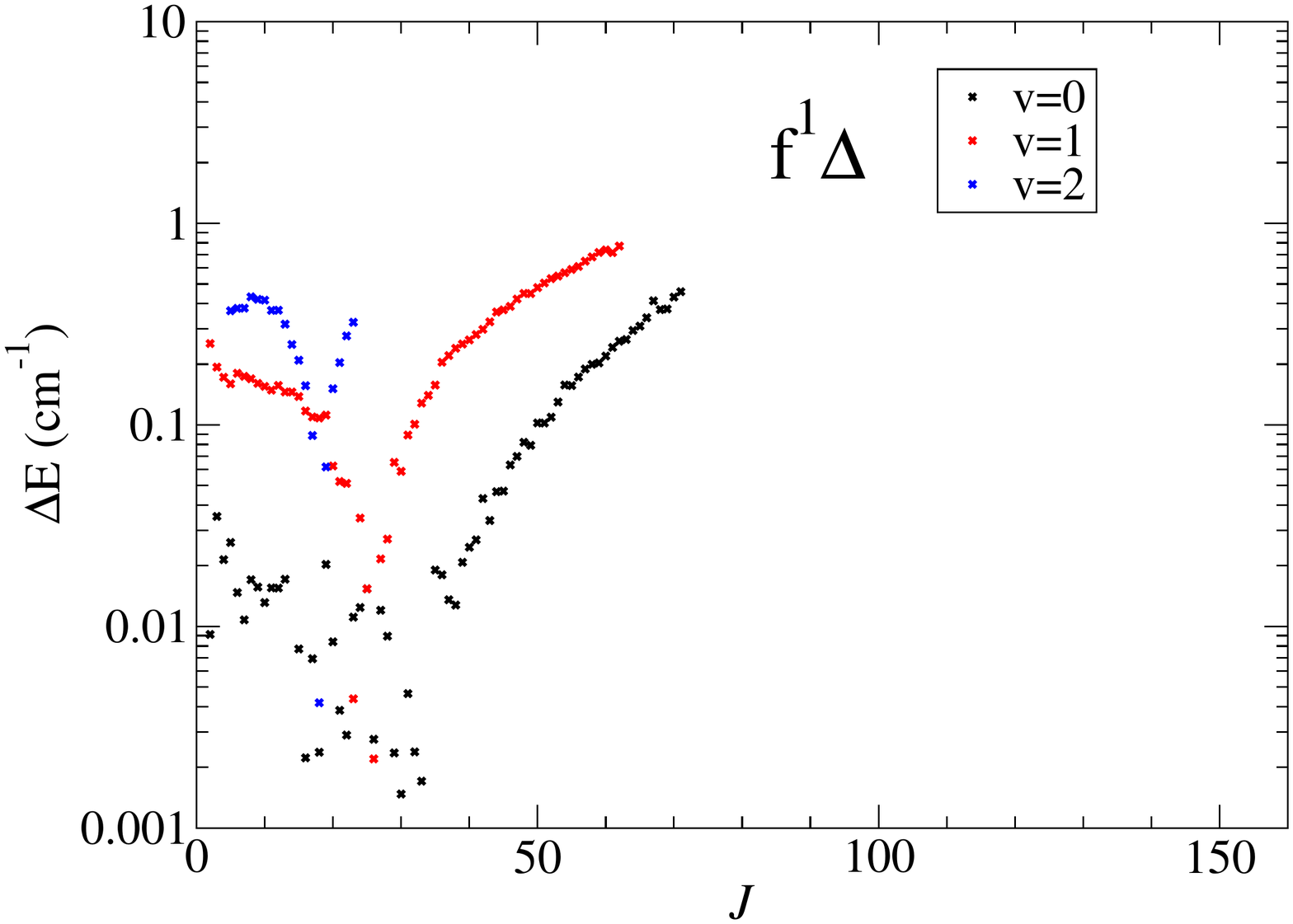}
\caption{\label{fig:Sch2} Visual comparison of the absolute energy difference between the  \Marvel\ experimentally-derived energy levels and those in the \citet{98Scxxxx.TiO} linelist for singlet states. Note the logarithmic vertical axis. }
\end{figure}

\Cref{fig:Sch1} compares the \Marvel\ energy levels against those derived by 
\citet{98Scxxxx.TiO} for the triplet states.  The \X{} and \A{} states have 
differences generally less than 0.01 \cm{} for $J < 50$, with larger errors for 
higher rotational levels. The \E{} state has significant errors up to 2 \cm{}; 
this is partially to be expected as a significant source of experimental data 
for this state post-dates Schwenke's work. Many of the \B{} 
state levels have quite high errors around 3 \cm{}. Most of the \B{} state data 
come from \citet{79HoGeMe.TiO}, so for the most part Schwenke and us should have 
used the same data. The error bars on these data are much smaller than 
differences in the energy levels. Schwenke reports some difficulty in the 
fitting, giving a RMSE of 0.743 \cm{} for these lines.  For the \C{} 
state, there are significant differences between Schwenke's fitted energies and 
the \Marvel\ energies; Schwenke himself reported a RMSE of 1.582 \cm{} between 
his fit and the experimental energy levels he used. This state is significantly 
affected by perturbations that are difficult to model theoretically and which 
have recently been analysed by \citet{03NaItDa.TiO}. 

\begin{figure}
\includegraphics[width=0.7\textwidth]{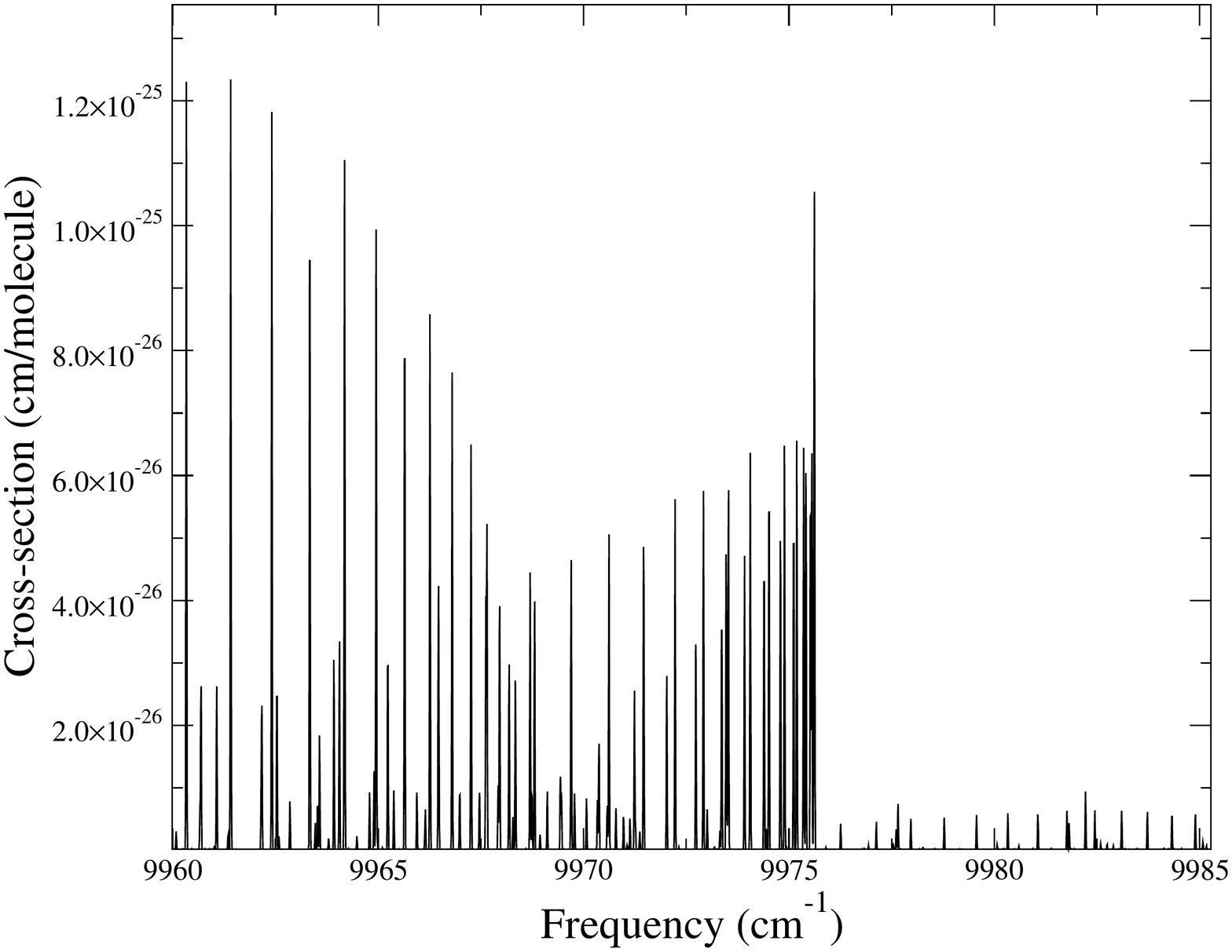}
\caption{\label{fig:bdbandhead} Simulated absorption cross-section from the \citet{98Scxxxx.TiO} line list at 300 K, $\delta v$ = 0.01 \cm{}. The \Sb{}--\Sd{} (1-0) bandhead experimentally is 9972.42 \cm{} \citep{80GaBrDa.TiO}.}
\end{figure}

\Cref{fig:Sch2} compares the \Marvel\ experimentally-derived energy levels and 
the fitted energy levels used in the \citet{98Scxxxx.TiO} linelist for singlet 
levels. The \Sd{}, \Sa{}, \Sc{} and \Sf{} levels seem reasonable; the deviation 
from the fitted Schwenke lines increases for larger $J$ in general. However, errors for the \Sb{} state are 
particularly high, around 3 \cm{}. Schwenke reports RMSE of 0.054 \cm{}. 
However, our predicted \Sb{}--\Sd{} band-heads reproduce experiment almost 
perfectly, whereas there are clear discrepancies between experiment and the 
Schwenke data (see \Cref{fig:bdbandhead}). We therefore conclude that there 
is an approximately 3 \cm{} off-set error in the \Sb{} state Schwenke energy 
levels. 

\subsection{Comparison with VALD}
\begin{figure}
\includegraphics[width=0.5\textwidth]{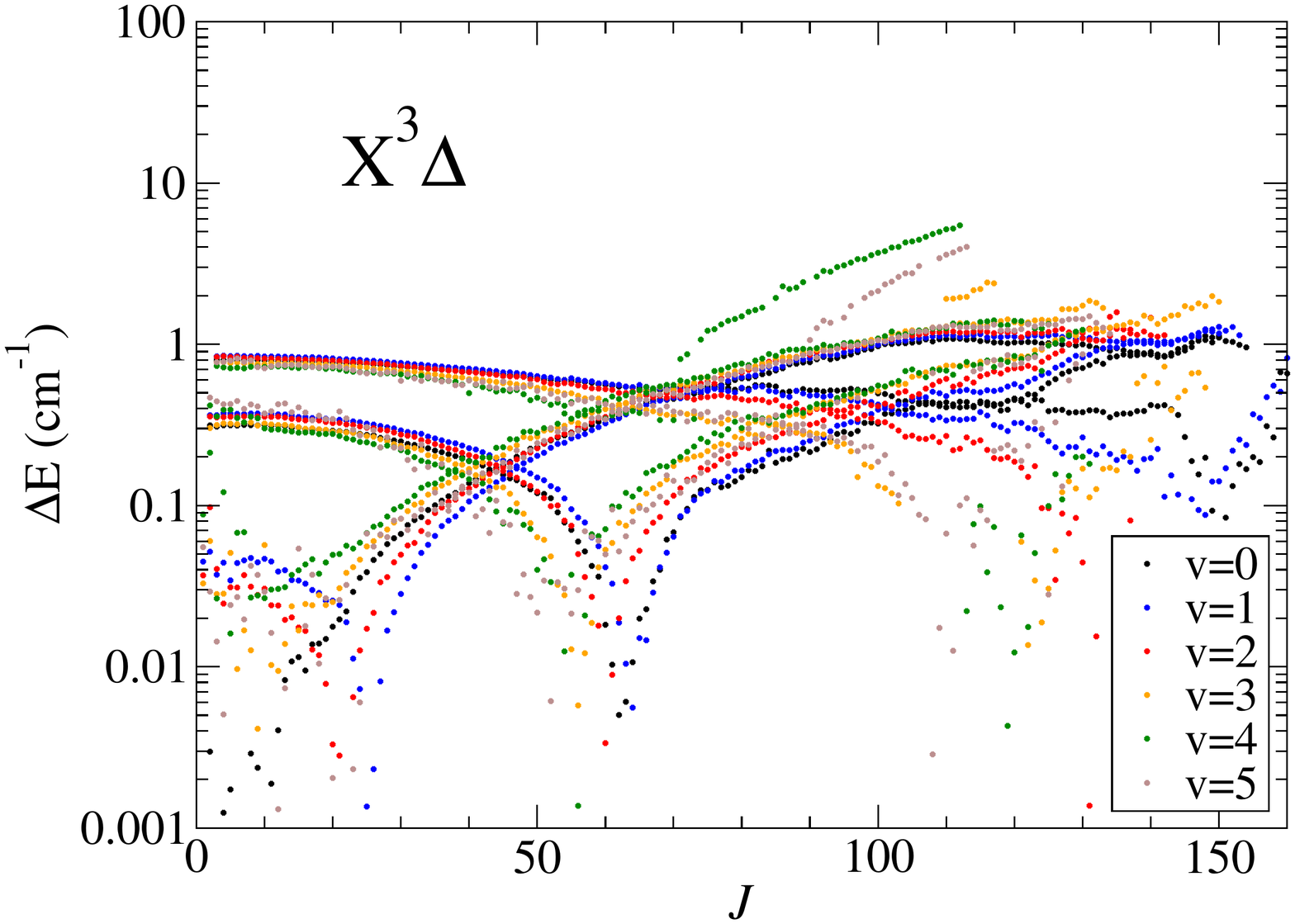}
\includegraphics[width=0.5\textwidth]{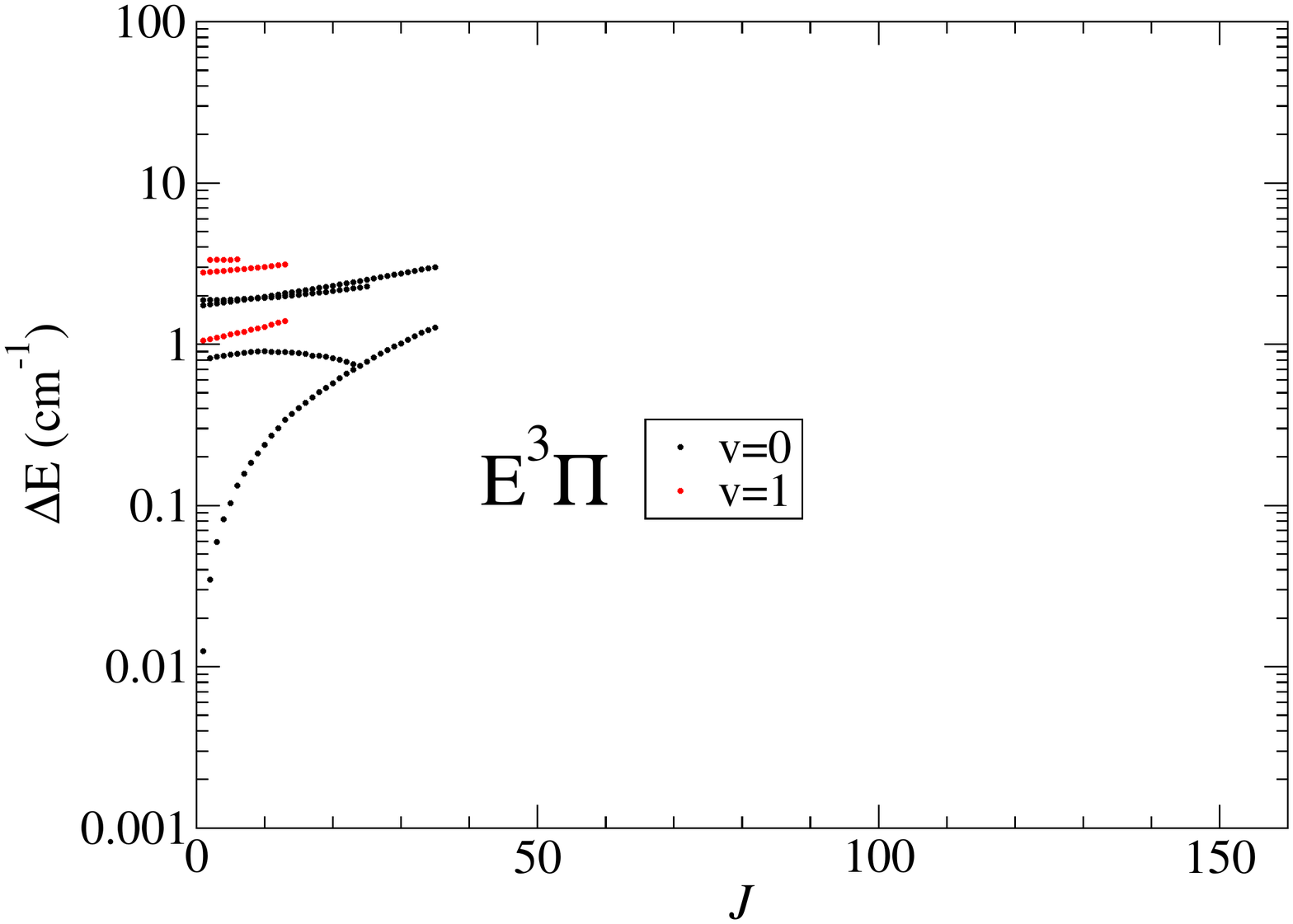}
\includegraphics[width=0.5\textwidth]{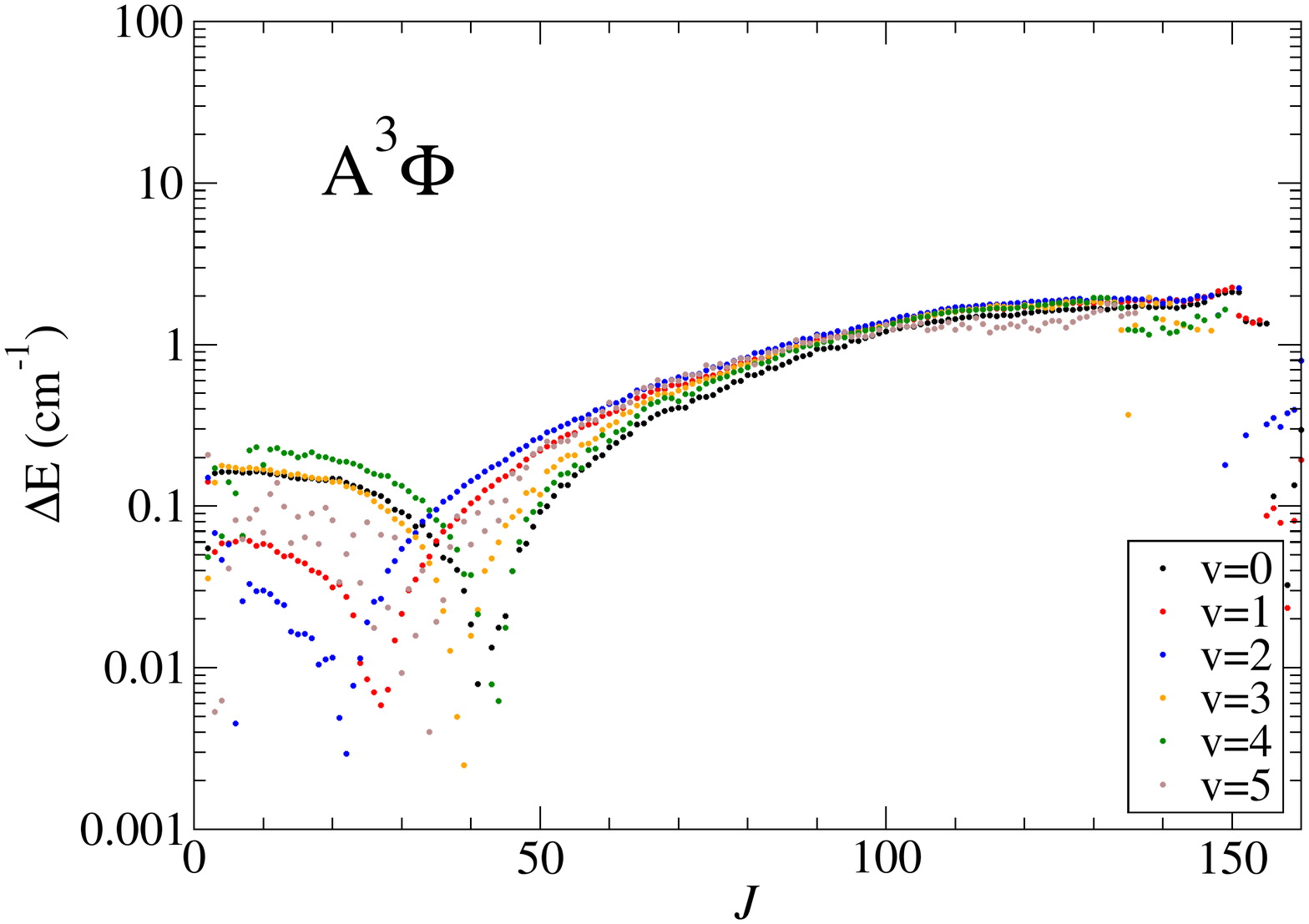}
\includegraphics[width=0.5\textwidth]{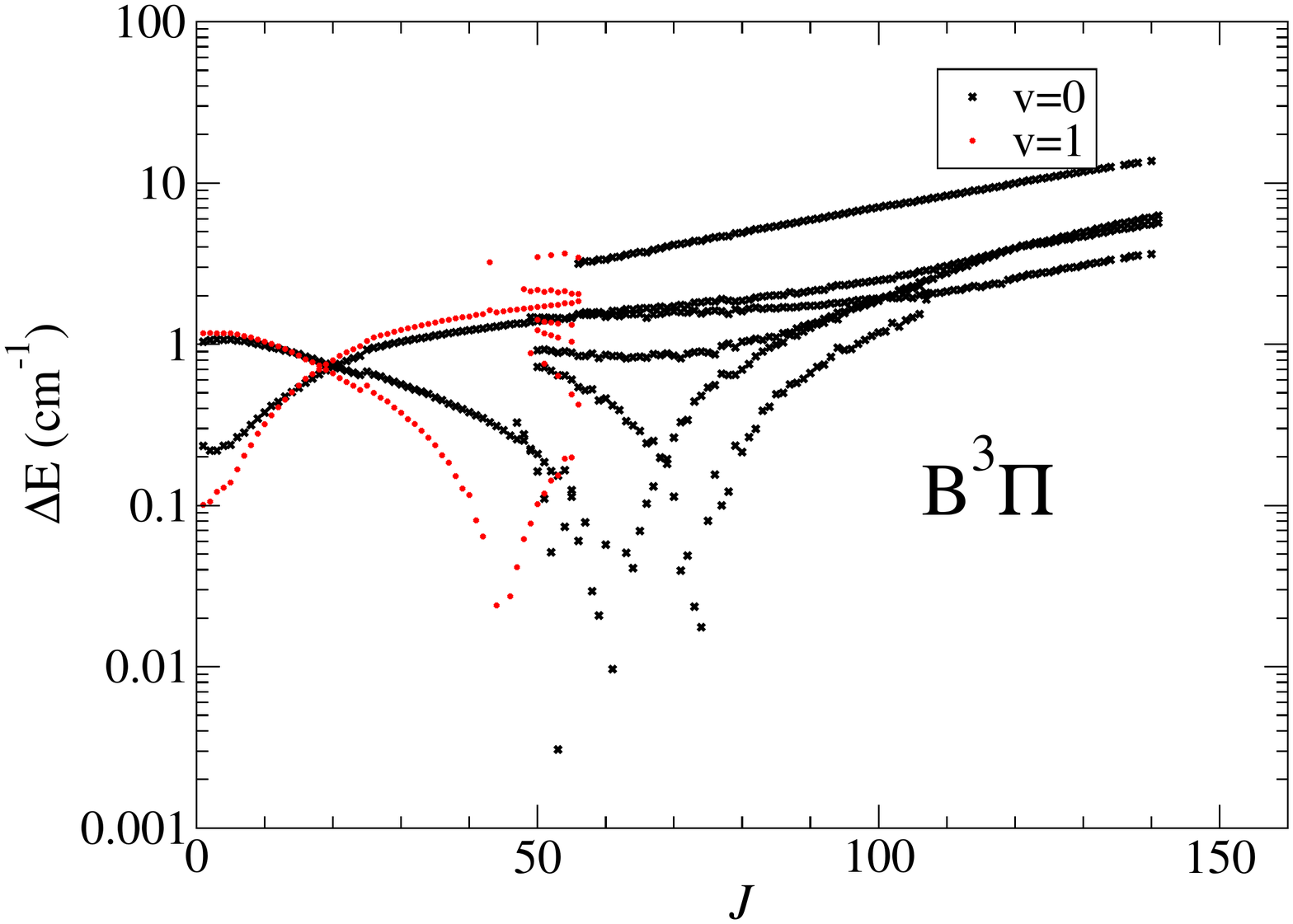}
\includegraphics[width=0.5\textwidth]{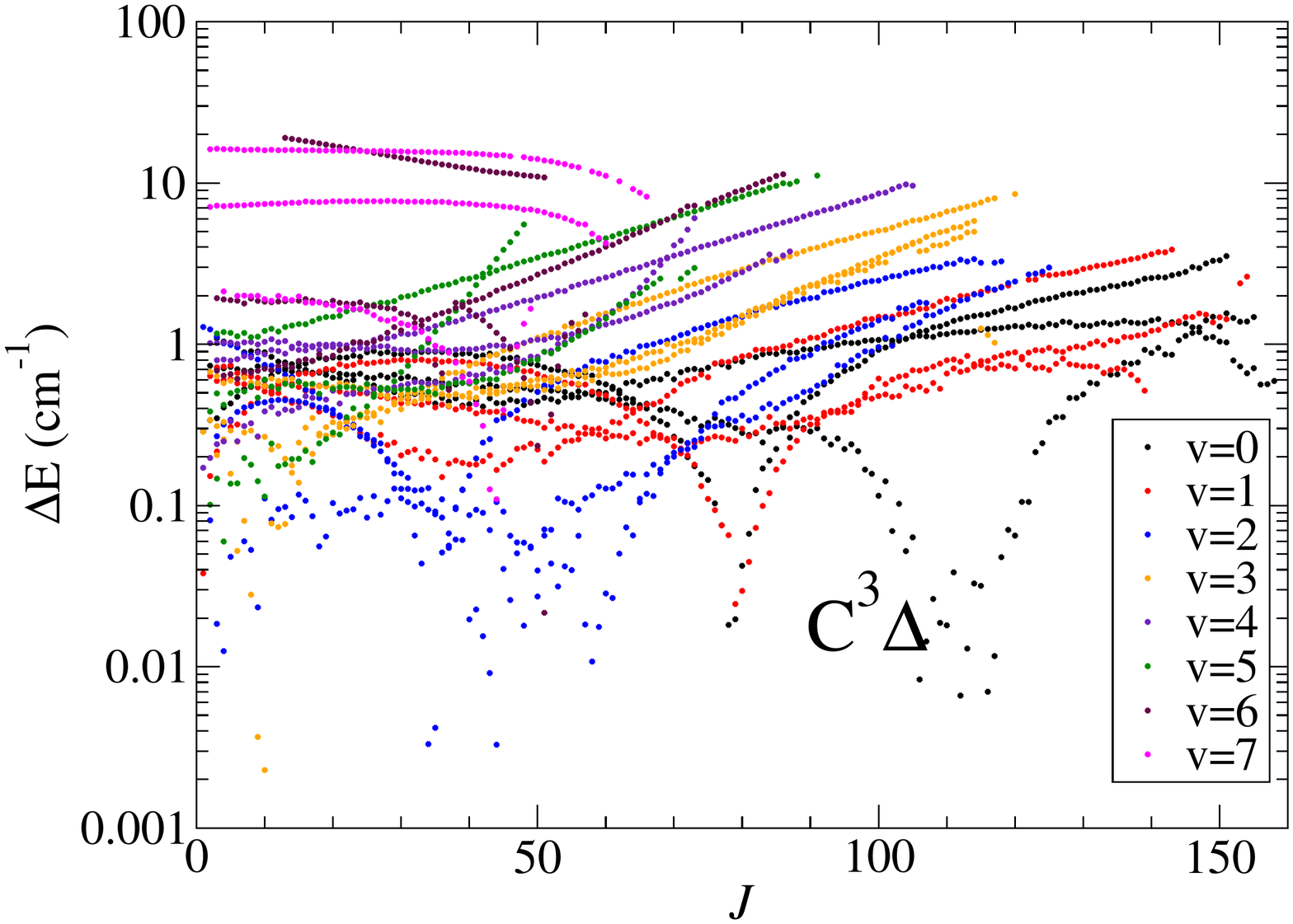}
\caption{\label{fig:Plez1} Visual comparison of the absolute energy difference between the \Marvel\ experimentally-derived energy levels and those in the \citet{98Plxxxx.TiO} linelist for triplet states. Note the logarithmic vertical axis and that the axis range is different from \Cref{fig:Sch1}. }
\end{figure}

\begin{figure}
\includegraphics[width=0.5\textwidth]{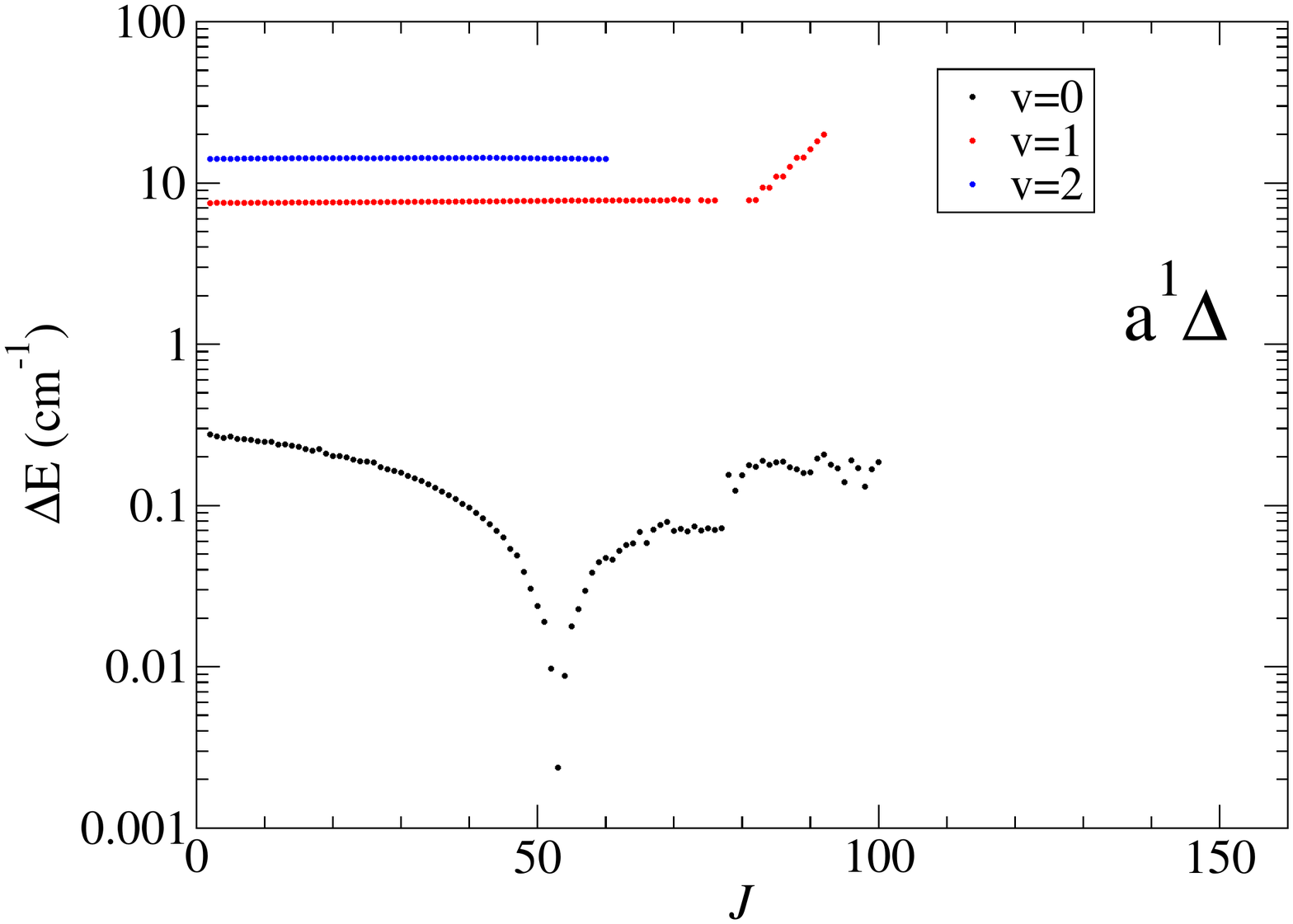}
\includegraphics[width=0.5\textwidth]{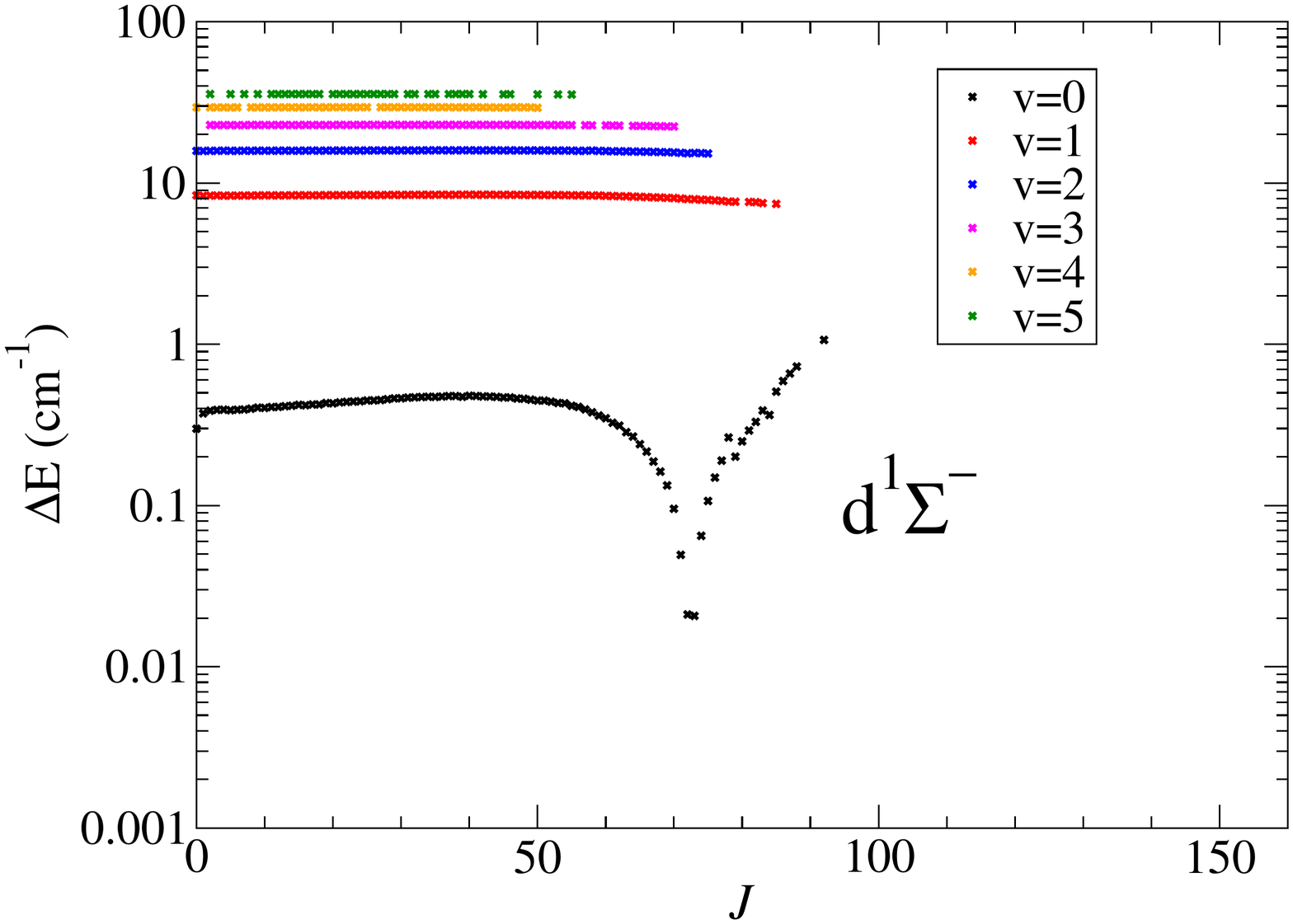}
\includegraphics[width=0.5\textwidth]{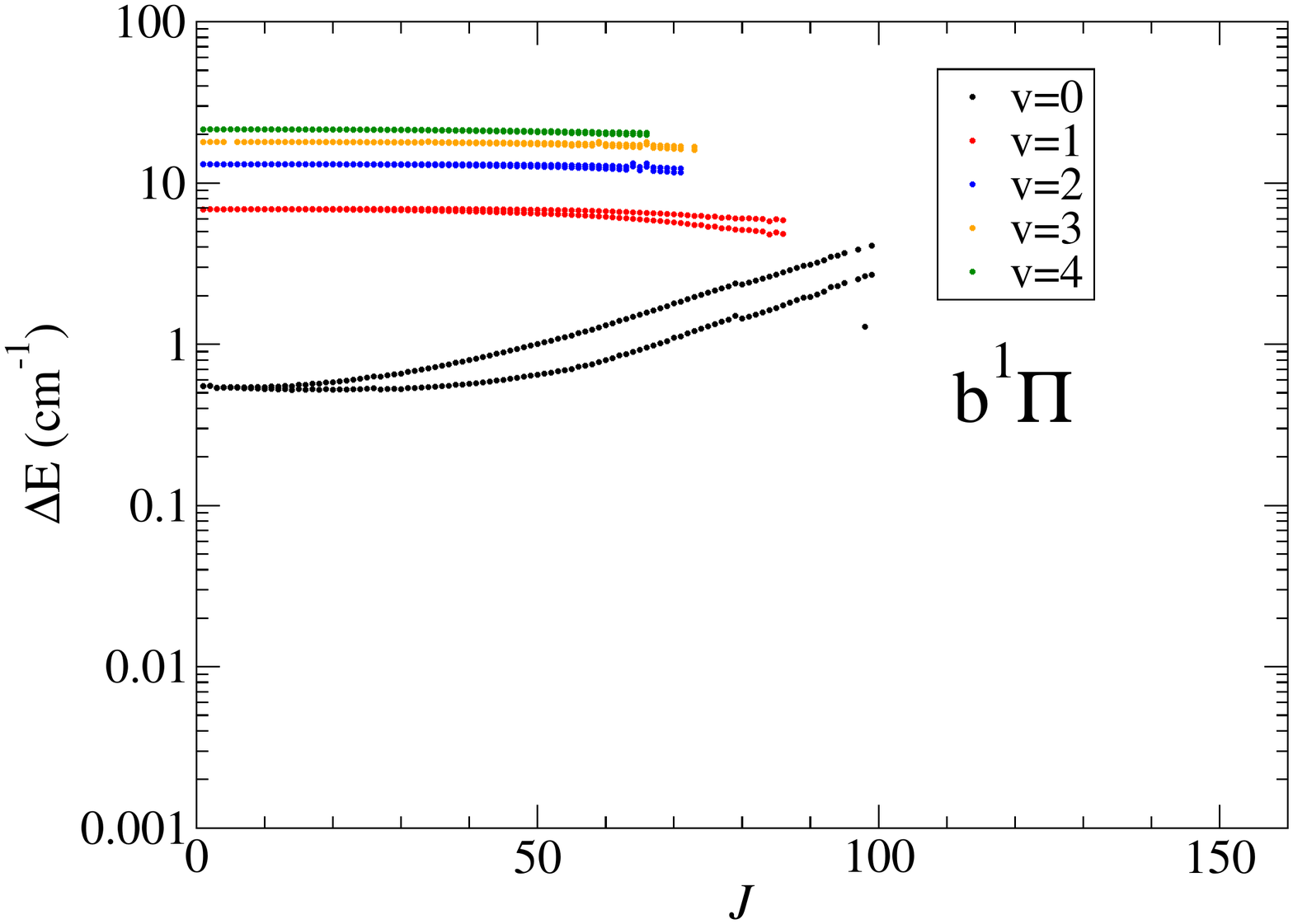}
\includegraphics[width=0.5\textwidth]{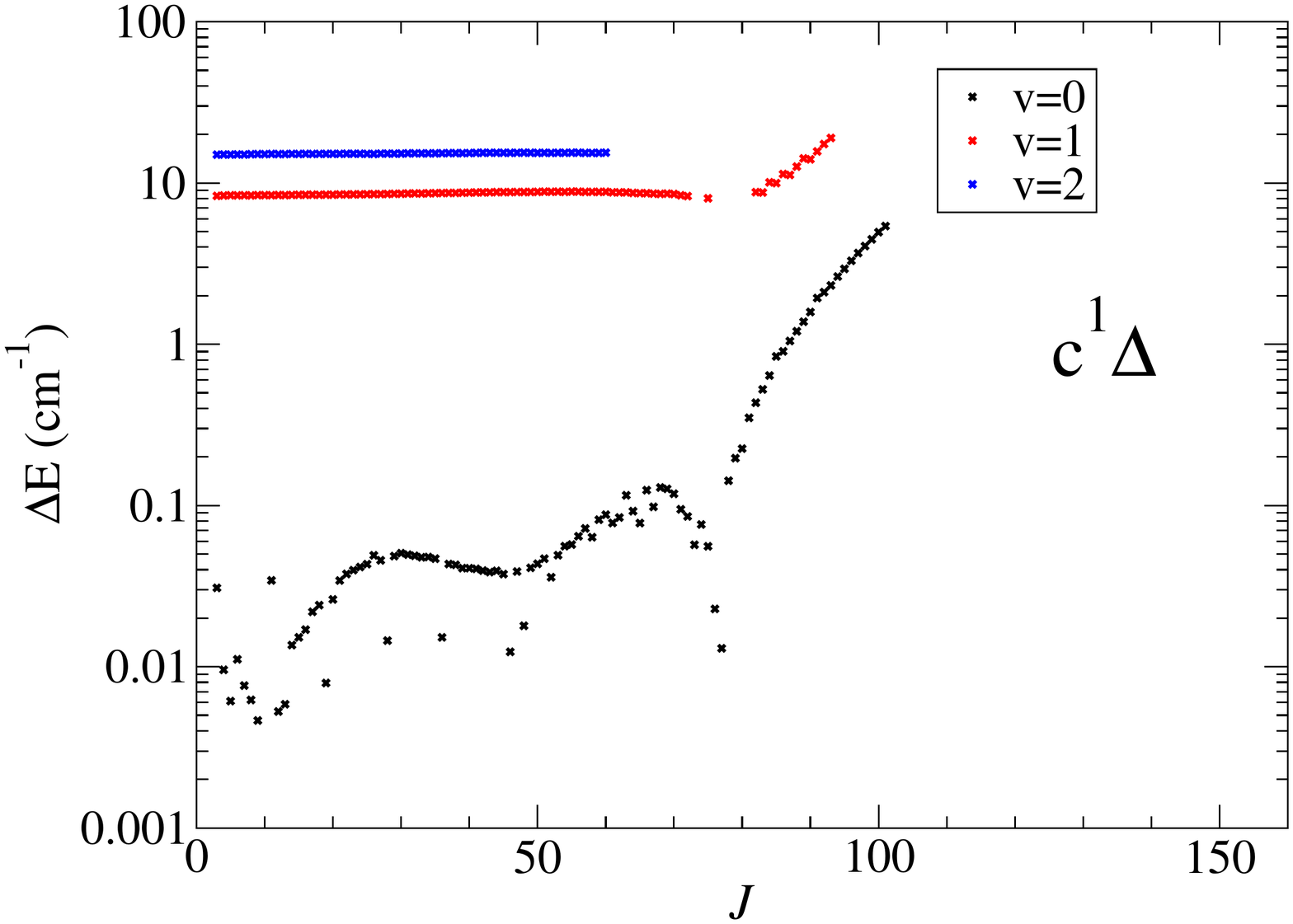}
\includegraphics[width=0.5\textwidth]{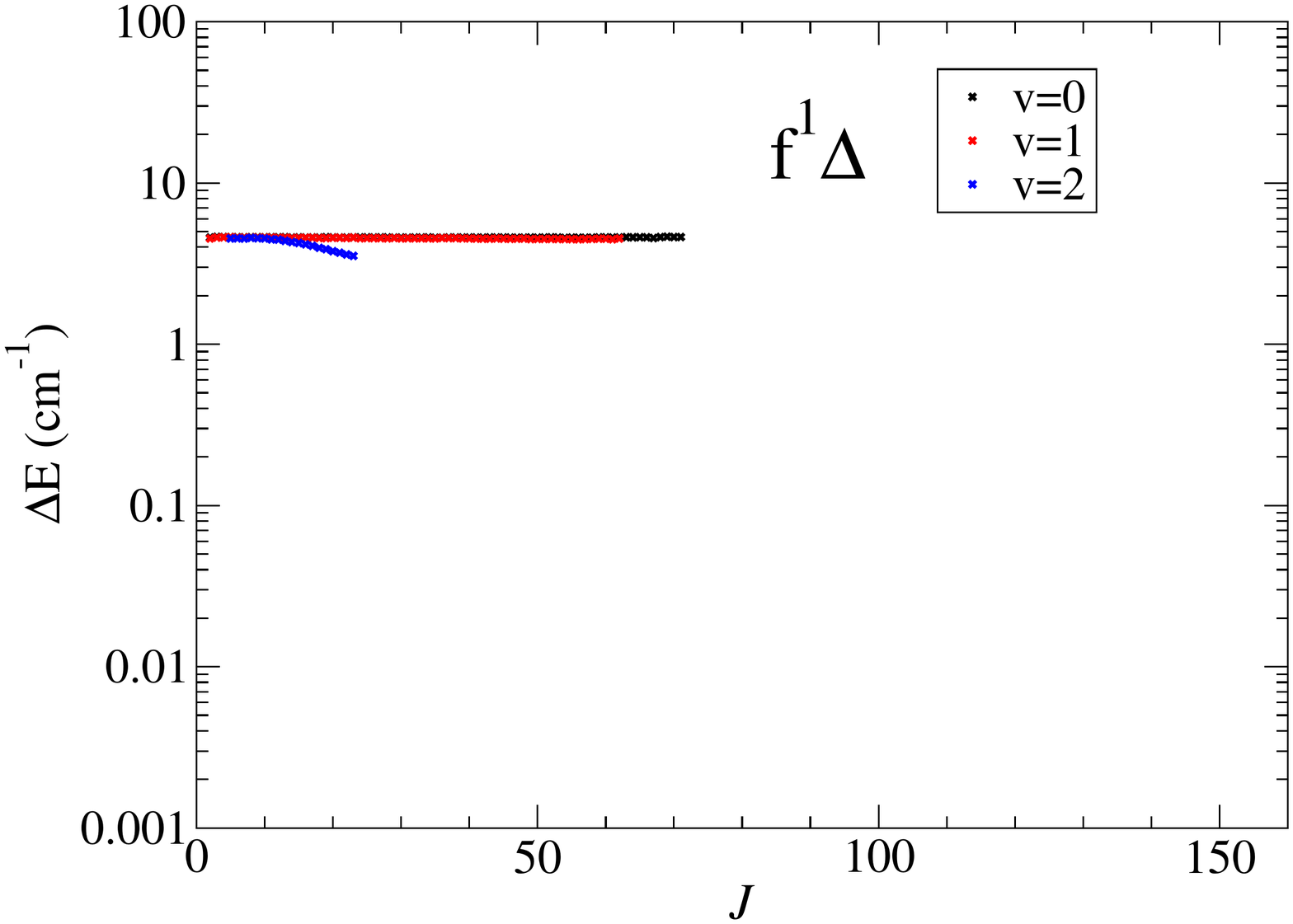}
\caption{\label{fig:Plez2} Visual comparison of the absolute energy difference between the  \Marvel\ experimentally-derived energy levels and those in the \citet{98Plxxxx.TiO} linelist for singlet states. Note the logarithmic vertical axis and that the axis range is different from \Cref{fig:Sch2}. }
\end{figure}

\Cref{fig:Plez1} and \Cref{fig:Plez2} show a visual comparison of the 2012 
version of the Plez TiO line list from the VALD database \citep{VALD3} vs \Marvel\ energy levels. For the triplets, we 
get results qualitatively similar to the Schwenke comparisons, though the errors 
are often about a factor of 10 larger (note the difference in the vertical scale 
between the Plez and Schwenke comparisons). However, for the singlets it is 
clear that the vibrational spacings within some singlet states is incorrect. The 
Phillips experimental frequencies (for which the most recent version of this 
line list is fitted) may have been correctly reproduced. However, other 
experimental data would not be due to these erroneous vibrational frequencies. 
The \Marvel\ energies will thus allow a more thorough understanding of the whole 
spectrum of TiO.

\subsection{Future Directions}
\subsubsection{Recommended Experiments}
The experimental coverage of rovibronic bands in TiO is extensive. However, the complexity of the electronic structure of this species and its importance in understanding, modelling and interpreting the spectroscopy and opacity of cool stars and hot Jupiter exoplanets means that extra experimental data are always welcome. We would like to direct experimentalists towards some key transitions for which data are not yet available, and for which our experience with \abinitio\ computations \citep{jt632,jt623,jt599,jtCrH} on these species leads us to conclude that they will not be calculated to satisfactory accuracy. 

The \D{} state has been identified by \citet{97BaMeMe.TiO} using fluorescence from a very high $^3\Pi$ state but its spectrum has not been rotationally resolved or measured with high accuracy. For the purposes of absorption spectroscopy of astrophysical objects, further data are probably not critical as this state does not contribute to any allowed absorption bands from the electronic states with significant thermal population at 5000 K, nor does it appear to be a strong perturber of the other states. However, it will contribute to weak background absorption and, more importantly, the partition function of TiO.  

Rotationally-resolved data involving higher vibrational excitations of the \B{} and \E{} electronic states are both  achievable (given the detection of band-heads), and valuable for constraining the shape of the potential energy curves of these states. 

Hints from experimental observations,  e.g., $^1\Pi$ state near 22,300 \cm{} by \citet{03NaItDa.TiO}, \abinitio\ evidence and results from similar diatomic species strongly suggest that experimental identification of electronic states between 20,000 \cm{} and 30,000 \cm{} is not complete for singlet states. Targeted (non-absorption) experiments, perhaps two-photon ones, are probably required to map out this region more thoroughly. This means that understanding \TiO\ absorption in the UV and bluer region of the visible spectra may be currently incomplete. This is of most relevance to transit spectroscopy of hot Jupiters around stars with strong UV fluxes.

\section{Conclusions}

We have collated all suitable available assigned TiO experimental data. We have 
used over 48,000 assigned transitions to produce 10,564 energy levels. These span 
11 electronic states, and 84 total rovibronic bands. 

The Supplementary Information to this paper contains three main files: 48Ti-16O.marvel.inp, which 
contains the final input data of spectroscopic transitions in \Marvel\ format, 
48Ti-16O.energies, 
which contains the sorted energies in the main component, and 
48Ti-16O\_FFN\_ca\_33.energies, which contains the relative energies in the free-floating 
network incorporating the \Sc{} $v$=3 and \Sa{} $v$=3 states. There is also three zip folders containing sorted folders and files with predicted transition frequencies using the \Marvel\ energies. 

The data collated here assists with the evaluation of the partition function for \TiO. However, there are two other electronic states, the \D{} and \SGamma{} states, which high quality theory \citep{10MiMaxx.TiO} predicts exist below 20,000 \cm{} that have not been experimentally characterised in rotationally-resolved spectra. Further, in many cases only a small number of vibrational levels have \Marvel\ data. Therefore, we will defer the detailed evaluation of an updated recommended partition function for the upcoming \TiO\ linelist paper \citep{jtTiO} that will produce an extensive spectroscopic model incorporating a large number of vibrational levels in  all low-lying electronic states of \TiO. 

The \Marvel\ energy level data is going to be immediately useful in the construction of the new ExoMol line list for TiO \citep{jtTiO}. The energy levels presented here will allow the accurate refinement of the potential energy curves and coupling constants, i.e. the spectroscopic model, in order to maximise the quality of the predicted energy levels. The refinement process is particularly important for transition metal diatomics due to the complexity of the electronic states and the insufficient accuracy of even modern \abinitio\ methods \citep{jt632}.

Finally, we note that a major part of this work was performed by 16 and 17 year old pupils from the Highams Park
School in London, as part of a project known as ORBYTS (Original Research By Young Twinkle
Students). Two other \Marvel\ studies on astronomically important molecules, methane  \citep{jtCH4Marvel}  and acetylene \citep{jtC2H2Marvel}), were undertaken as part of the same project and
will be published elsewhere. \citet{ORBYTSed} discusses our experiences of working with school
children to perform high-level research.

\section*{Acknowledgments}
We would like to thank Bob Kurucz, Mohamed Ahmed, Sheila Smith, Tim Morris, Jon Barker, Fawad Sheikh, Highams Park School and Researchers in Schools for support and helpful discussions. 

We thank Claude Amiot, Thomas DeVore, Bob Kurucz and Amanda Ross for providing data. 

This work has been supported by  the UK Science, Technology and Facilities Council (STFC) under grant ST/M001334 and the European Research Council under ERC Advanced Investigator Project 267219 and ERC grant number 320360.  The authors acknowledge the use of the UCL Legion High Performance Computing Facility (Legion@UCL), and associated support services, in the completion of this work. 

The work performed in Hungary was supported by the NKFIH (grant no. K119658). The collaboration between the London and Budapest teams received support from COST action CM1405, MOLIM: Molecules in Motion.
\bibliographystyle{apj}

\end{document}